# New insights into tackling large amplitude oscillatory shear (LAOS): an analytic LAOS approach (aLAOS)

Pengguang Wang, Jiatong Xu, Ziyu Zhou, Hongbin Zhang*

*Advanced Rheology Institute, Department of Polymer Science and Engineering, School of Chemistry and Chemical Engineering, Shanghai Jiao Tong University, Shanghai 200240, China*

*Author to whom correspondence should be addressed; email: hbzhang@sjtu.edu.cn

**Synopsis**

Large amplitude oscillatory shear (LAOS) has been widely applied for performing rheological analysis of complex fluids, and the Fourier transform (FT) rheology is considered one of the mainstays in LAOS since FT is an essential mathematic tool that is applied in numerous fields of signal analysis. However, the challenge of understanding FT, diverse mathematical frameworks, and complicated data processing render existing LAOS methods hard to be effectively used. As a result, their accessibility and implementation are limited in practice. Thus, the development of novel LAOS methods remains an unmet need. In this context, a newly analytic LAOS (aLAOS) approach was deduced here to tackle LAOS by the reverse use of the methodology of the FT rheology. More specifically, the rheological signals and measures can be precisely reconstructed based on the established equations involving Fourier coefficients instead of processing numerous discrete points of the Lissajous curves. Furthermore, considering the acquired results from the LAOS tests on the various typical yield stress fluids, it was demonstrated that the aLAOS could unify the mainstream LAOS methods into one Fourier coefficient-based framework and visually distinguish the contribution of arbitrary higher harmonics. Consequently, the proposed approach is arising as a promising candidate for the convenient treatment of LAOS, which could also highly promote the multi-field applications of LAOS.

## I. INTRODUCTION

Dynamic oscillatory shear tests are widely used to extensive study soft matter and fluids, which is crucial for the formulation, processing, and functionality of complex materials in wide fields, e.g., polymer melts and solutions, emulsions, suspensions, biomacromolecules, surfactants, dairy foods, and even foods for medical uses in dysphagia therapy [1-3]. Small amplitude oscillatory shear (SAOS) is regarded as a recognized way to investigate the linear viscoelastic behaviors of samples [4, 5]. More specifically, large and rapid deformations happen in most cases with nonlinear mechanical responses to render large amplitude oscillatory shear (LAOS) a sound approach to quantify the nonlinear viscoelastic behavior of fluids. Moreover, compared with SAOS, LAOS has a much higher resolution to distinguish fluids with similar microscopic and mesoscopic structures (e.g. linear or branched polymer topology) [6]. As a result, LAOS can provide deep insights for studying such types of structural diversities. Furthermore, distinct differences arise between the SAOS and LAOS methods in the experimental signal analysis. Thus, the development of innovative LAOS methods is necessary for thoroughly investigating nonlinear responses because of the low applicability of SAOS material functions in LAOS [6]. In addition, steady shear viscosity is helpful in polymer processing, while inherent limits exist: i) the polymer may not reach equilibrium in practice; ii) chemically cross-linked and structurally sensitive hydrogen-bonded samples are inapplicable; and iii) the analysis of the microstructure is limited and observing the elastic relaxation time scale is inhibited because of the steady state requirement [6]. Conversely, LAOS is accessible for independently changing stress/strain amplitude and frequency to permit a broad spectrum of conditions [7]. Additionally, the input signal of LAOS is smooth with no sudden jumps in speed or position to generate and control a relatively easy flow field [8].

Fourier transform (FT), which is a fundamental mathematic system for carrying out signal analysis, is universal in treating complex signals, such as images, voice, light, electromagnetic signals, and digital signals. Joseph Fourier proposed that a complex and time-dependent signal $s(t)$ can be represented in a spectrum $S(\omega)$ as the sum of infinite sinusoidal waves that are a set of orthogonal functions [9-11]. Particularly, the FT results demonstrate the periodic contributions to a time-dependent signal and present information on the amplitude and phase (or real and imaginary parts) at different frequencies. Then, the complex signal can be mathematically represented in the time domain and the frequency domain. Moreover, Fourier series provide a mathematical expression for a complicated signal, in terms of sines and cosines. FT rheology was then established and applied to quantify mechanical nonlinearities [11-13], which indicates that the oscillatory stress/strain signal can be investigated from two complementary coordinate frames [6]. The primary analytical approach is transforming the stress/strain versus time ($\sigma(t)/\gamma(t)$) in the time domain into Fourier series, where the nonlinearity of the fluids [14-16] and the yield stress determination of typical yield stress fluids (YSFs) [12, 13, 17, 18] can be evaluated by observing the ratio $I_3/I_1$ (the intensity ratio of the 3rd harmonic to the 1st harmonic) and $I_3/I_1\gamma_0^2$ (i.e. $Q$ value [19], $\gamma_0$: strain amplitude). Furthermore, the dependence of $\sigma(t)$ on strain $\gamma(t)$ and strain-rate $\dot{\gamma}(t)$ can be plotted in the deformation domain as Lissajous curves that are two-dimensional projections, which is the fundament for many LAOS methods [20-24].

A series of new and impactive LAOS methods have been established in the literature, such as FT rheology [6, 11-13], Chebyshev polynomials [22, 23], stress decomposition [20-22, 25], strain-stiffening $S$ and shear-thickening $T$ ratios [23, 24], dissipation ratio [26, 27], stress bifurcation [28], and sequence of physical processes [18, 29-33] (SPP, e.g. apparent cage modulus [18], flow curve from LAOS [18], transient moduli [34, 35], derivatives of transient moduli [32, 33], and other transient-moduli-based measures [36, 37]), which contribute to the research in diverse fields, such as cellulose nanomaterials [38, 39], foods [40], medical dysphagia therapy [3], and emulsions [41, 42]. However, it can be inferred that the wide application of LAOS methods in non-rheological fields is inhibited by certain limits [43]: i) a deep understanding of FT is still missing; ii) different

LAOS methods have different mathematical frameworks, meaning that each method has its own special data processing way; iii) thus, the data processing may be relatively difficult and time-consuming to be understood and realized; and iv) some curves deduced from LAOS methods may exist various types of noises depending on the resolution and sensitivity of the applied rheometer. Moreover, in the rheological field, the influence of higher harmonics on the extracted results from the LAOS-based methods still needs to be further interpreted and visualized. In addition, it should be noted that the FT and FT rheology have been applied in LAOS to remove noises and even harmonics to obtain smooth curves. For example, Ewoldt et al. [23] carried out the discrete FT operation and obtained smooth Lissajous and stress decomposition curves, as well as some nonlinear measures, along with the MITlaos software [24]. Moreover, Rogers et al. [44] used the FT to reduce noise in data, accompanied by the *oreo* software [45]. However, the analytical perspective still needs to be developed further, just as that in establishing analytic geometry.

Under this perspective, the present work is rooted in FT rheology since FT is mathematically rigorous and presents a set of orthogonal, periodic, derivable, and integrable functions over the infinite interval from $-\infty$ to $+\infty$ [9-11]. Considering that the mathematical descriptions of FT rheology can [11-13]: i) approximate the time-dependent, periodic, and distorted waveforms; ii) provide a specific function; iii) present a clear form containing simple sine and cosine waves; and iv) benefit the following mathematical derivations, herein, different from the traditional approach of using FT series to interpret the newly proposed measures/mathematical frameworks, the FT rheology was reversely used. Thereby, a new way was developed, called analytic LAOS approach (aLAOS), to give a new insight into LAOS and simplify the application of classic LAOS methods. The basic principle of aLAOS is using an equation involving Fourier coefficients to describe a distorted rheological signal instead of adopting discrete points of Lissajous curves to generate results. This equation can be applied to the subsequent mathematical derivations to obtain analytic solutions. LAOS tests on several YSFs were carried out to verify our aLAOS approach including Carbopol gel, Laponite suspension, xanthan gum solution, cellulose nanofiber suspension, ketchup, reflective silver paint, and yogurt. The contribution of a higher harmonic to the results from LAOS analyses was also visually demonstrated.

**II. THEORY**

In this section, the applied data are all from the oscillatory stress sweep test of 0.2 wt% Carbopol gel at 1 Hz, which are in direct accordance with the previously reported works in the literature [46-50].

**A. LAOS**

The LAOS and LAOS methods will be introduced in Sec. II A1. A certain limit in the application of LAOS methods in other fields has been put forward [43], which will be discussed in Sec. II A2.

*1. LAOS and LAOS methods*

The rheological response is linear if linear differential equations with constant coefficients can describe the correlation between stress and strain [6]. In dynamic oscillatory shear, the output sine waveform with amplitude-independent coefficients and phase shift exhibits linearity when a sinusoidal stress/strain field is applied. Nonlinear appears when a distorted output signal is observed under a sinusoidal field. Therefore, SAOS and LAOS are recognized as the perturbation and deviation of the equilibrium state to probe the linear and nonlinear viscoelasticity, respectively [6]. The SAOS functions (e.g., the storage modulus $G'(\omega)$ and loss modulus $G''(\omega)$) are influenced by the deformation frequency. For example, complex compliances and viscosities can be obtained from complex moduli [4]. In contrast, in LAOS, complex rheological responses appear and are not easily described. Thus, LAOS fascinates rheologists to propose different approaches to offer deeper insights since the first appearance of LAOS response of polymers in the 1960s [51-53].

LAOS is considered one of the most important parts of rheology to study nonlinear viscoelasticity since the applied amplitude and frequency can be independently controlled to separately adjust the flow strength and time scale [6]. In addition, LAOS takes out of the equilibrium state compared with SAOS [54]. Meanwhile, LAOS can include sequentially solidlike and liquidlike responses during a single period [18, 29-33]. Therefore, LAOS is regarded as an absorbing testing protocol for numerous industrial applications. Lissajous curves are also applied to investigate the transient rheological response of YSFs by plotting stress versus strain ($\sigma \sim \gamma$) and stress versus strain rate ($\sigma \sim \dot{\gamma}$) curves. Various methods have been proposed in the literature to describe, analyze, and explain the nonlinear viscoelastic behavior of samples via different analysis approaches [6, 18, 20-23, 25, 29-31, 54].

*2. A limit in the application of LAOS methods*

Different LAOS methods possess different rheological parameters to study and describe the category and intensity of the nonlinear viscoelastic behavior of fluids. The explanation of LAOS behavior accordingly depends on the underlying assumptions and the employed mathematical frameworks [44]. It has been indicated that the LAOS methods may have the limitation that they possess comparatively low accessibility in practical applications [43]. For example, the relationship between food LAOS behavior and sensory attribute still lacks in where almost no comprehensive work is published [43], while numerous descriptions in the literature show the urgency of performing large stress and strain research for oral processing [55-57]. This phenomenon may be attributed to the comparatively little cooperation between the engineering and food science disciplines, making thus LAOS an uncommonly used tool in food research [43]. Furthermore, although rheological codes are provided for LAOS tests to calculate parameters (e.g., MITlaos [23, 24] and *oreo* [45]), these are sometimes inapplicable depending on the rheometer brand. In most cases, MATLAB and facility with calculus and linear algebra are necessary to calculate LAOS results, which needs prior programming knowledge. Therefore, there exists a barrier between LAOS tools and fluid research because high level mathematics and programming may be secondary in many fields [43].

**B. FT rheology in aLAOS**

FT, which is an important mathematic technique widely applied in signal treatment, is ubiquitous in modern society for analyzing complex signals, such as music, voices, images, or digital medical imaging (e.g. magnetic resonance imaging [10]). Detailed information about FT and FT rheology was first discussed in Sec. II B1 including the concept and ability, the framework, and the application in LAOS (i.e. FT rheology). After that, the reverse use of FT rheology was conceived in Sec. II B2, which was the opposite of the traditional investigation approach [18, 20, 21, 23, 24, 29-33] by using FT coefficients to build the foundation for newly proposed LAOS methods.

*1. FT and FT rheology*

Generally, FT is an invertible, linear, complex transformation over the range from $-\infty$ to $+\infty$, possessing a set of orthogonal functions [6]. The results of FT contain the time domain and the frequency domain to mathematically present the treated complex signal. By providing a single input frequency $\omega$ for the 1st harmonic, the outputs from FT at frequencies besides $\omega$ are related to the nonlinearity responses from higher harmonics. The linear sum of the different waveforms in the time domain also leads to a linear sum in the frequency domain because of the mathematically linear operation of FT. FT separates a series of periodic compositions from time-dependent signals and provides both the amplitudes and phases (or real and imaginary parts) for these separated compositions with different frequencies [6].

Briefly speaking, the main abilities of FT can be described as the following statements [10]. i) FT separates random noises from the received signal to obtain valuable information. ii) FT decomposes a distorted signal into many sinusoidal waves with different frequencies and amplitudes. Then, the frequency domain of a complicated signal can be obtained with different amplitudes. iii) It should be denoted that any irregular signal or wave can be considered as the superposition of simple sine and cosine waves. In other words, an approximation of a complex signal can be realized by combining a series of sine waves with different frequencies ($\sin\omega t$, $\cos\omega t$, $\sin 3\omega t$, $\cos 3\omega t$, $\cdots$). In addition, Fourier series provide a mathematical expression for a complicated signal, in terms of sines and cosines. The incorporation of more harmonics yields a better superposition between the raw data and the FT-based result.

The Fourier series of real form on a continuous periodic function $f(x)$ with a frequency of $2\pi$ can be defined as follows [58]:

$$f(x) = \sum_{k=0}^{\infty}(a_k \cos kx + b_k \sin kx), \qquad (1)$$

$$a_k = \frac{1}{\pi}\int_{-\pi}^{\pi} f(x)\cos kx\, dx, \quad b_k = \frac{1}{\pi}\int_{-\pi}^{\pi} f(x)\sin kx\, dx, \qquad (2)$$

which is the base of FT. FT of any real or complex signal $f(x)$ (giving a frequency-dependent spectrum $\hat{f}(\omega)$ ) and the corresponding Fourier integral (calculating $f(x)$ from $\hat{f}(\omega)$ ) are described as Eq. 3 with clear mathematical forms [58]:

$$\hat{f}(\omega) = \int_{-\infty}^{\infty} f(x)e^{-i\omega x}dx, \quad f(x) = \frac{1}{2\pi}\int_{-\infty}^{\infty} \hat{f}(\omega)e^{i\omega x}d\omega. \quad (3)$$

For better understanding, FT can give the real form expansions of $f(x)$:

$$f(x) = \int_0^{\infty} (a_k \cos\omega x + b_k \sin\omega x)d\omega, \quad (4)$$

$$a_k = \frac{1}{\pi}\int_{-\infty}^{\infty} f(x)\cos\omega x dx, \quad b_k = \frac{1}{\pi}\int_{-\infty}^{\infty} f(x)\sin\omega x dx. \quad (5)$$

The half-sided FT is carried out in a semi-infinite domain when $x$ is in the range of $0 \sim +\infty$ in Eqs. 3 and 5, which is commonly applied for interpreting the results of experimental tests because data acquisition starts at a finite time.

For FT rheology, the form of Fourier series at the strain-controlled condition ($\gamma(t) = \gamma_0 \sin\omega t$) can be defined as follows [6, 11]:

$$\sigma(t) = \sum_{n=1}^{\infty} I_n \sin(n\omega t + \phi_n) = \gamma_0 \sum_{n=1}^{\infty} (G_n'' \cos n\omega t + G_n' \sin n\omega t), \quad (6)$$

where $I_n$ is $n$th harmonic magnitude, $\phi_n$ denotes phase angle, $\gamma_0$ refers to the strain amplitude, $\omega$ represents angular frequency, $t$ states time, and $n$ refers to $n$th harmonic. While both strain- and stress-controlled oscillatory shear conditions are adopted, strain-controlled tests are more common [44]. In addition, $G_1'$ and $G_1''$ can be generally considered as the storage modulus $G'$ and loss modulus $G''$ obtained from LAOS tests, respectively [6, 33]. For typical and idealized tests, only odd higher harmonics are supposed to occur because the rheological response of a sample is normally independent of the shear direction. Hence, the response must be an odd function. The emerging even harmonics are generally ascribed to the imperfection of the test [59-63], where even harmonics show relatively low amplitudes compared with the odd higher harmonics [6]. As a result, the Fourier series of FT rheology can be also described as a function of the odd harmonics:

$$\sigma(t) = \gamma_0 \sum_{n=1,odd}^{\infty} (G_n'' \cos n\omega t + G_n' \sin n\omega t). \quad (7)$$

Similarly, the Fourier series at the stress-controlled condition with a stress amplitude $\sigma_0$ and an input sinusoidal signal $\sigma = \sigma_0 \sin\omega t$, is [64]:

$$\gamma(t) = \gamma_0 \sum_{n=1}^{\infty} (a_n \cos n\omega t + b_n \sin n\omega t) \text{ or } \gamma(t) = \gamma_0 \sum_{n=1,odd}^{\infty} (a_n \cos n\omega t + b_n \sin n\omega t). \quad (8)$$

Interpolation can be used in the conversion between the distorted strain response (with a perfect input stress wave) and the distorted stress response (with a perfect input strain signal) [21, 23]. Furthermore, the Fourier series can be described as the power series by introducing Chebyshev polynomials ($x = \sin\omega t$ and $y = \cos\omega t$) [22]:

$$\begin{aligned}\sigma(t) &= \sum_{n=1}^{\infty} a_n' \cos^n \omega t + \sum_{n=0}^{\infty} b_{2n+1}' \sin^{(2n+1)} \omega t + \sin\omega t \cos\omega t \sum_{n=0}^{\infty} c_{2n}' \sin^{2n} \omega t \\ &= \sum_{n=1}^{\infty} a_n' y^n + \sum_{n=0}^{\infty} b_{2n+1}' x^{2n+1} + xy\sum_{n=0}^{\infty} c_{2n}' x^{2n}\end{aligned}. \quad (9)$$

When even harmonics are removed, the power series is as follows:

$$\sigma(t) = \sum_{n=1,odd}^{\infty} a'_n y^n + \sum_{n=1,odd}^{\infty} b'_n x^n \;. \tag{10}$$

Furthermore, the coefficients of the power series can be calculated by conducting polynomial fitting.

## 2. Reverse use of FT rheology in aLAOS

FT rheology has been widely applied to build a mathematical foundation for innovative LAOS methods [18, 20, 21, 23, 24, 29-33]. In more detail, the proposed measures are usually associated with the Fourier series and Fourier coefficients, which can be recognized as the forward use of FT rheology (decomposing a measure at the left side of an equation into terms at the right side of an equation). For example, strain-stiffening $S$ and shear-thickening $T$ ratios present the equations [23, 24]: $S = \frac{G'_L - G'_M}{G'_L} = \frac{4e_3 + \cdots}{e_1 + e_3 + \cdots}$ and $T = \frac{\eta'_L - \eta'_M}{\eta'_L} = \frac{4v_3 + \cdots}{v_1 + v_3 + \cdots}$, where $G'_M$ is the minimum-strain modulus, $G'_L$ refers to the large-strain modulus, $\eta'_M$ is the minimum-rate dynamic viscosity, $\eta'_L$ represents the large-rate dynamic viscosity, $e_n$ and $v_n$ refer to the Chebyshev coefficients. Then, the calculated $S$ and $T$ values can be decomposed into the Chebyshev coefficients.

It should be noticed that the concepts embodied in the MITlaos [23, 24] and *oreo* [45] software can be regarded as the rudiment of aLAOS. However, although the equations between the new measures and early Fourier/Chebyshev coefficients have been proposed in the literature [18, 20, 21, 23, 24, 29-33], the reverse use of these equations (combining the right side of an equation to generate the value of the measure at the left side of an equation) still lacks. Meanwhile, the influence of a higher harmonic on the results from the LAOS methods needs to be further investigated. In other words, the reverse utilization of FT rheology has been scarcely reported as a concept in the literature, i.e. taking advantage of the fact that Fourier coefficients have been given and then combining them to generate the results (e.g. $\frac{4e_3 + \cdots}{e_1 + e_3 + \cdots} = S$ and $\frac{4v_3 + \cdots}{v_1 + v_3 + \cdots} = T$).

Although the MITlaos software has had great success [23, 24], new passions need to be motivated. Here, FT rheology was used to generate a function to approximate the raw data (the output stress/strain signal), where the obtained function can be submitted to the mathematical frameworks of classic LAOS methods. It is foreseeable that for many fields, handling functions that possess trigonometric parts and Fourier coefficients is much more convenient and accessible than processing a group of many discrete points (raw data, i.e. raw Lissajous curves) by using different LAOS methods.

In addition, although FT rheology has been studied for several decades in LAOS, the influence of a higher harmonic on the results from classic LAOS methods can be further explored and visualized. By using FT rheology to offer a specific function containing the terms of all higher harmonics, the definite contribution of each higher harmonic can be described by functions when the FT-rheology-based specific function is submitted to classic LAOS methods. Thus, every higher harmonic can be investigated independently of other harmonics.

Adapting FT rheology as the fundament brings several advantages. The approximation of a distorted waveform can be realized by FT series via the superposition of a series of sine waves. The similarity between the raw data and the reconstructed results generated by FT rheology increases with the increase in the number of the introduced higher harmonics. Thus, the first advantage of applying FT rheology is that the reconstructed results are close to the raw data.

Accordingly, a complicated signal possessing a group of many points can be expressed, in terms of sines and cosines, which is the second and third advantages of applying FT rheology: the function of a complex signal can be clearly defined, and the function just contains simple trigonometric terms. To highlight the above-mentioned advantages, the series expansion from the Oldroyd 8-constant framework can be referred to, which demonstrates a complicated framework [65]. Then, the function given by FT rheology can be conveniently applied for the following mathematical derivations to generate aLAOS equations, which is the fourth advantage and the core strength of FT rheology. Therefore, FT rheology (Eqs. 6-8) was adopted as the fundament of this work.

## *3. Result*

The first step of the aLAOS approach was obtaining FT coefficients by treating the raw data from a LAOStress/LAOStrian test with FT rheology (Fig. 1). The LAOStress experiment on 0.2 wt% Carbopol gel was carried out containing sixty sampling points, where the lowest and highest stress amplitudes can be recognized as points 1 and 60, respectively. Accordingly, a series of perfect or distorted signals were obtained, and point 46 presented $G' \approx G''$. Dynamic moduli versus stress are demonstrated in Fig. 1(a) with marked points 1, 46, and 60. Then, the raw data from the stress sweep test were analyzed by the FT rheology, and the nonlinearities of all points were plotted in the $x$–$y$ projection FT fingerprint for each higher harmonic via stress versus $n$th harmonic (Fig. 1(b)). The normalized intensities of harmonics ($I_n/I_1$) are displayed in Fig. 1(b), where a visual "snapshot" of the start and increase in the harmonic intensity with the increase in the stress can be observed [66].

The normalized intensities of 3rd, 5th, and 7th harmonics ($I_3/I_1$, $I_5/I_1$, and $I_7/I_1$) versus the stress amplitude are depicted in Fig. 1(c), where point 46 of $G' \approx G''$ was denoted. The values of $I_3/I_1$, $I_5/I_1$, and $I_7/I_1$ approached zero at low stress levels. Moreover, these values gradually increased with the increase in the stress amplitude, which is directly associated with the structural transformation in the sample (in other words, the intracycle yield process within a Lissajous loop possesses many significantly higher harmonics [18]). After that, at high stress amplitudes, the Carbopol gel became a generally pure viscous fluid along with the decreased values of $I_3/I_1$, $I_5/I_1$, and $I_7/I_1$. The start of the growth for $I_3/I_1$ happened when $G'$ remained constant, and the peak of $I_3/I_1$ occurred after the crossover point of $G'$ and $G''$. Thus, the intensities of $I_3/I_1$, $I_5/I_1$, and $I_7/I_1$ were hard to be applied in the yield stress determination.

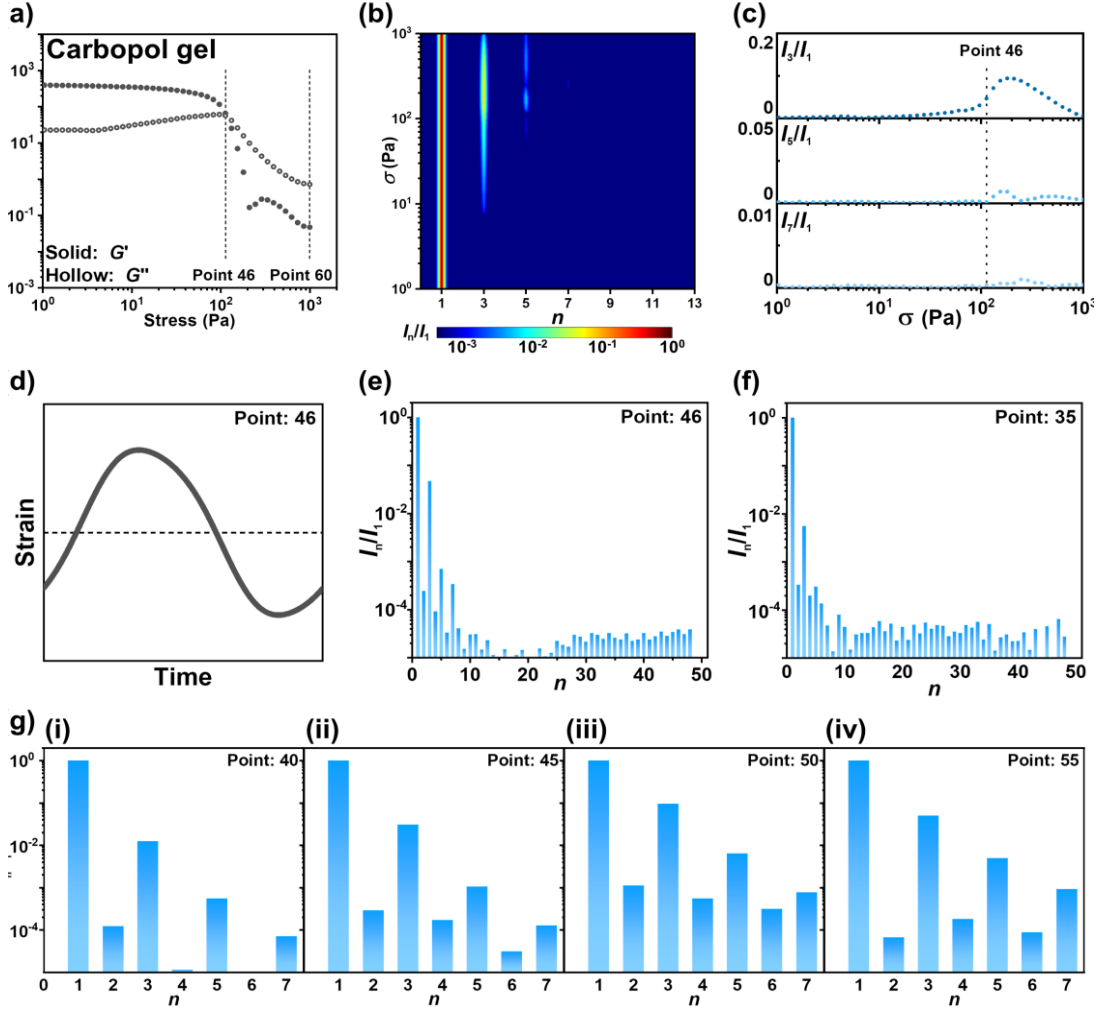

FIG. 1 Results from FT rheology analysis using the data from the oscillatory stress sweep test of 0.2 wt% Carbopol gel at 1 Hz, which will be applied for all afterward discussions in Sec. II (the first step of the aLAOS approach): (a) dynamic moduli of the amplitude sweep test; (b) stress-frequency projection ($\sigma$ versus ($n \cdot \omega$) plot); (c) the plot of normalized $I_3/I_1$, $I_5/I_1$, and $I_7/I_1$ values versus stress amplitude; (d) the rheological strain response by inputting a sinusoidal stress signal at point 46 of $G' \approx G''$ and (e) the corresponding FT spectrum ($I_1$~$I_{48}$) showing $I_n/I_1$ as a function of $n$th harmonic; (f) the FT spectrum ($I_1$~$I_{48}$) of point 35 in the SAOS region; and (g) the FT spectra ($I_1$~$I_7$) of (i) point 40, (ii) point 45, (iii) point 50, and (iv) point 55.

The whole stress sweep region was studied (Figs. 1(a)-1(c)), and point 46 of $G' \approx G''$ as a special point is adopted for further analysis and discussion (Figs. 1(d) and 1(e)). At point 46, when a perfect sinusoidal stress signal was input, a distorted strain signal was observed (Fig. 1(d), indicating the appeared nonlinearity and structural transformation of the sample. A perfect sine wave presents a single 1st harmonic in the frequency domain. However, the distorted sine wave in Fig. 1(d) generated infinitely higher harmonics. It is obvious that the contribution of a harmonic to the nonlinearity decreased with the increase in $n$th between $I_1$ and $I_9$. In addition, even harmonics are associated with the experiment's imperfection [59-63]. Meanwhile, the intensities of the even harmonics were smaller than odd harmonics, which is recognized as one of the typical properties of FT rheology. Point 35 in the SAOS region was additionally analyzed to give the results plotted in Fig. 1(e). Compared with point 46 in the nonlinear region, point 35 possessed much lower intensities of $I_3/I_1$, $I_5/I_1$, and $I_7/I_1$, which showed the linear behavior in the SAOS region.

Figure 1(g) shows in detail the change in the nonlinearity with the increase in the stress amplitude from point 40 (before yielding) to point 55 (after yielding). The values of $I_3/I_1$ and $I_5/I_1$ first increased and then decreased from point 40 to point 55, while $I_7/I_1$ increased as the stress amplitude increased.

More information on analyzing different samples by using FT rheology is provided in Appendix A.

*4. Summary*

Briefly, in Sec. II B, the superior abilities of FT and FT rheology were introduced. The reverse use of FT rheology was thus proposed.

FT is a mathematically rigorous system and possesses a set of orthogonal, periodic, derivable, and integrable functions over the infinite interval from $-\infty$ to $+\infty$ [9-11]. The FT rheology has been widely applied to study the nonlinearity of complex fluids as one of the pillars of LAOS. Thus, FT and FT rheology endow the aLAOS approach with a solid foundation and great convenience. The comparative advantages of applying FT rheology can be expressed as the following statements: i) the reconstructed signal is close to the raw data; ii) a group of many discrete points can be described by a function containing simple terms; iii) the function can be easily applied in following mathematical derivations.

The Fourier coefficients from FT rheology analysis were obtained by processing the raw data from an oscillatory stress sweep test, which is the first step of the aLAOS approach. In the next section, the results will be applied to reconstructing Lissajous curves: the second step of the aLAOS approach.

**C. aLAOS in Lissajous curve**

The Lissajous curve will be introduced in Sec. II C1. Then, raw elastic and viscous Lissajous curves, as well as reconstructed elastic and viscous Lissajous curves, will be demonstrated and discussed in detail in Sec. II C2 including the principle of analytic Lissajous curve (Sec. II C2.1), the situation of point 46 presenting $G' \approx G''$ (Secs. II C2.2 and II C2.4), the visualization of harmonic contributions (Sec. II C2.3 and II C2.4), the strain- and stress-controlled conditions (Sec. II C2.5), and the whole stress sweep range (Sec. II C2.6).

*1. Lissajous curve in LAOS*

In Figs. 2 and 3, the blue lines are the representative Lissajous curves based on point 46 at $G' \approx G''$. The Lissajous curve method presents an approach to visually demonstrate and explain nonlinear viscoelastic behavior by treating the raw data from LAOS tests [40]. The Lissajous curves can be divided into elastic and viscous Lissajous curves, which correspond to the plots of the stress versus strain $\sigma \sim \gamma$ and stress versus strain rate $\sigma \sim \dot{\gamma}$ [27]. On top of that, the Lissajous loops are recognized as the most basic and cogent method of investigating the nonlinearity of complex fluid through the emerging deviations from ellipse [67], which provides graphical representations for rapid qualitative evaluation [6]. These inherently visual plots provide qualitative discussions for quantitative material parameters, such as $S$ and $T$ ratios, as well as Fourier and Chebyshev coefficients. By changing strain amplitude and frequency, a series of elastic/viscous Lissajous curves are plotted together to generate a Pipkin diagram. After that, comparisons among these elastic/viscous Lissajous curves permit the evaluation of the nonlinear rheological response from the aspects of the relative extent and type based on the evolution of waveform shape [43].

The nonlinear LAOS behavior can be also plotted in a three-dimensional space, where the loop contains three axes: $\gamma$, $\dot{\gamma}$, and $\sigma$. The three-dimensional curve has three projections on the $\sigma \sim \gamma$, $\sigma \sim \dot{\gamma}$, and $\gamma \sim \dot{\gamma}$ planes. The projection on the $\sigma \sim \gamma$ plane is the elastic Lissajous curve and the $\sigma \sim \dot{\gamma}$ plane displays the viscous Lissajous curve, where the two projections can be compared with the purely elastic response and purely viscous response, respectively [68]. The intracycle information is also provided from the projections. For example, the $\sigma \sim \gamma$ loop area is the energy dissipation per unit volume per cycle and the $\sigma \sim \dot{\gamma}$ loop can collapse to the steady shear curve at low frequency. Furthermore, since the qualitative feature of the Lissajous curve analysis, parameters are proposed for quantitative evaluation as further LAOS behavior indicators.

## *2. Analytic Lissajous curve*
### *2.1. Principle*

Considering that the FT rheology has the ability to reconstruct a signal similar to the raw signal by using an explicit and simple function, the analytic Lissajous curve method can be further developed and submitted to different LAOS methods to offer a new perspective. Although the concept in MITlaos software [69] has shown the rudiment of aLAOS by introducing $I_3$, the influence of higher harmonics on the reconstructed Lissajous curves remains elusive. In addition, for example, introducing only 1st and 3rd harmonics will be proved insufficient for SPP by this work. It is foreseeable that major Lissajous-curve-based methods can be treated by aLAOS.

The reverse use of the FT rheology in the Lissajous curve (analytic Lissajous curve) to generate reconstructed Lissajous curves will be introduced. First, for the strain-controlled condition, the analytic Lissajous curve can be expressed as follows:

$$\sigma(t)=\gamma_0\sum_{n=1}^{\infty}(G_n''\cos n\omega t + G_n'\sin n\omega t), \quad \gamma(t)=\gamma_0\sin\omega t, \quad \dot{\gamma}(t)=\gamma_0\omega\cos\omega t. \tag{11}$$

For the stress-controlled condition, the functions are as follows:

$$\sigma(t)=\sigma_0\sin\omega t, \quad \gamma(t)=\gamma_0\sum_{n=1}^{\infty}(a_n\cos n\omega t + b_n\sin n\omega t), \quad \dot{\gamma}(t)=\gamma_0 n\omega\sum_{n=1}^{\infty}(-a_n\sin n\omega t + b_n\cos n\omega t). \tag{12}$$

Accordingly, a series of reconstructed elastic and viscous Lissajous curves can be generated by introducing different numbers of higher harmonics to plot $\sigma \sim \gamma$ and $\sigma \sim \dot{\gamma}$ curves. Figures 2 and 3 show the reconstructed Lissajous curves at the strain- and stress-controlled conditions, respectively. The raw Lissajous curves (blue lines) were plotted with the reconstructed Lissajous curves (red lines) for comparison.

First, the raw data were treated by FT rheology to obtain the Fourier coefficients. Then, the Fourier coefficients were submitted to the functions (Eqs. 11 and 12) to generate the reconstructed Lissajous curves. Thus, a series of reconstructed Lissajous curves were constructed by introducing different numbers of higher harmonics (e.g. $I_1$, $I_1$~$I_3$, $I_1$~$I_5$, $I_1$~$I_7$, $I_1$~$I_9$, and $I_1$~$I_{11}$, i.e. $G'_1$, $G''_1$, $G'_3$, $G''_3$, ⋯ ). Lastly, the contribution of each harmonic was separately calculated and visualized. For example, in Fig. 2, the purple lines are the visualized contributions of odd higher harmonics to the corresponding reconstructed Lissajous curves, and the green lines correspond to even higher harmonics. The equations used to show the contributions of harmonics were correspondingly denoted with the same color. The visualized contributions were translated and amplified for clarity, where the translation and amplification factors were demonstrated. The gray lines denote the lines that have been shown, and only the newly introduced harmonics are colored purple or green in the corresponding panel.

## 2.2. A sweep point at the strain-controlled condition

Figure 2 displays the results at the strain-controlled condition with a nonlinear stress signal and a perfect strain signal ($\gamma(t) = \gamma_0 \sin \omega t$ and $\dot{\gamma}(t) = \gamma_0 \omega \cos \omega t$) including elastic Lissajous curves $\sigma \sim \gamma$ (Fig. 2(a)) and viscous Lissajous curves $\sigma \sim \dot{\gamma}$ (Fig. 2(b)).

More specifically, the blue line in Fig. 2(a) (raw elastic Lissajous curve) shows the appeared nonlinearity of point 46 at $G' \approx G''$ through the emerging deviations from the ellipse to give a graphical representation for the qualitative evaluation of the structural transformation. The arisen nonlinearity can be interpreted by the cage model from SPP [18]. First, beginning at zero stress, a slightly increased strain caused the linear elastic cage deformation. Then, the modulus decreased with the increase in the strain, indicating the strain-softening of the sample. Finally, until the instantaneous strain rate reached zero ($\gamma = \gamma_0$), the cage instantaneously reformed. When the applied stress/strain amplitude was big enough, the cage broke, and static yielding occurred. Once yielded, the steady shear flow began.

The red lines in Fig. 2(a) indicate that by introducing only $I_1$, the reconstructed elastic Lissajous curve presents a standard ellipse shape and is similar to the linear viscoelastic behaviors in SAOS. Meanwhile, the deviation arose between the raw and reconstructed elastic Lissajous curves. Therefore, the reconstructed elastic Lissajous curve based on only $I_1$ failed to describe the nonlinearity. By adding $I_2$ and $I_3$, the extent of deviation decreased, reflecting that the nonlinearity was approximately expressed based on $I_1$~$I_3$, which can be regarded as one of the keys of MITlaos [23, 24].

A similar phenomenon happened in Fig. 2(b). The reconstructed viscous Lissajous curve exhibited a standard ellipse shape by introducing only $I_1$ and was distorted by introducing higher harmonics. Meanwhile, the proximity between the raw and reconstructed viscous Lissajous curves increased with the increase in the number of the added higher harmonics.

To sum up, the FT-rheology-based aLAOS approach is accessible for reconstructing elastic and viscous Lissajous curves. The reconstructed Lissajous curves can show high similarities to raw Lissajous curves at the strain-controlled condition by introducing $I_1$ and $I_3$.

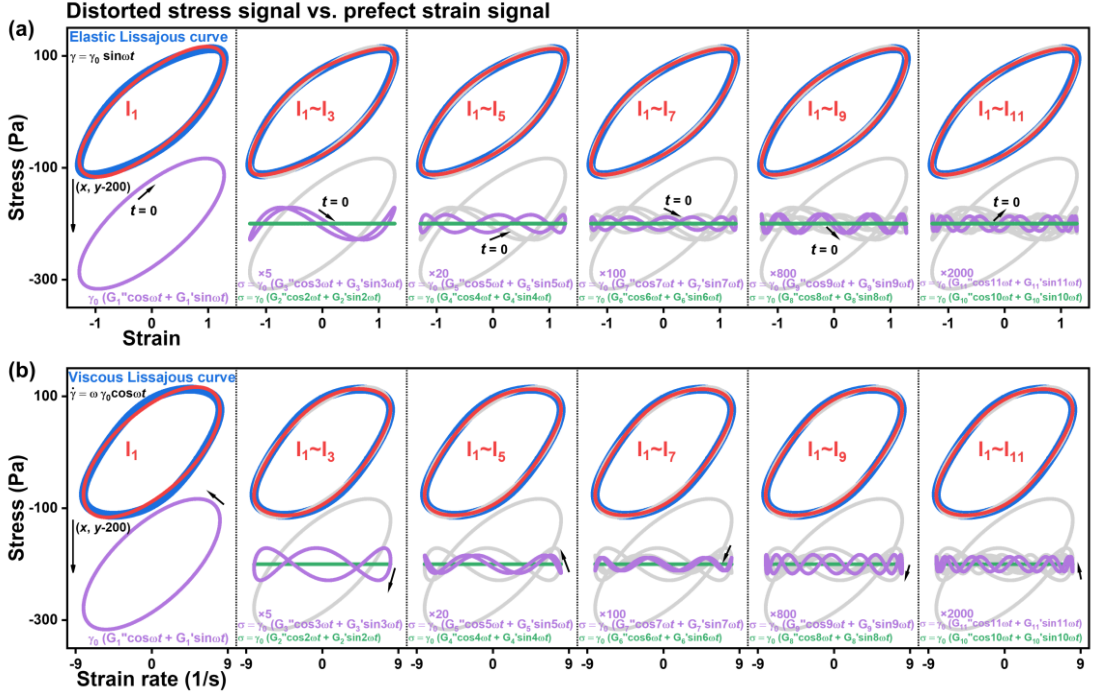

FIG. 2 The comparison between the raw and reconstructed Lissajous curves at point 46 ($G' \approx G''$, strain-controlled condition): (a) elastic Lissajous curves $\sigma \sim \gamma$ and (b) viscous Lissajous curves $\sigma \sim \dot{\gamma}$. Blue lines: raw Lissajous curves. Red lines: reconstructed Lissajous curves. Purple lines: visualized contributions of odd harmonics. Green lines: visualized contributions of even harmonics. Gray lines: the lines that have been shown.

## 2.3. Visualized harmonic contribution

In the first and second panels of Fig. 2, the demonstrated purple lines of $I_1$ and $I_3$ visually indicated the corresponding contributions to the reconstructed elastic (Fig. 2(a)) and viscous (Fig. 2(b)) Lissajous curves, where $I_1$ provides the dominate contribution and is much higher than that of $I_3$. In addition, the contribution of $I_2$ was negligible because the even harmonics showed the dependence of the rheological responses on the shear direction that was impossible for most fluids, which can be attributed to the imperfection of the teat [59-63]. Therefore, even harmonics presented low amplitudes [6]. When $I_4$ and $I_5$ were introduced, the reconstructed elastic and viscous Lissajous curves were slightly changed, and the visualized contributions of $I_5$ (the purple lines in the third panels of Fig. 2(a) and Fig. 2(b)) were lower than $I_3$. This indicates the crucial contribution of $I_3$ to nonlinearity. In addition, the contribution of higher harmonics decreased with the increase in the order. Likewise, a high similarity was realized between the raw and $I_1$~$I_5$-based Lissajous curves.

The detailed contributions of higher harmonics to reconstructed elastic and viscous Lissajous curves were also visually displayed. The impact of higher harmonics on the reconstructed Lissajous curves varied with the intracycle strain and strain rate regions, where several positions in the top half of the reconstructed elastic and viscous Lissajous curves were focused on. These positions were divided into two parts. More specifically, one part in the reconstructed elastic Lissajous curve (the point of maximum strain, near maximum strain region, and the point of zero strain) and the other part in the reconstructed viscous Lissajous curve (the point of maximum strain rate, near maximum strain rate region, and the point of zero strain rate).

In Fig. 2(a), compared with $I_1$, $I_3$ increased the intensity of the maximum strain point and maintained the zero strain point approximately constant. For the near maximum strain region, $I_3$ first decreased and then increased the intensity of the reconstructed elastic Lissajous curve when the strain was swept from zero strain to the maximum strain. As a result, the $I_1$-based elastic Lissajous curve became closer to the raw elastic Lissajous curve by introducing $I_3$. Furthermore, $I_5$ presented no contribution to the maximum strain point and decreased the intensity of the zero strain point. For the near maximum strain region, the contribution of $I_5$ was also clearly distinguished in the third panel. In a word, $I_3$ and $I_5$ induced the smooth upper right corner of the $I_1$-based ellipse close to a spindle shape and adjusted the maximum and zero strain points approximate to the corresponding two points in the raw elastic Lissajous curve.

Similarly, in Fig. 2(b), the contribution of $I_3$ did not change the intensity of the maximum strain rate point and elevated the zero strain rate point. In the near maximum strain rate region, $I_3$ first increased and then decreased the intensity of the curve when the strain rate was swept from the zero strain rate to the maximum strain rate. Therefore, the $I_1$~$I_3$-based viscous Lissajous curve was closer to the raw curve than the $I_1$-based curve. In the same way, $I_5$ generally maintained the zero strain rate point unchanged and decreased the intensity of the maximum strain rate point. Thus, $I_3$ and $I_5$ adjusted the $I_1$-based elliptic viscous Lissajous curve closer to a parallelogram, along with more precise coordinates for the maximum and zero strain rate points.

Briefly, the aLAOS approach can specifically define, precisely calculate, and visually demonstrate the contributions of higher harmonics to the reconstructed Lissajous curves. On top of that, aLAOS provides another perspective for understanding Lissajous curves.

### *2.4. A sweep point at a stress-controlled condition*

Figure 3 presents the situation at the stress-controlled condition presenting a perfect stress signal ($\sigma(t) = \sigma_0 \sin \omega t$) and a distorted strain signal including raw and reconstructed elastic Lissajous curves (Fig. 3(a)), as well as raw and reconstructed viscous Lissajous curves (Fig. 3(b)). Similar to the conclusion given in Fig. 2, the red lines in Fig. 3 exhibited standard ellipse shapes by introducing $I_1$ and are distorted by adding higher harmonics. The reconstructed Lissajous curves become closer to the corresponding raw Lissajous curves with the increase in the number of the higher harmonics introduced. A high similarity between the raw and the reconstructed Lissajous curves was achieved by introducing $I_1$~$I_3$.

Similarly, the contributions of harmonics to the reconstructed Lissajous curves at the stress-controlled condition were also analyzed. The contributions were analyzed by observing several positions including the maximum strain and strain rate points, maximum and zero stress points, and near maximum stress region.

Briefly, in Fig. 3, $I_1$ provided two elliptic shapes for reconstructing Lissajous curves. Particularly, in Fig. 3(a), $I_3$ shifted the maximum strain point to the right and maintained the zero stress point unchanged. Meanwhile, $I_3$ offset the maximum stress point and the curve in the near maximum stress region to the left. $I_5$ had the same influence on the maximum strain and stress points, as well as the zero stress point as $I_3$, whereas $I_5$ first shifted the curve in the near maximum stress region to the left and then to the right when the strain was swept from the maximum stress point to the zero stress point. In Fig. 3(b), $I_3$ offset the maximum strain rate point and zero stress point to the right. Meanwhile, $I_3$ remained the maximum stress point constant. $I_3$ first offset the curve in the near maximum stress region to the left and then to the right during the downward strain rate sweep. $I_5$ influenced also the maximum strain rate and stress points the same as $I_3$, while the zero stress point was shifted to the left. The contribution of $I_5$ to the curve in the near maximum stress region was further demonstrated. As a result, $I_3$ and $I_5$ adjusted the $I_1$-based elliptic curves closer to a spindle shape and a parallelogram shape for the reconstructed elastic and viscous Lissajous curves, respectively. Meanwhile, several special points in the reconstructed Lissajous curves were adjusted closer to those in the raw ones.

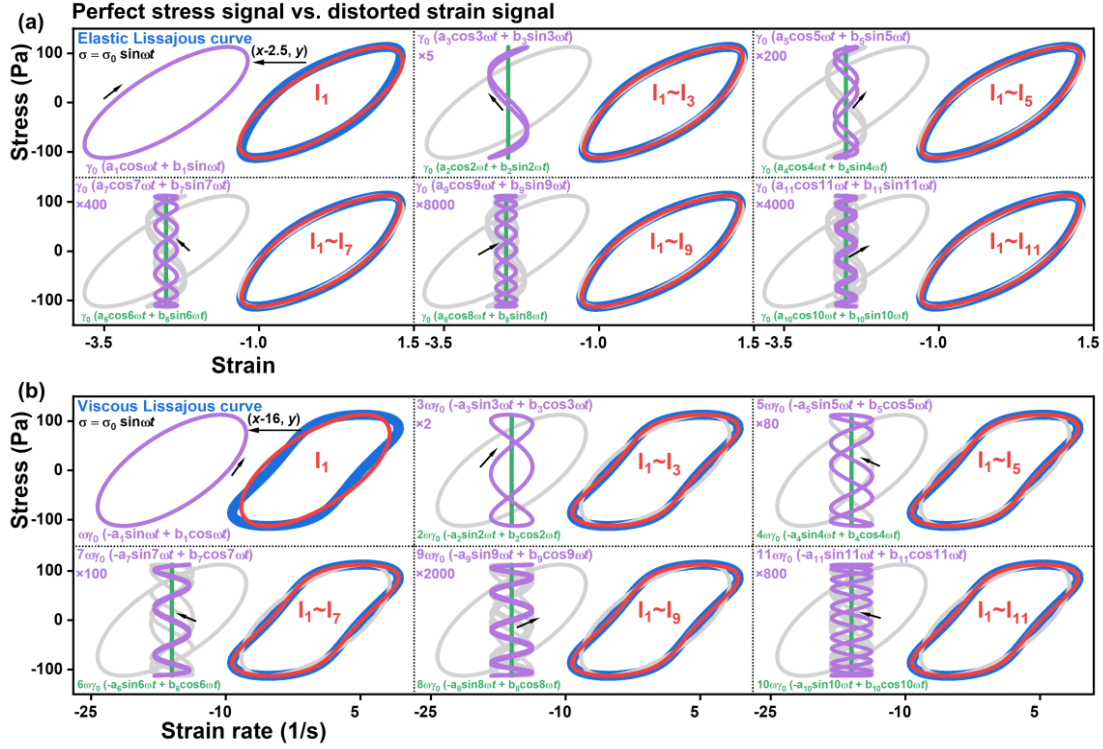

FIG. 3 Comparison between the raw and reconstructed Lissajous curves at point 46 ($G' \approx G''$, stress-controlled condition): (a) elastic Lissajous curves $\sigma \sim \gamma$ and (b) viscous Lissajous curves $\sigma \sim \dot{\gamma}$. Blue lines: raw Lissajous curves. Red lines: reconstructed Lissajous curves. Purple lines: visualized contributions of odd harmonics. Green lines: visualized contributions of even harmonics. Gray lines: the lines that have been shown.

## *2.5. Strain- and stress-controlled conditions*

While both strain- and stress-controlled oscillatory shear conditions are applied, strain-controlled tests are more common [44]. In Fig. 2, interpolation was used to translate the raw data at the stress-controlled condition to the data at the strain-controlled condition to generate comparable results.

Although the rheological response of the materials is surely associated with the stress/shear history [70, 71], the impact of strain- and stress-controlled conditions on the Lissajous curve shapes seem rarely mentioned in rheological reviews [54, 72-75]. In addition, except for the raised extra peaks, it has been demonstrated in the literature that the difference between the stress- and strain-controlled conditions can be ascribed to a high similarity between the two kinds of elastic Lissajous curves [76, 77]. Therefore, in this work, the stress- and strain-controlled conditions were generally considered to have no significant impact on the raw elastic Lissajous curve.

Meanwhile, based on Eqs. 11 and 12, aLAOS is independent of the strain- or stress-controlled condition. Both the strain- and stress-controlled tests on a sample can be carried out to generate precise results. However, investigating a sample at both the strain- and stress-controlled conditions will undoubtedly increase the complexity, which is the opposite of the purpose of this work: simplifying classic LAOS methods and promoting the application of LAOS methods in many fields.

Meanwhile, by using interpolation, comparable Lissajous curves at the strain/stress-controlled condition can be generated with equivalent stress and strain amplitudes, which is difficult when tests at both the strain- and stress-controlled conditions are carried out. Therefore, interpolation was applied in this work to generate comparable results at both the strain- and stress-controlled conditions.

Furthermore, the differences between the viscous Lissajous curves (Figs. 2(b) and 3(b)) can be explained based on Eqs. 11 and 12. The strain-controlled condition presents a perfect sinusoidal strain signal, reflecting that the derivative of the strain signal with time, the strain rate, has a perfect cosinoidal waveform (Eq. 11). Therefore, the viscous Lissajous curves are the plots of the distorted stress signal versus perfect cosinoidal wave. By contrast, at the stress-controlled condition, a sinusoidal stress signal was input, and a distorted strain signal was observed, leading to a more distorted strain rate waveform (Eq. 12). In other words, the derivative of the strain with time, the strain rate, amplifies the influence of nonlinearity (higher harmonics). Then, the viscous Lissajous curves are the plots of the stress signal versus the strain rate signal with a significantly distorted waveform. Briefly, the viscous Lissajous curves at the stress-controlled condition become more deviated from the elliptic shape than those at the strain-controlled condition, which originates from the amplification of the nonlinearity during taking derivatives.

### 2.6. Whole sweep process

Furthermore, a series of points of the stress sweep tests were analyzed, where the results are plotted in Fig. 4. For ease of comparison, the normalization of each curve was carried out. The raw/reconstructed Lissajous curves were arranged across the two-dimensional parameter space of curve type and point number (i.e. stress amplitude). The curve types were clearly denoted at the bottom of Fig. 4.

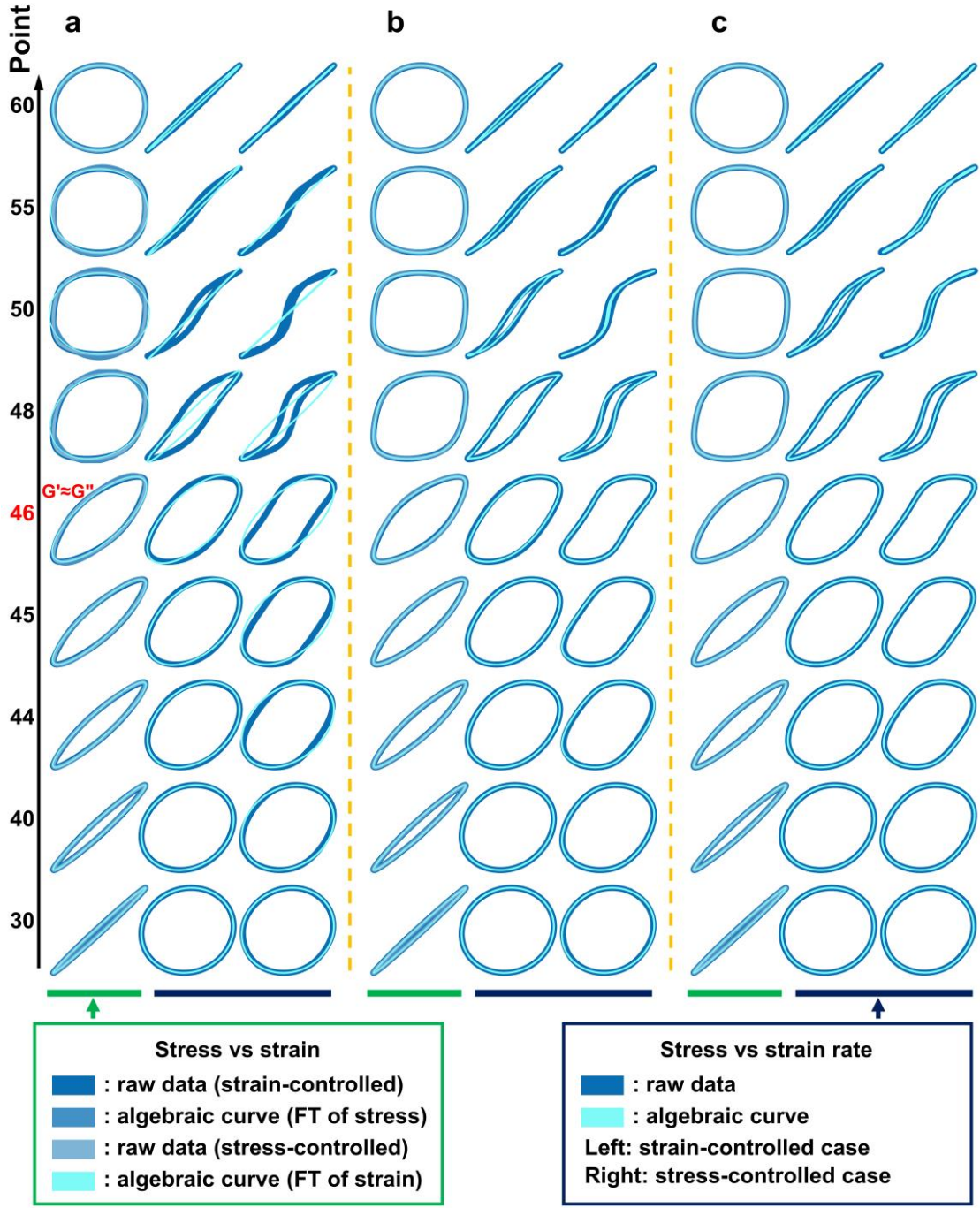

FIG. 4 Comparison between the normalized raw and reconstructed Lissajous curves at different stress sweep points. (a) Raw and reconstructed Lissajous curves by introducing $I_1$: left column: elastic Lissajous curves at the stress- and strain-controlled conditions; middle column: viscous Lissajous curves at the strain-controlled condition; right column: viscous Lissajous curves at the stress-controlled condition. Raw and reconstructed Lissajous curves by introducing (b) $I_1$~$I_3$ and (c) $I_1$~$I_{11}$. Red number: the point of $G' \approx G''$.

The raw Lissajous curves are shown in Figs. 4a, 4b, and 4c. Figures. 4a, 4b, and 4c also depict the $I_1$-based, $I_1$~$I_3$-based, and $I_1$~$I_{11}$-based Lissajous curves, respectively. In more detail, for example, the left column in Fig. 4(a) contains four types of elastic Lissajous curves: the raw curves at the strain-controlled condition, the $I_1$-based curves based on the FT of distorted stress signals, the raw curves at the stress-controlled condition, and the $I_1$-based curves based on the FT of distorted strain signals. Moreover, the middle column in Fig. 4(a) consists of two types of viscous Lissajous curves, namely the raw and $I_1$-based curves at the strain-controlled condition. Lastly, the right column in Fig. 4(a) includes the raw and $I_1$-based viscous Lissajous curves at the stress-controlled condition.

The rheological behavior was linear viscoelastic at low point numbers, as demonstrated by the elliptic shapes of the Lissajous curves. Then, the openings of these ellipses gradually became larger and smaller for the elastic and viscous Lissajous curves, respectively, along with the increasingly nonlinear and pseudoplastic rheological responses. The beginning of nonlinearity can be readily observed by visual tests from the gradually distorted elliptic shape with the increase in the point number. When the stress amplitude was further raised, the elastic Lissajous curve became more rectangular, possessing rounded corners and then circular. The viscous Lissajous curves happened shear thinning at high stress level. Therefore, 0.2 wt% Carbopol gel is a shear-thinning viscoelastic fluid.

As can be observed in Fig. 4a, the $I_1$-based Lissajous curves correspond well with the raw Lissajous curves at the SAOS region, indicating the linear rheological response. At extremely high stress amplitude (point 60), this sample behaved like a purely viscous material. Meanwhile, the raw and reconstructed Lissajous curves were similar to each other, where the elastic and viscous Lissajous curves present circular and linear shapes, respectively. However, deviations between the raw and reconstructed Lissajous curves are shown between points 44 and 55. Thus, the $I_1$-based Lissajous curves can accurately mimic the rheological responses at low and extremely high stress levels.

In Fig. 4(b), the $I_1$~$I_3$-based Lissajous curves show high similarities to the raw Lissajous curves throughout the whole stress sweep range, no matter at the strain-controlled condition or the stress-controlled condition, demonstrating the dominant contribution of $I_3$ to the nonlinearity. Small deviations between the raw and reconstructed Lissajous curves of points 48 and 50 arose, which were addressed by introducing more higher harmonics. In Fig 4(c), the deviations of all the Lissajous curves are negligible. As a result, a series of raw Lissajous curves were successfully reconstructed by introducing 1st~11th harmonics via aLAOS approach. Furthermore, although the visualization of higher harmonic contributions was not carried out in Fig 4, this can be done for each stress sweep point the same as that in Sec. II C2.3.

In brief, in this section, a series of representative points were systematically analyzed by the aLAOS approach. The Lissajous curves were reconstructed based on $I_1$, $I_1$~$I_3$, and $I_1$~$I_{11}$, which were thoroughly compared with the raw Lissajous curves. $I_1$~$I_3$-based Lissajous curves can approximate the raw Lissajous curves throughout the whole stress amplitude range, while $I_1$~$I_{11}$-based Lissajous curves almost restore the original rheological responses. Meanwhile, the aLAOS approach was proven to be accessible to strain- and stress-controlled conditions.

Appendix B provided more information on analyzing different samples by using the aLAOS approach in the Lissajous curves.

*3. Summary*

The Lissajous curve and the application of aLAOS approach in the Lissajous curve were introduced and discussed in Sec. II C. The principle of aLAOS in the Lissajous curve was first proposed in Sec. II C2.1 (Eqs. 11 and 12) based on the FT rheology. Then, in Figs. 2 and 3, a series of reconstructed Lissajous curves were discussed. The $I_1$~$I_{11}$-based Lissajous curves showed high similarities to the raw Lissajous curves at both the strain- and stress-controlled conditions (Secs. II C2.2 and 2.4). The contributions of higher harmonics to the reconstructed Lissajous curves were defined, visually demonstrated, and discussed in Secs. II C2.3 and 2.4.

Then, the difference between the strain- and stress-controlled conditions was explained in Sec. II C2.5. Interpolation was thus applied to translate the data at the stress-controlled condition to the data at the strain-controlled condition since the aLAOS approach was proposed to simplify the LAOS analyses. Therefore, the strain/stress-controlled condition was not distinguished in this work.

Furthermore, a series of stress sweep points were analyzed by introducing $I_1$, $I_1$~$I_3$, and $I_1$~$I_{11}$. $I_1$~$I_{11}$-based Lissajous curves almost reproduced the original rheological responses. Thus, the aLAOS approach is not just capable of one specific point but is universal throughout the whole range of an oscillatory stress sweep test. Meanwhile, both the strain- and stress-controlled conditions were discussed and proved applicable for reconstructing Lissajous curves.

In a word, the aLAOS approach is accessible for reconstructing raw Lissajous curves using a few equations and visually showing the harmonic contributions based on Eqs. 11 and 12. Thus, the aLAOS approach can replace the raw data with a group of many discrete points by using continuous, smooth, derivable, and integrable functions (Eqs. 11 and 12).

**D. aLAOS in stress decomposition**
*1. Stress decomposition*
*1.1. Principle*

In the linear region, $G'$ and $G''$ represent the elastic response and the viscous energy dissipation, respectively, while the stress is difficult to be unambiguously decomposed in LAOS [20]. As was mentioned above, the FT rheology presents the frequency-domain and time-domain representations of the rheological response to describe nonlinear viscoelastic behaviors via intensity and phase angle of higher harmonics. It does not directly give the function of the input stress/strain waveform to the output strain/stress signal [20].

Cho et al. [20] proposed an approach to decompose the distorted oscillatory stress signal $\sigma(\gamma,\dot{\gamma}/\omega)$ into elastic and viscous stresses:

$$\sigma(\gamma,\dot{\gamma}/\omega) = \frac{\sigma(\gamma,\dot{\gamma}/\omega) - \sigma(-\gamma,\dot{\gamma}/\omega)}{2} + \frac{\sigma(\gamma,\dot{\gamma}/\omega) - \sigma(\gamma,-\dot{\gamma}/\omega)}{2}. \tag{13}$$

Here, the strain signal of $\gamma(t)=\gamma_0 \sin \omega t$ and $\dot{\gamma}(t)=\gamma_0 \omega \cos \omega t$ are input. As a result, the total stress is recognized as the superposition of the elastic part $\sigma'(\gamma,\gamma_0)$ and viscous part $\sigma''(\dot{\gamma}/\omega,\gamma_0)$. The corresponding forms of $\sigma'(\gamma,\gamma_0)$ and $\sigma''(\dot{\gamma}/\omega,\gamma_0)$ are as follows:

$$\sigma'(x,\gamma_0)=\frac{\sigma(\gamma,\dot{\gamma}/\omega)-\sigma(-\gamma,\dot{\gamma}/\omega)}{2}, \ \sigma''(y,\gamma_0)=\frac{\sigma(\gamma,\dot{\gamma}/\omega)-\sigma(\gamma,-\dot{\gamma}/\omega)}{2}. \qquad (14)$$

Then, we have:

$$\oint \sigma'(\gamma,\gamma_0)d\gamma=0, \ \oint \sigma''(\dot{\gamma}/\omega,\gamma_0)d\dot{\gamma}/\omega=0. \qquad (15)$$

It is noted that $\sigma'(\gamma,\gamma_0)$ is an odd function for $\gamma$ and even for $\dot{\gamma}$, whereas $\sigma''(\dot{\gamma}/\omega,\gamma_0)$ is the opposite. Stress decomposition is applicable in the strain-controlled condition since each point ($\sigma(\gamma,\dot{\gamma}/\omega)$) has the partners ($\sigma(-\gamma,\dot{\gamma}/\omega)$ and $\sigma(\gamma,-\dot{\gamma}/\omega)$). On the contrary, in the stress-controlled condition, the partner disappears, which can be solved by using interpolation [21, 23].

Yu et al. [21] present the other description of stress decomposition, which is generally equal to the stress decomposition of Cho et al.:

$$\sigma'(\gamma,\gamma_0)=\frac{\sigma(\gamma,\dot{\gamma}/\omega)+\sigma(\gamma,-\dot{\gamma}/\omega)}{2}, \ \sigma''(\dot{\gamma}/\omega,\gamma_0)=\frac{\sigma(\gamma,\dot{\gamma}/\omega)+\sigma(-\gamma,\dot{\gamma}/\omega)}{2}. \qquad (16)$$

The assumption applied for the two methods [20, 21]:

$$\sigma'(\gamma,\gamma_0)=\sum_{n=1,odd}^{\infty} q'_n \gamma^n, \ \sigma''(\dot{\gamma}/\omega,\gamma_0)=\sum_{n=1,odd}^{\infty} p'_n \dot{\gamma}/\omega^n. \qquad (17)$$

Actually, $q'_n=b'_n$ and $p'_n=a'_n$ hold, where $a'_n$ and $b'_n$ are the coefficients of the power series (Eq. 10) by using Chebyshev polynomials.

Cho et al. [20] and Yu et al. [21] have described the relationship between the stress decomposition and Chebyshev coefficients. In the next two sections, the correlation between stress decomposition and Fourier coefficients, as well as the relationship between the two types of stress decompositions, will be discussed.

### 1.2. First type of stress decomposition and Fourier coefficients

First, the stress decomposition of Cho et al. [20] was proved to be closely associated with FT rheology and directly depended on the Fourier coefficients. At the strain-controlled condition, FT rheology gives the following expression of a distorted stress signal:

$$\sigma(t)=\gamma_0 \sum_{n=1}^{\infty}(G''_n \cos n\omega t + G'_n \sin n\omega t). \qquad (18)$$

For the elastic part $\sigma'(\gamma,\gamma_0)$, two angels should be found, $\theta_1$ and $\theta_2$ ($\theta=\omega t$), to satisfy the following situation:

$$\gamma_{\theta 1}=\gamma(\theta_1)=\gamma_0 \sin \theta_1 = -\gamma_0 \sin \theta_2 = \gamma(\theta_2)=\gamma_{\theta 2}, \ \dot{\gamma}_{\theta 1}=\dot{\gamma}(\theta_1)=\gamma_0 \omega \cos \theta_1=\gamma_0 \omega \cos \theta_2=\dot{\gamma}(\theta_2)=\dot{\gamma}_{\theta 2}, \qquad (19)$$

where $\gamma(t)=\gamma_0 \sin \omega t$ and $\dot{\gamma}(t)=\gamma_0 \omega \cos \omega t$. It is obvious that $\sigma(\gamma_{\theta 1},\dot{\gamma}/\omega)$ and $\sigma(\gamma_{\theta 2},\dot{\gamma}/\omega)$ are the partners for the stress decomposition from Cho et al. [20]. Then, the equation, $\theta_1+\theta_2=2\pi$, can be obtained. After that, based on Eqs. 14, 18, and 19, the value of $\sigma'(\gamma,\gamma_0)$ at $\gamma_{\theta 1}=\gamma_0 \sin \theta_1$ is calculated as following derivations:

$$\begin{aligned}
\sigma'(\gamma_{\theta 1},\gamma_0) &= \frac{\sigma(\gamma_{\theta 1},\dot{\gamma}_{\theta 1}) - \sigma(\gamma_{\theta 2},\dot{\gamma}_{\theta 2})}{2} = (\gamma_0 \sum_{n=1}^{\infty}(G_n''\cos n\theta_1 + G_n'\sin n\theta_1) - \gamma_0 \sum_{n=1}^{\infty}(G_n''\cos n\theta_2 + G_n'\sin n\theta_2))\bigg/2 \\
&= (\gamma_0 \sum_{n=1}^{\infty}(G_n''\cos n\theta_1 + G_n'\sin n\theta_1) - \gamma_0 \sum_{n=1}^{\infty}(G_n''\cos n(2\pi - \theta_1) + G_n'\sin n(2\pi - \theta_1)))\bigg/2 \\
&= (\gamma_0 \sum_{n=1}^{\infty}(G_n''\cos n\theta_1 + G_n'\sin n\theta_1) - \gamma_0 \sum_{n=1}^{\infty}((+G_n''\cos n\theta_1) + (-G_n'\sin n\theta_1)))\bigg/2 \\
&= \gamma_0 \sum_{n=1}^{\infty} G_n' \sin n\theta_1
\end{aligned} \quad (20)$$

From Eq. 20, the specific expression of the elastic part value $\sigma'(\gamma_{\theta 1},\gamma_0)$ generates the following equation:

$$\sigma'(\gamma,\gamma_0) = \gamma_0 \sum_{n=1}^{\infty} G_n' \sin n\omega t \quad . \tag{21}$$

Therefore, it is clear that the elastic part is closely related to the odd functions (sine waves) in Eq. 18 given by FT rheology and dominated by the Fourier coefficients. As can be seen, even higher harmonics are included in $\sigma'(\gamma,\gamma_0)$ based on the stress decomposition from Cho et al. [20], which can be simply removed based on Eq. 21.

Similarly, for the viscous part $\sigma''(\dot{\gamma},\gamma_0)$, the derivation process is done as follows:

$$\gamma(\theta_3) = \gamma_0 \sin\theta_3 = \gamma_0 \sin\theta_4 = \gamma(\theta_4), \quad \dot{\gamma}(\theta_3) = \gamma_0\omega\cos\theta_3 = -\gamma_0\omega\cos\theta_4 = \dot{\gamma}(\theta_4), \tag{22}$$

$$\theta_3 + \theta_4 = \pi \text{ or } \theta_3 + \theta_4 = 2\pi + \pi, \tag{23}$$

$$\begin{aligned}
\sigma''(\dot{\gamma}_{\theta 3},\gamma_0) &= \frac{\sigma(\gamma_{\theta 3},\dot{\gamma}_{\theta 3}) - \sigma(\gamma_{\theta 4},\dot{\gamma}_{\theta 4})}{2} \\
&= (\gamma_0 \sum_{n=1}^{\infty}(G_n''\cos n\theta_3 + G_n'\sin n\theta_3) - \gamma_0 \sum_{n=1}^{\infty}(G_n''\cos n(\pi - \theta_3) + G_n'\sin n(\pi - \theta_3)))\bigg/2 \\
&= (\gamma_0 \sum_{n=1}^{\infty}(G_n''\cos n\theta_3 + G_n'\sin n\theta_3) - \gamma_0 \sum_{n=1}^{\infty}(((-1)^n G_n''\cos n\theta_3) + ((-1)^{n+1} G_n'\sin n\theta_3)))\bigg/2 \\
&= \gamma_0(G''\cos\theta_3 + G_2'\sin 2\theta_3 + G_3''\cos 3\theta_3 + G_4'\sin 4\theta_3 + \cdots) \\
&= \gamma_0(\sum_{n=1,odd}^{\infty} G_n''\cos n\theta_3 + \sum_{n=2,even}^{\infty} G_n'\sin n\theta_3)
\end{aligned} \quad (24)$$

Accordingly, the specific expression of the viscous part value $\sigma''(\dot{\gamma}_{\theta 3},\gamma_0)$ gives the following equations:

$$\sigma''(\dot{\gamma},\gamma_0) = \gamma_0(\sum_{n=1,odd}^{\infty} G_n''\cos n\omega t + \sum_{n=2,even}^{\infty} G_n'\sin n\omega t) \quad . \tag{25}$$

Surprisingly, the viscous part is dependent on both the partial even (cosine waves) and odd (sine waves) terms in Eq. 18. The even terms in Eq. 25 coincide with the conclusions given by Cho et al. [20] and Yu et al. [21], reflecting that the intensities of odd harmonics in the viscous part are symmetric about the origin. However, the arisen odd functions in Eq. 25 indicate that the contributions of even harmonics on the viscous part are symmetric about the viscous stress axis. Furthermore, the even higher harmonics influence the viscous stress via $G_n'$. Therefore, the application of FT rheology in stress decomposition specifically demonstrates the influence of even harmonics. Meanwhile, Eq. 25 indicates that the viscous part from the stress decomposition of Cho et al. [20] is dependent on the Fourier coefficients.

### *1.3. Second type of stress decomposition and Fourier coefficients*

In the same way, the stress decomposition of Yu et al. [21] can be interpreted by FT rheology to demonstrate the relationship between the stress decomposition and Fourier coefficients. The same as the derivation process displayed in the previous section, based on Eqs. 16 and 18, the calculation of the elastic part $\sigma'(\gamma,\gamma_0)$ at the situation of the stress decomposition of Yu et al. [21] was carried out and briefly shown as follows:

$$\theta_1 + \theta_2 = \pi \text{ or } \theta_1 + \theta_2 = 2\pi + \pi, \tag{26}$$

$$\sigma'(\gamma_{\theta 1},\gamma_0) = \frac{\sigma(\gamma_{\theta 1},\dot{\gamma}_{\theta 1}) + \sigma(\gamma_{\theta 2},\dot{\gamma}_{\theta 2})}{2}$$
$$= \left(\gamma_0 \sum_{n=1}^{\infty}(G_n'' \cos n\theta_1 + G_n' \sin n\theta_1) + \gamma_0 \sum_{n=1}^{\infty}(((-1)^n G_n'' \cos n\theta_3) + ((-1)^{n+1} G_n' \sin n\theta_3)))\bigg/2 \cdot \tag{27}$$
$$= \gamma_0 (\sum_{n=1,odd}^{\infty} G_n' \sin n\theta_1 + \sum_{n=2,even}^{\infty} G_n'' \cos n\theta_1)$$

Then, the elastic part $\sigma'(\gamma,\gamma_0)$ is described as the following equation:

$$\sigma'(\gamma,\gamma_0) = \gamma_0 \sum_{n=1,odd}^{\infty} G_n' \sin n\omega t + \sum_{n=2,even}^{\infty} G_n'' \cos n\omega t \cdot \tag{28}$$

Similar to the situation of Eq. 25 obtained based on the stress decomposition of Cho et al. [20], the gained elastic part based on the stress decomposition of Yu et al. [21] contains both the partial even (cosine waves) and odd (sine waves) functions in Eq. 28. The odd terms in Eq. 28 match the conclusions provided by Cho et al. [20] and Yu et al. [21], indicating that the contributions of odd harmonics to the elastic part are symmetric about the origin. By contrast, the appeared even functions in Eq. 28 reflect that the intensities of the even harmonics in the elastic part are symmetric about the elastic stress axis. Furthermore, the even higher harmonics influence the elastic stress via $G_n''$.

The viscous part $\sigma''(\dot{\gamma},\gamma_0)$ is also calculated, which is briefly demonstrated as follows:

$$\theta_3 + \theta_4 = 2\pi, \tag{29}$$

$$\sigma''(\dot{\gamma}_{\theta 3},\gamma_0) = \frac{\sigma(\gamma_{\theta 3},\dot{\gamma}_{\theta 3}) + \sigma(\gamma_{\theta 4},\dot{\gamma}_{\theta 4})}{2}$$
$$= \left(\gamma_0 \sum_{n=1}^{\infty}(G_n'' \cos n\theta_3 + G_n' \sin n\theta_3) + \gamma_0 \sum_{n=1}^{\infty}((+G_n'' \cos n\theta_3) + (-G_n' \sin n\theta_3)))\bigg/2, \tag{30}$$
$$= \gamma_0 \sum_{n=1}^{\infty} G_n'' \cos n\theta_3$$

$$\sigma''(\dot{\gamma},\gamma_0) = \gamma_0 \sum_{n=1}^{\infty} G_n'' \cos n\omega t \cdot \tag{31}$$

Similar to the situation in Eq. 21, it can be seen that the viscous part is closely associated with the even functions (cosine waves) in Eq. 31 and is dominated by the Fourier coefficients.

### 1.4. Summary

To sum up, the stress decomposition from Cho et al. [20] in the form of Fourier series is described as Eq. 32:

$$\sigma'(\gamma,\gamma_0) = \gamma_0 \sum_{n=1}^{\infty} G_n' \sin n\omega t, \quad \sigma''(\dot{\gamma},\gamma_0) = \gamma_0 \sum_{n=1,odd}^{\infty} G_n'' \cos n\omega t + \gamma_0 \sum_{n=2,even}^{\infty} G_n' \sin n\omega t \cdot \tag{32}$$

Similarly, the stress decomposition from Yu et al. [21] can be expressed as Eq. 33:

$$\sigma'(\gamma,\gamma_0) = \gamma_0 \sum_{n=1,odd}^{\infty} G'_n \sin n\omega t + \sum_{n=2,even}^{\infty} G''_n \cos n\omega t \text{, } \sigma''(\dot{\gamma},\gamma_0) = \gamma_0 \sum_{n=1}^{\infty} G''_n \cos n\omega t \cdot \qquad (33)$$

When even harmonics are removed, Eqs. 32 and 33 are both transformed into the following functions:

$$\sigma'(\gamma,\gamma_0) = \gamma_0 \sum_{n=1,odd}^{\infty} G'_n \sin n\omega t \text{, } \sigma''(\dot{\gamma},\gamma_0) = \gamma_0 \sum_{n=1,odd}^{\infty} G''_n \cos n\omega t \text{,} \qquad (34)$$

which is the relationship provided by both Cho et al. [20] and Yu et al. [21].

Therefore, in Sec. II 1, the two kinds of stress decomposition proposed by Cho et al. [20] and Yu et al. [21] were introduced and discussed. When even harmonics, which are attributed to the imperfection of the experiment [59-63], are removed, the two methods are equivalent (Eq. 34). The main difference between the two stress decomposition methods of Cho et al. [20] and Yu et al. [21] is that even harmonics influence the results from the two methods in different ways, which can be clearly demonstrated by comparing Eqs. 32 with 33. Accordingly, this section shows how even harmonics affect the elastic and viscous parts, which is less considered by Cho et al. [20] and Yu et al. [21]. Meanwhile, it is proved that the results from the two methods surely present differences because of the even harmonics. Although the intensities of the even harmonics are relatively lower than those of the odd harmonics, this discussion is meaningful because both the stress decomposition methods process the raw data that inherently possesses even harmonics. The removement of even harmonics can be done by using the aLAOS approach. Most importantly, the full expressions of the two stress decomposition methods (Eqs. 32 and 33) have been given based on the Fourier series by using Fourier coefficients, which is seldom mentioned in the literature on stress decomposition [20, 21]. In a word, this section benefits the application of the aLAOS approach in stress decomposition.

## 2. Analytic stress decomposition
### 2.1. Principle

The two stress decomposition methods have been discussed in the previous section and are fully expressed in the form of Fourier series by using Fourier coefficients. Here, the use of the aLAOS approach in stress decomposition (the analytic stress decomposition), adopting Eq. 32 derived in this work by using the framework from Cho et al. [20], is as the following expressions:

$$\gamma_0 \sum_{n=1}^{\infty} G'_n \sin n\omega t = \sigma'(\gamma,\gamma_0) \text{, } \gamma_0 \sum_{n=1,odd}^{\infty} G''_n \cos n\omega t + \gamma_0 \sum_{n=2,even}^{\infty} G'_n \sin n\omega t = \sigma''(\dot{\gamma},\gamma_0) \cdot \qquad (35)$$

It is clear that Eq. 35 is the reverse use of Eq. 32. Eq. 35 allows the combination of Fourier coefficients to generate a series of reconstructed curves.

### 2.2. Output result from one point

Figure 5 correspondingly demonstrates the reconstructed elastic and viscous parts from the analytic stress decomposition along with the elastic and viscous stress curves from stress decomposition. Figure 5(a) displays the plots of the separated elastic and viscous stress curves by treating the raw elastic and viscous Lissajous curves of point 49 with stress decomposition. The use of aLAOS approach in stress decomposition was visually shown in Fig. 5(b). Figure 5(c) includes the raw and reconstructed elastic parts $\sigma' \sim \gamma$ while figure 5(d) contains the raw and reconstructed viscous parts $\sigma'' \sim \dot{\gamma}$.

As shown in Fig. 5(a), black lines are the Lissajous curves of point 49, and blue lines correspond to the results from stress decomposition (it is noted that in the following the blue lines refer to the "raw results" that were obtained using the reported LAOS methods to process the raw data). The elastic and viscous stress curves denote the appeared nonlinearity of the sample at point 49. Obviously, the elastic stress $\sigma'$ was an odd function of strain $\gamma$, and the viscous stress $\sigma''$ was an odd function of strain rate $\dot{\gamma}$. Thus, the nonlinear viscoelastic response at point 49 observed in LAOS experiments was successfully decomposed into physically meaningful elastic and viscous parts based on stress decomposition, which indicated that the stress signal is a function of $\gamma$ and $\dot{\gamma}$ instead of the original time-domain description $\sigma(t)$.

The second derivative of the elastic stress curve provides a visual "snapshot" of either strain-stiffening ($d^2\sigma'/d^2\gamma > 0$), strain-softening ($d^2\sigma'/d^2\gamma < 0$), or linear elastic behavior ($d^2\sigma'/d^2\gamma = 0$) by observing the curvature. In the same way, the shear-thickening and shear-thinning behaviors can be evaluated via the positive ($d^2\sigma''/d^2\gamma > 0$) and negative ($d^2\sigma''/d^2\gamma < 0$) curvatures of the viscous stress curve, respectively. Moreover, the curvature is associated with the third-order Chebyshev coefficients, where $e_3$ and $v_3 > 0$, as well as $e_3$ and $v_3 < 0$ often leads to the positive curvature and negative curvature since the intensity of each Chebyshev coefficient typically decays monotonically with the order *n*. Figure 5(a) visually denotes the strain-stiffening and shear-thinning behaviors at point 49. During the strain swept from zero to the maximum strain, the elastic stress first slowly increased and then sharply, indicating the strain-stiffening behavior. Similarly, during the strain rate change from zero to the maximum strain rate, the viscous stress increased with a negative curvature, reflecting the shear-thinning behavior.

Figure 5(b) clearly shows that the analytic stress decomposition possesses just four parts: the data given by the rheometer, the data obtained via FT rheology, elementary functions with straightforward forms, and negligible terms. Accordingly, the analytic stress decomposition brings convenience, whereas the stress decomposition needs to process the many points in Lissajous curves one by one. As the analytic stress decomposition requires higher harmonics that consist of infinite numbers of order, it is obvious that the difficulty arises when many higher harmonics should be introduced. Therefore, it is meaningful and important to estimate the number of the higher harmonics that need to be adopted to give the approximate elastic and viscous stress curves, which is also displayed in Fig. 5(b). Generally speaking, the reconstructed elastic and viscous parts by introducing different numbers of higher harmonic intensities showed that $I_1 \sim I_3$, $I_1 \sim I_5$, and $I_1 \sim I_9$ can be used for qualitative, approximate, and quantitative evaluations, respectively, which will be discussed in the following statements.

In Fig. 5(c), comparisons were made by plotting the elastic stress curve (blue lines) together with the reconstructed elastic parts (red lines). Figure 5(c) gives a graphical representation of the similarity of the results from the analytic stress decomposition to the raw curve from stress decomposition. The red line in the first panel of Fig. 5(c) indicated that by introducing only $I_1$, the reconstructed elastic part presented a linear shape and was similar to the linear viscoelastic behaviors in the SAOS region. Meanwhile, a deviation arose between the raw and reconstructed elastic parts. Therefore, the $I_1$-based elastic part cannot describe the nonlinear rheological behavior. By introducing $I_2$ and $I_3$, a higher similarity emerged, reflecting that the raw elastic part was approximately mimicked, which accords with the conclusion of MITlaos [23, 24]. The situation in Fig. 5(d) was similar to that in Fig. 5(c). The reconstructed viscous part presented linearity based on only $I_1$ and was distorted by introducing $I_2$ and $I_3$. Meanwhile, the proximity between the raw and reconstructed viscous parts was obviously promoted.

Furthermore, as shown in Figs. 5(c) and 5(d), the incorporation of more harmonics improves the superposition between the raw and reconstructed stress decomposition results. To sum up, the analytic stress decomposition can establish reconstructed elastic and viscous parts when the sample is at the nonlinear region. The raw and reconstructed curves can show good overlaps by introducing $I_1$ and $I_3$. Figure 5 depicts that $I_1 \sim I_3$, $I_1 \sim I_5$, and $I_1 \sim I_9$ are accessible for yielding qualitative, approximate, and quantitative evaluations, respectively.

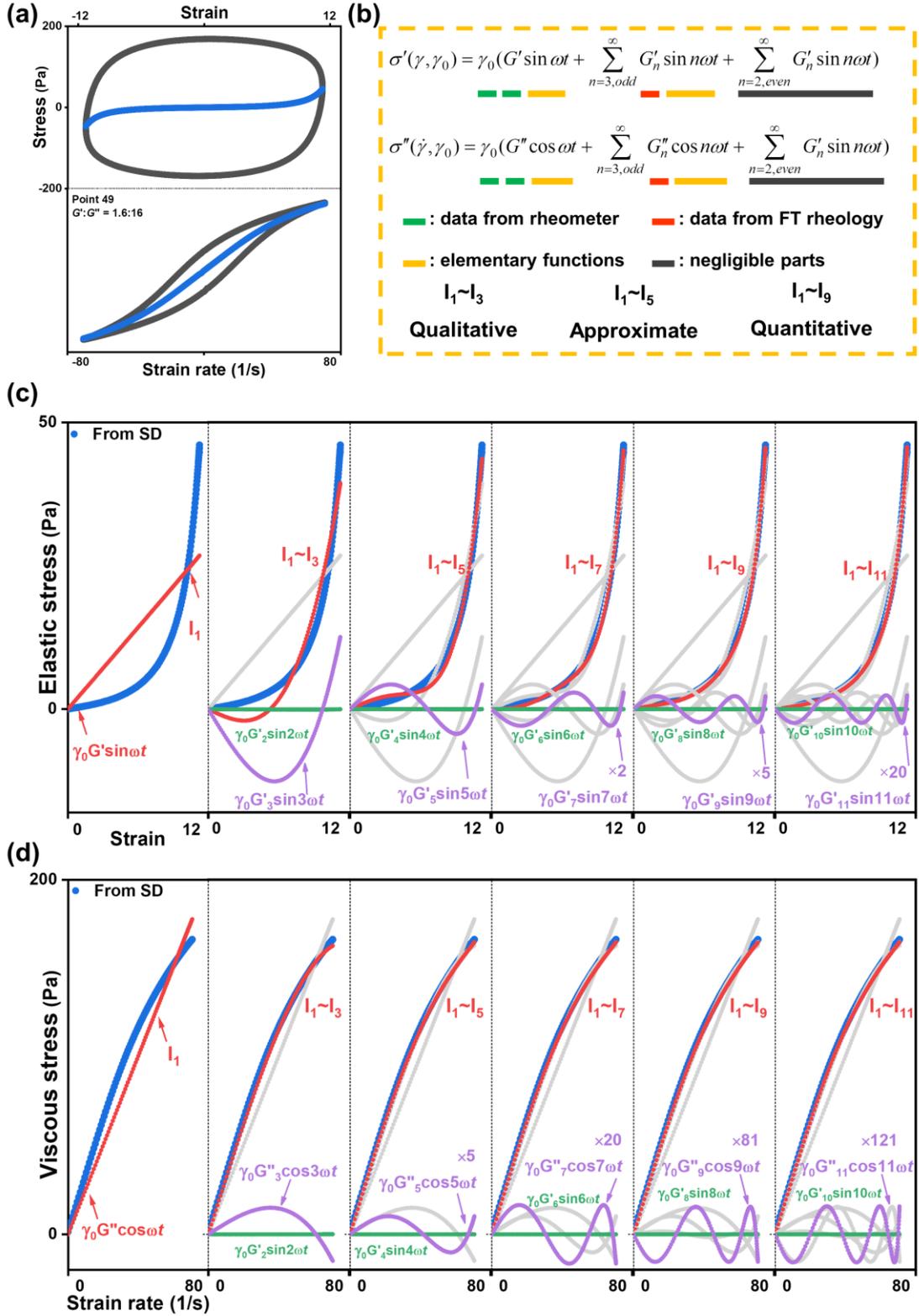

FIG. 5 Comparison between the elastic and viscous parts from stress decomposition and the reconstructed elastic and viscous parts from the aLAOS approach in stress decomposition: (a) the raw elastic and viscous Lissajous curves of point 49 along with the raw elastic and viscous parts; (b) the analytic stress decomposition; (c) the plot of the raw and reconstructed elastic stresses versus strain; and (d) the plot of the raw and reconstructed viscous stresses versus strain rate. Blue lines:

raw results. Red lines: reconstructed results. Purple lines: visualized contributions of odd harmonics. Green lines: visualized contributions of even higher harmonics. Gray lines: the curves that have been shown.

### *2.3. Visualized harmonic contribution*

This section discusses the visualized contributions of harmonics (purple and green lines in Figs. 5(c) and 5(d)). From the 1st~3rd panels in Figs. 5(c) and 5(d), respectively, the demonstrated contributions of $I_1$ and $I_3$ visually show that $I_1$ provides the dominant intensity contributions to the elastic and viscous parts, while $I_3$ offers crucial contributions to the curvature of the reconstructed curves. $I_3$ determines the concavity of each curve and corresponds to the physical interpretation of the intra-cycle nonlinearity. Meanwhile, the contributions of $I_2$ and $I_4$ are negligible. When $I_5$ was introduced, the reconstructed viscous part slightly changed, whereas the reconstructed elastic part became much closer to the raw elastic part. The contribution of a higher harmonic decreased with the increase in the order, and $I_1$~$I_9$-based elastic and viscous parts presented high similarities to the raw results.

The impact of harmonics on the reconstructed elastic parts was studied at different positions including the maximum strain point, middle strain region, and near zero strain region. In Fig. 5(c), based on the $I_1$-based elastic part, $I_3$ greatly promoted the intensity of the maximum strain point and apparently lowered the curves in the middle and near zero strain regions. $I_5$ further adjusted the curve shape by elevating the maximum strain point and the curve in the near zero strain region, as well as decreasing the intensity of the curve in the middle strain region. However, the $I_1$~$I_9$-based elastic part remained deviations from the blue line, which was eliminated by introducing $I_7$ through increasing the intensities of the maximum strain point and the curve in the near zero strain region, as well as lowering the curve in the middle strain region. Thus, a high similarity appeared in the fourth panel, while the $I_1$~$I_9$-based elastic part and the blue line showed good superposition. For the purple lines in Fig. 5(d), harmonics elevated the maximum strain point. Furthermore, the $I_1$~$I_3$-based viscous part showed a good overlap with the blue line.

Briefly, the aLAOS approach in stress decomposition provides a tool to clearly define, precisely calculate, and visually demonstrate the influences of harmonics on the reconstructed elastic and viscous parts. As a result, it can be argued that the aLAOS approach provides another aspect of understanding the decomposed elastic and viscous stresses.

### *2.4. Whole sweep process*

The universality of the analytic stress decomposition was further studied. First, a series of points of the stress sweep test shown in Fig. 1(a) was analyzed by stress decomposition and the results were plotted in Fig. 6. After that, the reconstructed results from the analytic stress decomposition are displayed in Fig. 7, where different numbers of harmonic intensities were introduced including $I_1$, $I_1$~$I_3$, $I_1$~$I_5$, and $I_1$~$I_9$.

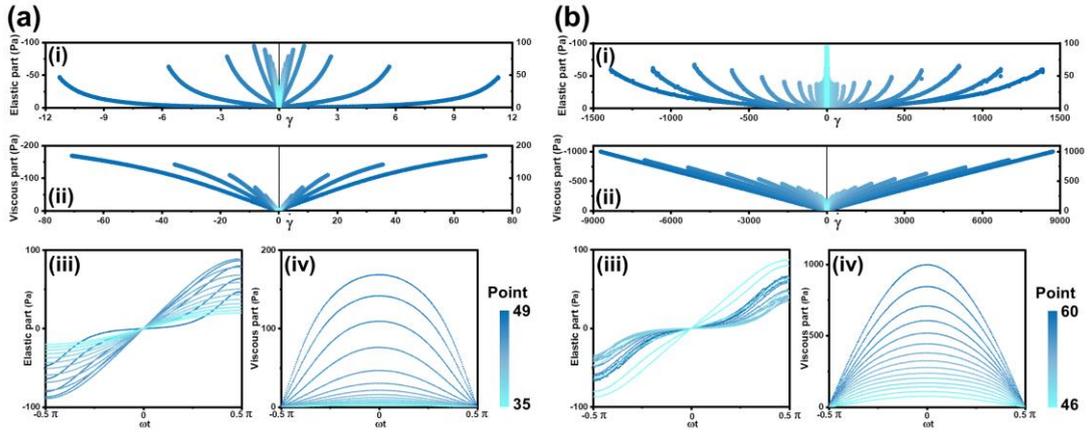

FIG. 6 Raw elastic and viscous parts of different points. (a) The results of points 35 ~ 49: (i) elastic stress versus strain; (ii) viscous stress versus strain rate; (iii) elastic and (iv) viscous stresses versus angle. (b) The results of points 46 ~ 60.

The stress decomposition results of points 35 ~ 49 were displayed in Fig. 6(a), while the results of points 46 ~ 60 are demonstrated in Fig. 6(b). At low stress levels, the elastic (Fig. 6(a.i)) and viscous (Fig. 6(a.ii)) stress curves showed linearity. As the stress amplitude increased, the slopes of the elastic (Figs. 6(a.i) and 6(b.i)) and viscous (Figs. 6(a.ii) and 6(b.ii)) stress curves decreased along with the sharply increased intensity of the viscous stress. Meanwhile, the elastic and viscous stress curves gradually bent, as demonstrated in Figs. 6(a.i) and 6(a.ii). Furthermore, the beginning of the nonlinearity was readily observed by the visual investigation from the gradually distorted linear shape when the point number increased. In the post-yield region in Figs. 6(b.i) and 6(b.ii), the intensity of the elastic stress curve almost remained unchanged. The elastic stress curves were higher and lower than the viscous stress curves at low and high stress levels, respectively, reflecting the manifestation of a solid-liquid transition during the oscillatory stress sweep. At an extremely high stress level, the viscous stress curve was almost a linear line, indicating the linear viscous behavior of the sample.

Although the curves of the viscous stress versus angle were all similar to the sinusoidal functions (Figs. 6(a.iv) and 6(b.iv)), the elastic stress curves were gradually distorted with the increase in the stress amplitude (Figs. 6(a.iii) and 6(b.iii)). Meanwhile, the nonlinearity of the elastic stress was much larger than the viscous stress and total stress, denoting that the stress decomposition is more sensitive to nonlinearity than FT rheology that only considers the nonlinearity of the total stress [20].

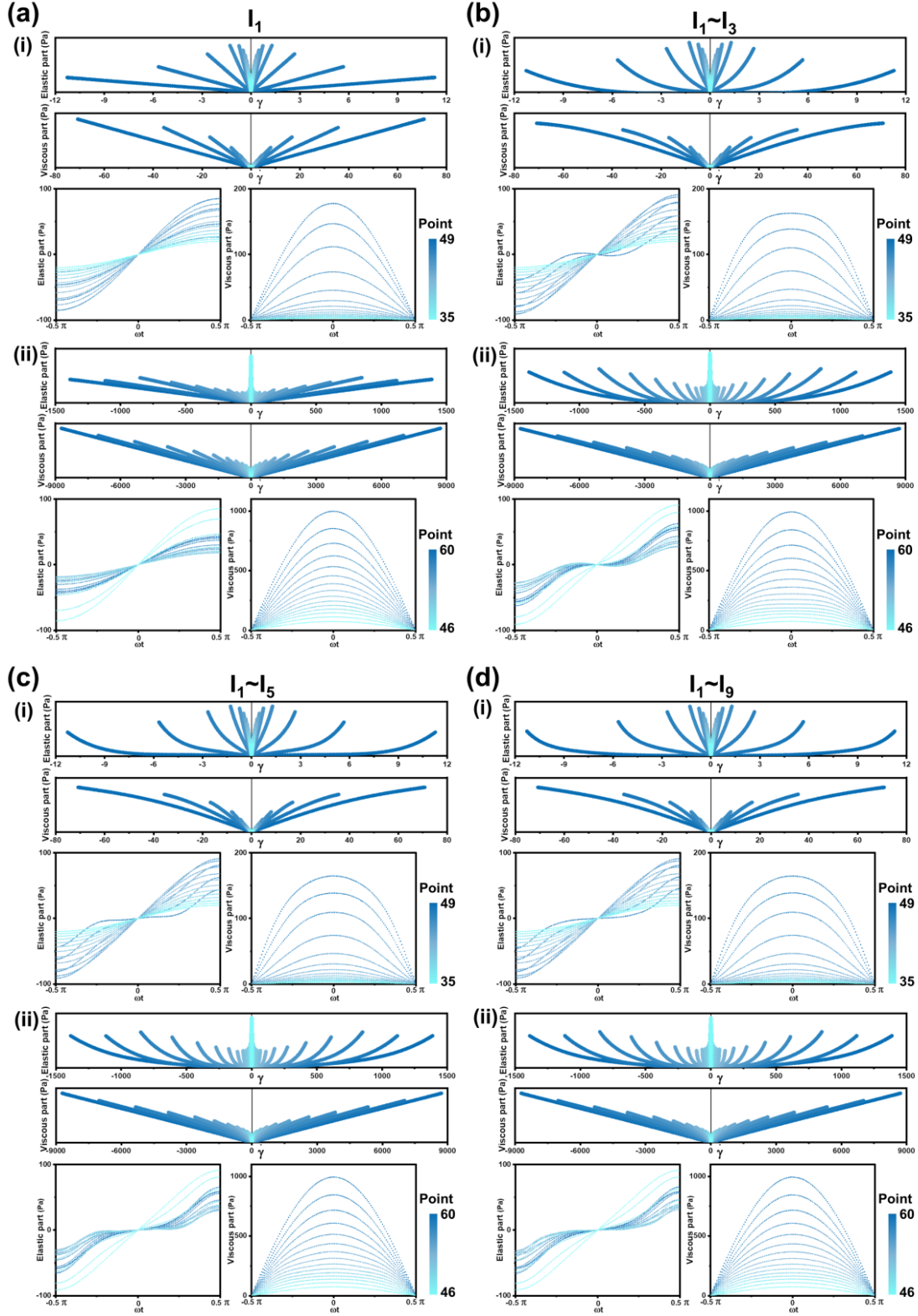

FIG. 7 Reconstructed elastic and viscous parts by introducing different numbers of higher harmonics. (a) $I_1$-based results: (i) points 35 ~ 49 containing the plots of elastic stress versus strain, viscous stress versus strain rate, as well as elastic and viscous stresses versus angle; (ii) points 46 ~ 60. (b) $I_1$~$I_3$-based results. (c) $I_1$~$I_5$-based results. (b) $I_1$~$I_9$-based results. The corresponding $y$-axis scales are set the same as those in Fig. 6.

$I_1$-, $I_1\sim I_3$-, $I_1\sim I_5$-, and $I_1\sim I_9$-based elastic and viscous parts were shown in Figs. 7a, 7b, 7c, and 7d, respectively. As shown in Fig. 7a, the $I_1$-based elastic and viscous parts showed the linear rheological response, where the plots of elastic and viscous stresses versus angle possessed all the sinusoidal curves. In Fig. 7(b), the $I_1\sim I_3$-based elastic and viscous parts indicated nonlinearities, demonstrating the contribution of $I_3$ to the nonlinearity of the reconstructed curves. High similarities appeared in Fig. 7 when 1st~9th harmonics were introduced.

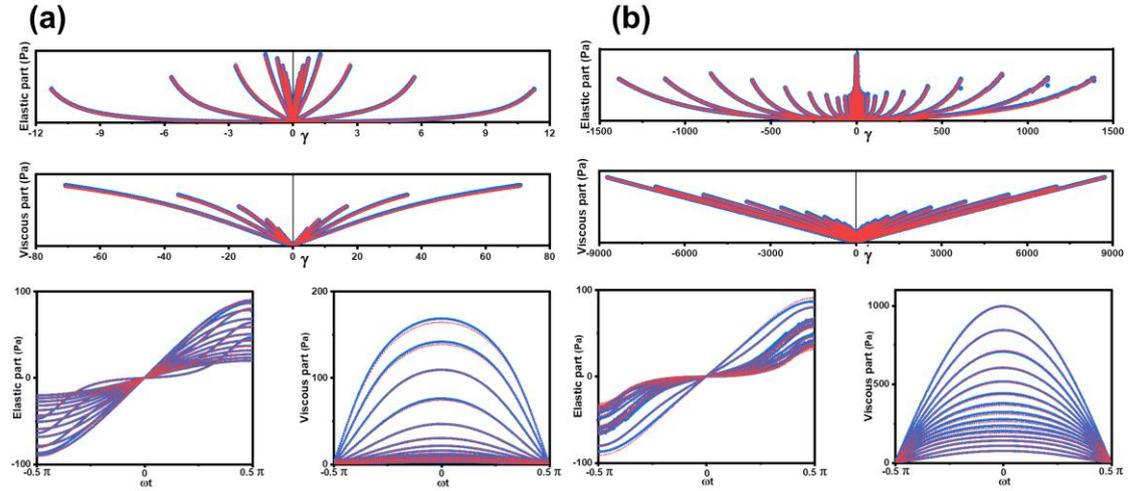

FIG. 8 Comparison between the raw and reconstructed stress decomposition results. (a) The results of points 35 ~ 49 containing the plots of elastic stress versus strain, viscous stress versus strain rate, as well as the elastic and viscous stresses versus angle. (b) The results of points 46 ~ 60. Blue lines: raw results. Red lines: reconstructed results. The corresponding $y$-axis scales are set the same as those in Fig. 6.

Figure 8 compares the $I_1\sim I_9$-based elastic and viscous parts (red lines) with the raw elastic and viscous stress curves (blue lines). Good overlaps were shown between the raw and reconstructed results no matter for points 35 ~ 49 (SAOS region ~ LAOS region) or points 46 ~ 60 (LAOS region) when point 46 presents $G' \approx G''$. Both the intensities and curvatures of these reconstructed results corresponded well with the raw curves.

Briefly, in this section, a series of sequential points (from the SAOS region to the LAOS region) was thoroughly investigated by stress decomposition (Fig. 6) and the analytic stress decomposition (Fig. 7). Based on the FT coefficients, four kinds of analytic results (Fig. 7) were obtained by using $I_1$, $I_1\sim I_3$, $I_1\sim I_5$, and $I_1\sim I_9$. Good superpositions were shown in Fig. 8 between the raw and $I_1\sim I_9$-based stress decomposition results. Compared with Fig. 5 showing a specific point, figures 6, 7, and 8 prove that the use of aLAOS approach in stress decomposition (Eq. 35) is also applicable on a larger scale.

Appendix C provided further detailed information.

## 3. Summary

In short, based on FT rheology and Fourier coefficients, the reconstructed stress decomposition results can be given instead of using raw data. Point 49 as a specific situation was studied in detail (Fig. 5), and the investigation on a larger scale was further carried out and displayed in Figs. 6, 7, and 8. From the extracted results, it was demonstrated that the analytic stress decomposition is accessible for both the SAOS and LAOS regions, reflecting that $I_1 \sim I_3$, $I_1 \sim I_5$, and $I_1 \sim I_9$ are capable of giving qualitative, approximate, and quantitative evaluations, respectively. Meanwhile, the full expressions of the two stress decomposition methods (Eqs. 32 and 33) were given based on Fourier coefficients. The influence of harmonics on the stress decomposition results was further visually demonstrated, which offers another perspective to interpret the stress decomposition method.

Stress decomposition is directly related to the FT rheology and Fourier series. When even harmonic terms are excluded, analytic stress decomposition possesses simple and beautiful mathematical expressions. Based on Eqs. 25 and 28, the contributions of even harmonics to the stress decomposition results are clearly defined for the two methods of Cho et al. [20] and Yu et al. [21]. The two methods are equivalent by removing even harmonics, where the analytic elastic and viscous parts contain purely elastic- and viscous-dependent responses, respectively. By using equations, there is no need for searching the partner of each point ($\sigma(\gamma, \dot{\gamma})$) to conquer the issue in which the corresponding partner may not always exist. Furthermore, the data processing procedure would be significantly simplified by introducing several Fourier coefficients and trigonometric functions rather than handling many points one by one. The aLAOS approach in stress decomposition presents a near-optimal basis set to express the fluctuation in continuous functions $\sigma'(\gamma)$ and $\sigma''(\dot{\gamma})$, which allows the probe of rheological responses to arbitrary intermediate points.

### E. aLAOS in strain-stiffening $S$ and shear-thickening $T$ ratios

The stress decomposition discussed above provides a visual "snapshot" of either strain-stiffening/strain-softening or shear-thickening/shear-thinning. The quantitative evaluation method of these nonlinear behaviors was further proposed, which was named the strain-stiffening $S$ ratio and shear-thickening $T$ ratio [23, 24]. In addition, based on the stress decomposition established by Yu et al. [21], a yield stress determination method was proposed and named stress bifurcation [28], evaluating the start and end of the solid-liquid transition during an oscillatory stress/strain sweep. Accordingly, the $S$ and $T$ ratios will be first discussed in this section, and stress bifurcation will be afterward studied in the next section.

### 1. $S$ and $T$ ratios in LAOS

FT rheology is commonly used for quantifying LAOS tests, which is a sensitive measure of nonlinearity. However, FT rheology concerns less about the clear physical explanation of the higher harmonics [23, 24]. Meanwhile, the output $G'$ and $G''$ values from the rheometer by using $I_1$, i.e. the Fourier coefficients of $G'_1$ and $G''_1$, are arbitrarily used as the viscoelastic moduli in the LAOS region and often unable to observe the plenty nonlinearities in the raw data. Therefore, the *S* and *T* ratios are proposed along with a set of material measures to quantify the nonlinearity, which offers a complete and low-dimensional framework for the physical description of deviations from the linear rheological response [23, 24]. This method separately characterizes the elastic and viscous nonlinearities, which generates a geometric representation for the adjectives of elastic stiffening/softening and viscous thickening/thinning that may be qualitatively familiar and poorly defined. Additionally, the provided measures are sensitive to contrasting different complex fluids.

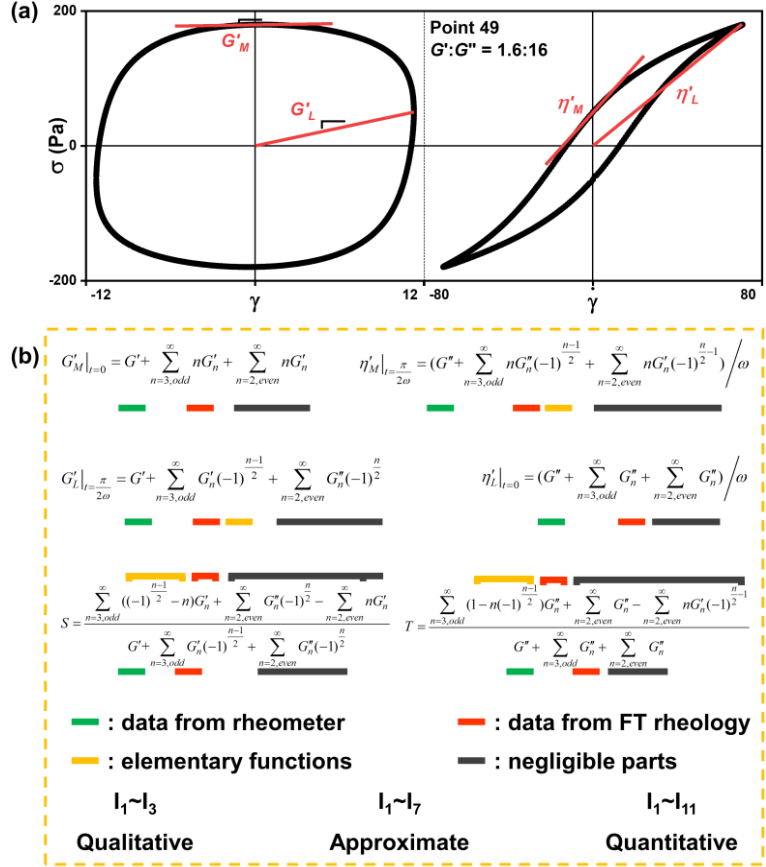

FIG. 9 aLAOS approach in strain-stiffening *S* and shear-thickening *T* ratios: (a) the definitions of *S* and *T* ratios of point 49 with $G':G''$=1:10; (b) the principle of the use of the aLAOS approach in *S* and *T* ratios, where an approximate criterion can be generally given.

To capture the local rheological behavior in the Lissajous curves, four elastic and viscous moduli were geometrically defined as in Fig. 9(a) and associated with the Fourier and Chebyshev coefficients as follows [23, 24]:

$$G'_M = \left.\frac{d\sigma}{d\gamma}\right|_{\gamma=0} = \sum_{n=1,odd} nG'_n = e_1 - 3e_3 + \cdots, \quad G'_L = \left.\frac{\sigma}{\gamma}\right|_{\gamma=\pm\gamma_0} = \sum_{n=1,odd} (-1)^{\frac{n-1}{2}} G'_n = e_1 + e_3 + \cdots, \quad (37)$$

$$\eta'_M = \left.\frac{d\sigma}{d\dot\gamma}\right|_{\dot\gamma=0} = \frac{1}{\omega}\sum_{n=1,odd} (-1)^{\frac{n-1}{2}} nG''_n = v_1 - 3v_3 + \cdots, \quad \eta'_L = \left.\frac{\sigma}{\dot\gamma}\right|_{\dot\gamma=\pm\dot\gamma_0} = \frac{1}{\omega}\sum_{n=1,odd} G''_n = v_1 + v_3 + \cdots, \quad (38)$$

where $G'_M$ is the minimum-strain modulus, $G'_L$ denotes the large-strain moduli, $\eta'_M$ is the minimum-rate dynamic viscosity, $\eta'_L$ represents the large-rate dynamic viscosity, $e_n$ and $v_n$ refer to the $n$th-order elastic and viscous Chebyshev coefficients, respectively.

The minimum-strain modulus $G'_M$ is the slope at zero transient strain, and the large-strain modulus $G'_L$ stands for the slope of the line between the origin and the maximum strain point (Eq. 37), which represents two natural approaches to calculate the elastic modulus (the left panel in Fig. 9(a)). For the oscillatory test with $\gamma = \gamma_0 \sin\omega t$, the zero strain point possesses a local maximum value of strain rate, while the maximum strain point presents a zero strain-rate [23, 24]. The maximum and zero strain rates reflect the locally constant viscous contribution and no viscous contribution to the total stress, respectively. Therefore, at the zero strain, the change in the stress is related only to the elastic deformation. Accordingly, the residual stress at the maximum strain point is induced only by the elastic behavior of the sample. Meanwhile, Eq. 37 directly indicates that $G'_M$ and $G'_L$ are distinct from $G'$ and $G''$ in the nonlinear region [23, 24]. Using the two parameters from the elastic Lissajous curves of Fig. 9(a), the sample is intracycle strain-stiffening at the high strain amplitude since $G'_L > G'_M$.

The dissipated energy per cycle per unit volume is directly associated with the first-order viscous modulus ($E_d = \pi \gamma_0^2 G''_1$), which represents the total contribution and hardly differentiates the local change of viscous dissipation during an oscillatory shear cycle. Thus, the minimum-rate dynamic viscosity at the zero strain rate point $\eta'_M$ and the large-rate dynamic viscosity at the maximum strain rate point $\eta'_L$ are proposed (Eq. 38, the right panel in Fig. 9(a)), indicating the transient viscous dissipation contribution [23, 24].

The four alternative parameters of the elastic modulus and dynamic viscosity were also applied for quantifying the nonlinearities within Lissajous curves. The intracycle strain stiffening and shear thickening responses are denoted by $S$ and $T$ ratios [23, 24]:

$$S = \frac{G'_L - G'_M}{G'_L} = \frac{4e_3 + \cdots}{e_1 + e_3 + \cdots}, \quad T = \frac{\eta'_L - \eta'_M}{\eta'_L} = \frac{4v_3 + \cdots}{v_1 + v_3 + \cdots}. \quad (39)$$

Accordingly, the intracycle rheological behavior of a complex fluid can be interpreted as follows: the linear elastic response ($S = 0$); strain stiffening ($S > 0$); strain softening ($S < 0$); linear viscous response ($T = 0$); shear thickening ($T > 0$); and shear thinning ($T < 0$) [23, 24].

## *2. Analytic LAOS approach in S and T ratios*
## *2.1. Principle*

The $S$ and $T$ ratios have been introduced in the previous section. Here, the use of the aLAOS approach in $S$ and $T$ ratios (analytic $S$ and $T$ ratios), adopting Eqs. 37~39 from the method of the $S$ and $T$ ratios [23, 24], will be introduced as following statements.

First, the rheological response refers to Eq. 11 in Sec. II C2.1: $\gamma_0 \sum_{n=1}^{\infty} (G''_n \cos n\omega t + G'_n \sin n\omega t) = \sigma(t)$ and $\gamma(t) = \gamma_0 \sin\omega t$. $G'_M$ is calculated as the following equation:

$$G'_M = \left.\frac{d\sigma}{d\gamma}\right|_{\gamma=0} = \left.\frac{\sum_{n=1}^{\infty} n(G'_n \cos n\omega t - G''_n \sin n\omega t)}{\cos \omega t}\right|_{t=0 \text{ or } \frac{\pi}{\omega}} . \tag{40}$$

Then, two situations of the $G'_M$ values were obtained at $t=0$ and $t=\pi/\omega$, respectively:

$$G'_M\big|_{t=0} = G'_1 + 2G'_2 + 3G'_3 + \cdots = \sum_{n=1,odd}^{\infty} nG'_n + \sum_{n=2,even}^{\infty} nG'_n , \tag{41}$$

$$G'_M\big|_{t=\frac{\pi}{\omega}} = G'_1 - 2G'_2 + 3G'_3 + \cdots = \sum_{n=1,odd}^{\infty} nG'_n - \sum_{n=2,even}^{\infty} nG'_n . \tag{42}$$

From Eqs. 41 and 42, the influence of even harmonics on the $G'_M$ value is clearly demonstrated. Meanwhile, Eqs. 41 and 42 are equivalent if even harmonics are removed, which also accords with the conclusion given by Eq. 37.

Finally, the measures of $G'_L$, $\eta'_M$, and $\eta'_L$ were calculated the same as that of $G'_M$ as follows:

For $G'_L$, the situation is:

$$G'_L = \left.\frac{\sigma}{\gamma}\right|_{\gamma=\pm\gamma_0} = \left.\frac{\sum_{n=1}^{\infty}(G''_n \cos n\omega t + G'_n \sin n\omega t)}{\sin \omega t}\right|_{t=\frac{\pi}{2\omega} \text{ or } \frac{3\pi}{2\omega}} , \tag{43}$$

$$G'_L\big|_{t=\frac{\pi}{2\omega}} = G'_1 - G''_2 - G'_3 + G''_4 + \cdots = \sum_{n=1,odd}^{\infty} G'_n(-1)^{\frac{n-1}{2}} + \sum_{n=2,even}^{\infty} G''_n(-1)^{\frac{n}{2}} , \tag{44}$$

$$G'_L\big|_{t=\frac{3\pi}{2\omega}} = G'_1 + G''_2 - G'_3 - G''_4 + \cdots = \sum_{n=1,odd}^{\infty} G'_n(-1)^{\frac{n-1}{2}} - \sum_{n=2,even}^{\infty} G''_n(-1)^{\frac{n}{2}} . \tag{45}$$

For $\eta'_M$, the calculation is:

$$\eta'_M = \left.\frac{d\sigma}{d\dot\gamma}\right|_{\dot\gamma=0} = \left.\frac{\sum_{n=1}^{\infty} n(G'_n \cos n\omega t - G''_n \sin n\omega t)}{-\omega \sin \omega t}\right|_{t=\frac{\pi}{2\omega} \text{ or } \frac{3\pi}{2\omega}} , \tag{46}$$

$$\eta'_M\big|_{t=\frac{\pi}{2\omega}} = \frac{G''_1 + 2G'_2 - 3G''_3 - 4G'_4 + \cdots}{\omega} = \sum_{n=1,odd}^{\infty} \frac{n}{\omega} G''_n(-1)^{\frac{n-1}{2}} + \sum_{n=2,even}^{\infty} \frac{n}{\omega} G'_n(-1)^{\frac{n}{2}-1} , \tag{47}$$

$$\eta'_M\big|_{t=\frac{3\pi}{2\omega}} = \frac{G''_1 - 2G'_2 - 3G''_3 + 4G'_4 + \cdots}{\omega} = \sum_{n=1,odd}^{\infty} \frac{n}{\omega} G''_n(-1)^{\frac{n-1}{2}} - \sum_{n=2,even}^{\infty} \frac{n}{\omega} G'_n(-1)^{\frac{n}{2}-1} . \tag{48}$$

For $\eta'_L$, the process is:

$$\eta'_L = \left.\frac{\sigma}{\dot\gamma}\right|_{\dot\gamma=\pm\dot\gamma_0} = \left.\frac{\sum_{n=1}^{\infty}(G''_n \cos n\omega t + G'_n \sin n\omega t)}{\omega \cos \omega t}\right|_{t=0 \text{ or } \frac{\pi}{\omega}} , \tag{49}$$

$$\eta'_L\big|_{t=0} = \frac{G''_1 + G''_2 + G''_3 + \cdots}{\omega} = \sum_{n=1,odd}^{\infty} \frac{G''_n}{\omega} + \sum_{n=2,even}^{\infty} \frac{G''_n}{\omega} , \tag{50}$$

$$\eta'_L\big|_{t=\frac{\pi}{\omega}} = \frac{G''_1 - G''_2 + G''_3 - G''_4 + \cdots}{\omega} = \sum_{n=1,odd}^{\infty} \frac{G''_n}{\omega} - \sum_{n=2,even}^{\infty} \frac{G''_n}{\omega} . \tag{51}$$

Briefly, the contributions of even harmonics to $G'_L$, $\eta'_M$, and $\eta'_L$ values are demonstrated in Eqs. 44, 45, 47, 48, 50, and 51, where the equations coincide with Eqs. 37 and 38 when even harmonics are expelled. It is noticed that even harmonics influence the $G'_L$ value through all the viscous moduli ($G''_n$) while they affect $\eta'_M$ value via all the elastic moduli ($G'_n$).

For calculating $S$ and $T$ ratios, the situation in Fig. 9(a) is accepted ($G'_M$ at $\omega t = 0$, $G'_L$ at $\omega t = 0.5\pi$, $\eta'_M$ at $\omega t = 0.5\pi$, and $\eta'_L$ at $\omega t = 0$), which is represented as follows:

$$S = \frac{G'_L - G'_M}{G'_L} = \frac{\sum_{n=1,odd}^{\infty}((-1)^{\frac{n-1}{2}} - n)G'_n + \sum_{n=2,even}^{\infty} G''_n(-1)^{\frac{n}{2}} - \sum_{n=2,even}^{\infty} nG'_n}{\sum_{n=1,odd}^{\infty} G'_n(-1)^{\frac{n-1}{2}} + \sum_{n=2,even}^{\infty} G''_n(-1)^{\frac{n}{2}}}, \quad (52)$$

$$T = \frac{\eta'_L - \eta'_M}{\eta'_L} = \frac{\sum_{n=1,odd}^{\infty}(1 - n(-1)^{\frac{n-1}{2}})G''_n + \sum_{n=2,even}^{\infty} G''_n - \sum_{n=2,even}^{\infty} nG'_n(-1)^{\frac{n}{2}-1}}{\sum_{n=1}^{\infty} G''_n}. \quad (53)$$

When even harmonics are removed, Eqs. 52 and 53 are transformed into the following expression:

$$S = \frac{\sum_{n=1,odd}^{\infty}((-1)^{\frac{n-1}{2}} - n)G'_n}{\sum_{n=1,odd}^{\infty} G'_n(-1)^{\frac{n-1}{2}}}, \quad T = \frac{\sum_{n=1,odd}^{\infty}(1 - n(-1)^{\frac{n-1}{2}})G''_n}{\sum_{n=1,odd}^{\infty} G''_n}. \quad (54)$$

that is equivalent to Eq. 39 given by Ewoldt et al. [23, 24].

The analytic $S$ and $T$ ratios are based on the reverse use of Eq. 53 and give the following Eq. 55:

$$\frac{\sum_{n=1,odd}^{\infty}((-1)^{\frac{n-1}{2}} - n)G'_n + \sum_{n=2,even}^{\infty} G''_n(-1)^{\frac{n}{2}} - \sum_{n=2,even}^{\infty} nG'_n}{\sum_{n=1,odd}^{\infty} G'_n(-1)^{\frac{n-1}{2}} + \sum_{n=2,even}^{\infty} G''_n(-1)^{\frac{n}{2}}} = S, \quad \frac{\sum_{n=1,odd}^{\infty}(1 - n(-1)^{\frac{n-1}{2}})G''_n + \sum_{n=2,even}^{\infty} G''_n - \sum_{n=2,even}^{\infty} nG'_n(-1)^{\frac{n}{2}-1}}{\sum_{n=1}^{\infty} G''_n} = T. \quad (55)$$

As a result, the aLAOS approach allows for reconstructing $S$ and $T$ ratios by using Fourier coefficients accompanied with convenience by replacing many linear fitting processes.

Figure 9(b) demonstrates that the analytic $S$ and $T$ ratios possess a simple and efficient framework by using four parts: the results from the rheometer, the results from FT rheology, elementary functions, and negligible parts. As Eq. 55 needs infinite harmonics, it is important to use finite Fourier coefficients for the result generation. Generally, the reconstructed $S$ and $T$ ratios (Fig. 10) demonstrate that $I_1 \sim I_3$, $I_1 \sim I_7$, and $I_1 \sim I_{11}$ can generate qualitative, approximate, and quantitative results (Fig. 9(b)), respectively.

In this section, the analytic $S$ and $T$ ratios were discussed, where the contributions of even harmonics were shown. Moreover, the equations of $G'_M$, $G'_L$, $\eta'_M$, $\eta'_L$, $S$, and $T$ were given based on Fourier coefficients, which coincides with those of Ewoldt et al. [23, 24] if even harmonics are removed. After that, Eq. 55 allows the combination of Fourier coefficients to generate reconstructed $S$ and $T$ ratios.

### 2.2. Whole sweep process

Based on the data shown in Fig. 1, figure 10 shows the raw (blue points) and reconstructed (red points) values of $G'_M$, $G'_L$, $\eta'_M$, $\eta'_L$, $S$, and $T$, where comparisons are made. Different kinds of reconstructed values were calculated by introducing different numbers of harmonic intensities (i.e. $I_1$, $I_1 \sim I_3$, $I_1 \sim I_5$, $I_1 \sim I_7$, $I_1 \sim I_9$, and $I_1 \sim I_{11}$). For example, figure 10(a) is the plot of $G'_M$ versus stress amplitude, where the different panels involve different numbers of harmonics. In addition, the blue and red points, as well as the purple and green bars, represent the raw and reconstructed $G'_M$ values, as well as the visualized odd and even harmonic contributions, respectively. In the same way, the $G'_L$, $\eta'_M$, $\eta'_L$, $S$, and $T$ values were also represented in Fig. 10.

In the SAOS region, the $G'_M$ and $G'_L$ values converged (Figs. 10(a) and 10(b)) while the $\eta'_M$ and $\eta'_L$ values suffered from experimental noise (Figs. 10(c) and 10(d)). The start point of the solid-liquid transition was visually inspected by the sharply decreased values of $G'_M$ and $G'_L$ as the stress amplitude increased. $G'_M$ and $G'_L$ values decreased with the increase in the stress amplitude, whereas $\eta'_M$ and $\eta'_L$ values first decreased and then increased.

The intracycle nonlinearity can further be quantified by $S$ (Fig. 10(e)) and $T$ (Fig. 10(f)) values. The $S$ values were all positive in the LAOS region, reflecting the intracycle strain stiffening behavior. Meanwhile, $S$ was equal to zero in the SAOS region and remained generally constant at extremely high stress levels. For the $T$ ratio, figure 10(f) shows that the $T$ values are positive and negative before and after point 46 of $G' \approx G''$, indicating the shear thickening and shear thinning behaviors, respectively. Moreover, during the stress sweep process, the $T$ value generally first remained a small positive value, then sharply decreased to a large negative value, and finally significantly increased to the zero value. These rich behaviors in the rheological response of the sample were thus revealed and physically described by the $S$ and $T$ ratios.

The red points (reconstructed results) in Fig. 10 were further analyzed. In the first panels of Figs. 10(e) and 10(f), respectively, the reconstructed $S$ and $T$ values were equal to zero, indicating the linear viscoelastic behavior. When $I_2$ and $I_3$ were introduced, these values roughly mimicked the trends of the raw $S$ and $T$ values. By further introducing $I_4$ and $I_5$, approximations between the raw and reconstructed results were realized, where the deviations of the reconstructed values to the raw results still existed. Moreover, the $I_1$~$I_{11}$-based $S$ and $T$ values showed good superpositions with the raw results. Therefore, the incorporation of more harmonics leads to better overlap between the raw and the reconstructed $S$ and $T$ values. As a result, a general criterion was concluded and demonstrated at the bottom of Fig. 9(b), indicating that $I_1$~$I_3$, $I_1$~$I_7$, and $I_1$~$I_{11}$ can generate qualitative, approximate, and quantitative results, respectively. The aLAOS approach in $S$ and $T$ ratios is accessible for giving the reconstructed $S$ and $T$ values with high similarities to the raw results from the method of $S$ and $T$ ratios.

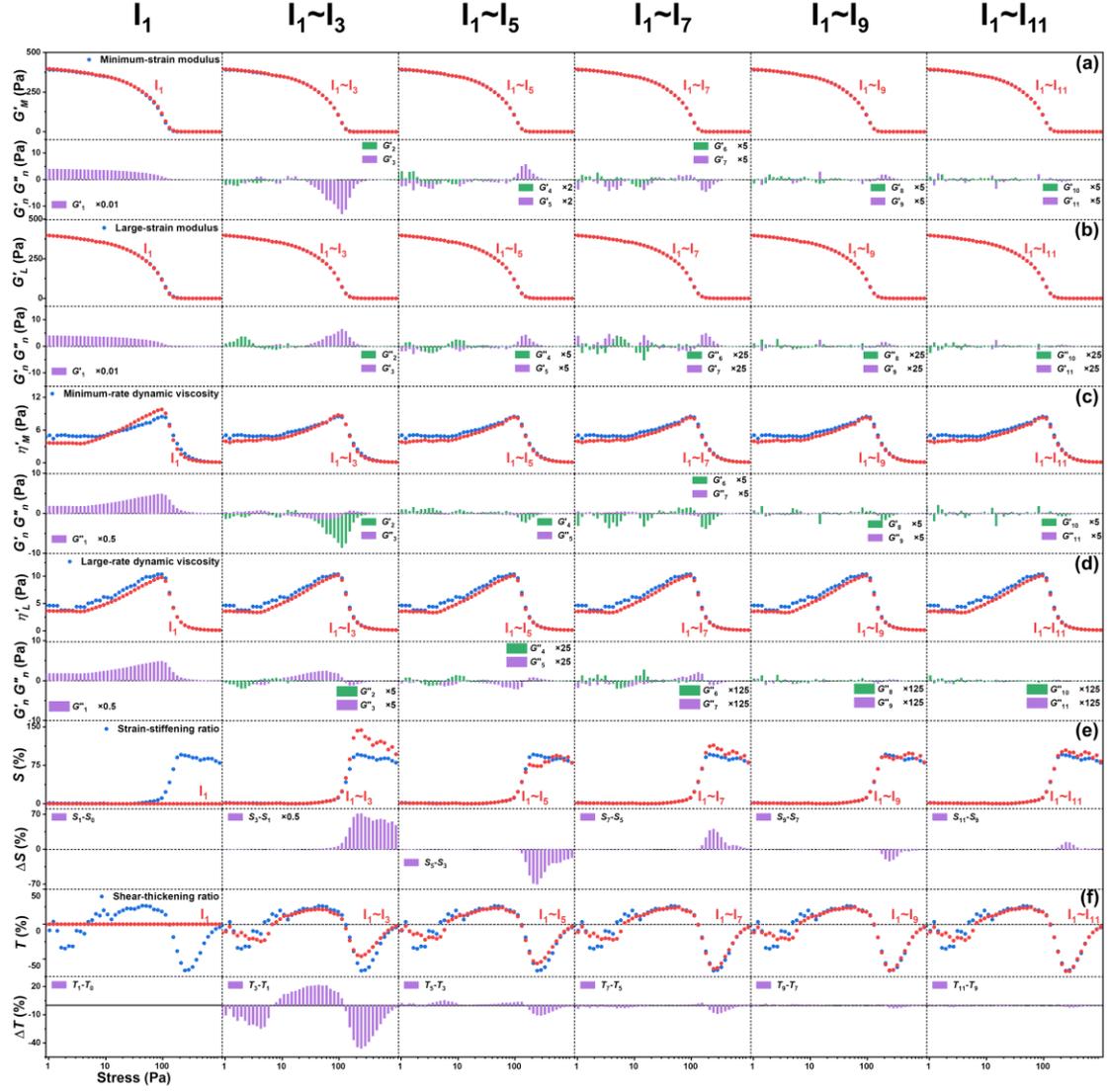

FIG. 10 Comparison between the raw and reconstructed results of *S* and *T* ratios. (a) The plot of the raw and reconstructed minimum-strain moduli $G'_M$ versus stress amplitude, where different kinds of results are generated based on different numbers of harmonics. (b-f) The plots of $G'_L$, $\eta'_M$, $\eta'_L$, *S*, and *T* values: (b) large-strain moduli $G'_L$; (c) minimum-rate dynamic viscosity $\eta'_M$; (d) large-rate dynamic viscosity $\eta'_L$; (e) strain-stiffening ratio *S*; and (f) shear-thickening ratio *T*. Blue points: raw results. Red points: reconstructed results. Purple bars: visualized contributions of odd harmonics. Green bars: visualized contributions of even harmonics.

## *2.3. Visualized harmonic contribution*

The visualized contributions of harmonics were represented as the purple and green bars in Fig. 10. In Fig. 10, $I_1$ offered the dominant contributions to the $G'_M$, $G'_L$, $\eta'_M$, and $\eta'_L$ values whereas $I_1$ presented no contribution to the *S* and *T* values. By contrast, $I_3$ had a negligible influence on the $G'_M$, $G'_L$, $\eta'_M$, and $\eta'_L$ values but dominated the trends of *S* and *T* values, which was closely associated with the physical interpretation of the intra-cycle nonlinearity.

The contributions of $I_2$ and $I_4$ were negligible for $G'_M$, $G'_L$, and $\eta'_L$ values. However, for the $\eta'_M$ values, the impact of $I_2$ and $I_4$ was more significant than $I_3$ and $I_5$, respectively. This phenomenon originates from the fact that the $\eta'_M$ value adopts the elastic modulus of $I_2$ and the viscous modulus of $I_3$, where the elastic contributions are overall higher than the viscous contributions in the SAOS region. Therefore, when the raw data was processed by the method of $S$ and $T$ ratios, even harmonics were inherently important for the $\eta'_M$ value. Furthermore, compared with $I_3$ and $I_5$, $I_2$ and $I_4$ provided obvious contributions to the $G'_M$, $G'_L$, $\eta'_M$, and $\eta'_L$ values in the SAOS region. The fluctuation of the $T$ values at a very low stress level can be attributed to the even harmonics.

For the $S$ values, the contributions of $I_4$ and $I_5$ were comparable to those of $I_2$ and $I_3$, while the contributions of $I_6$ and $I_7$ were much lower than those of $I_2$ and $I_3$ (Fig. 1(c)). On the contrary, for the $T$ values, the contributions of $I_4$ and $I_5$ were negligible compared with $I_2$ and $I_3$. However, $I_4 \sim I_{11}$ were still important to approximate the raw $T$ values. Furthermore, the contributions of higher harmonics to the $S$ values presented the alternating positive and negative values. In other words, $I_2 \sim I_3$, $I_6 \sim I_7$, and $I_{10} \sim I_{11}$ corresponded to positive contributions, while $I_6 \sim I_7$ and $I_{10} \sim I_{11}$ presented negative contributions. For the $T$ values in the LAOS region, all higher harmonics presented negative contributions. Accordingly, this result also indicates how the Fourier coefficients affect the reconstructed $S$ and $T$ values to mimic the raw results. As a result, the $I_1 \sim I_{11}$-based outcomes showed high similarities to the raw results. Further detailed information was provided in Appendix D.

## 3. Summary

In brief, the $S$ and $T$ ratios were introduced in Sec. II E1. Then, the principle of the aLAOS approach in $S$ and $T$ ratios was discussed in Sec. II E2.1, which was also displayed in Fig. 9. The equations of $G'_M$ (Eqs. 41 and 42), $G'_L$ (Eqs. 44 and 45), $\eta'_M$ (Eqs. 47 and 48), $\eta'_L$ (Eqs. 50 and 51), $S$ (Eq. 52), and $T$ (Eq. 53) were clearly given in the form of Fourier coefficients. Accordingly, the method of $S$ and $T$ ratios was directly related to the FT rheology and Fourier coefficients. From the above-mentioned equations, the contributions of even harmonics to the values of $G'_M$, $G'_L$, $\eta'_M$, $\eta'_L$, $S$, and $T$ were directly shown, which indicates that $G'_L$ is influenced by the viscous moduli ($G''_n$) of even harmonics, whereas even harmonics affect $\eta'_M$ via elastic moduli ($G'_n$). In addition, these equations coincide with those of Ewoldt et al. [23, 24] when even harmonics are removed (Eq. 54).

Then, the reconstructed $S$ and $T$ values were obtained by combining the Fourier coefficients in Fig. 10 and Sec. II E2.2. The aLAOS approach in $S$ and $T$ ratios offers the reconstructed values with high similarities to the raw values. Figure 10 leads to a general criterion, indicating that $I_1 \sim I_3$, $I_1 \sim I_7$, and $I_1 \sim I_{11}$ can carry out qualitative, approximate, and quantitative evaluations, respectively. Finally, the visualization of the harmonic contributions is displayed in Fig. 10 based on Eq. 55, which shows that $I_1$ provides the dominant contributions to $G'_M$, $G'_L$, $\eta'_M$, and $\eta'_L$ while $I_1$ offers no contribution to $S$ and $T$. On the contrary, $I_3$ had a negligible influence on $G'_M$, $G'_L$, $\eta'_M$, and $\eta'_L$ but dominated the $S$ and $T$ values. Meanwhile, the visualization reflects how the Fourier coefficients adjust the reconstructed *S* and *T* values to mimic the raw values.

Briefly, the aLAOS approach in *S* and *T* ratios is rational and provides another perspective to understand the *S* and *T* ratios. Undoubtedly, the proposed method brings convenience by using some parameters to calculate *S* and *T* ratios instead of processing linear fittings.

**F. aLAOS in stress bifurcation**

The stress decomposition has been introduced for interpreting the arisen nonlinearity in the LAOS region. Then, the method of *S* and *T* ratios to evaluate the strain stiffening/softening and shear thickening/thinning behaviors was further introduced. After that, based on the stress decomposition proposed by Yu et al. [21], a yield stress determination method was further proposed and named stress bifurcation [28], evaluating the start and end of the solid-liquid transition during an oscillatory stress/strain sweep. Therefore, in this section, the use of the aLAOS approach in stress bifurcation will be examined.

*1. Stress bifurcation in LAOS*

The yield stress is recognized as an essential parameter of YSFs, which is crucial for the design and application of complex fluids in many fields, such as colloidal gels, paints, cosmetics, wet concrete, and foods, and even foods for special medical purposes in dysphagia therapy [1-3]. However, yield stress determination relies on the test procedure including methods, equipment, and conditions [2]. Obvious differences may appear among these yield stresses. These differences are inherent because, for example, the yield stress determination is associated with both the stress amplitude and time, where the difference between the stress amplitude and the true yield stress dominates the time of the solid-liquid transition [28, 78-80]. The yield stress determination methods can be categorized as follows: steady shear, transient shear, and dynamic oscillatory shear methods. It has been indicated that each method has advantages and deficiencies (for more information, please refer to Yu et al. [28]).

The main function of the stress bifurcation is to determine the onset and end of a solid-liquid transition, which helps to understand the yielding behavior in LAOS. The general geometric average method proposed by Yu et al. [21] was adopted as the base of the stress bifurcation, which has been discussed in Sec. II D.

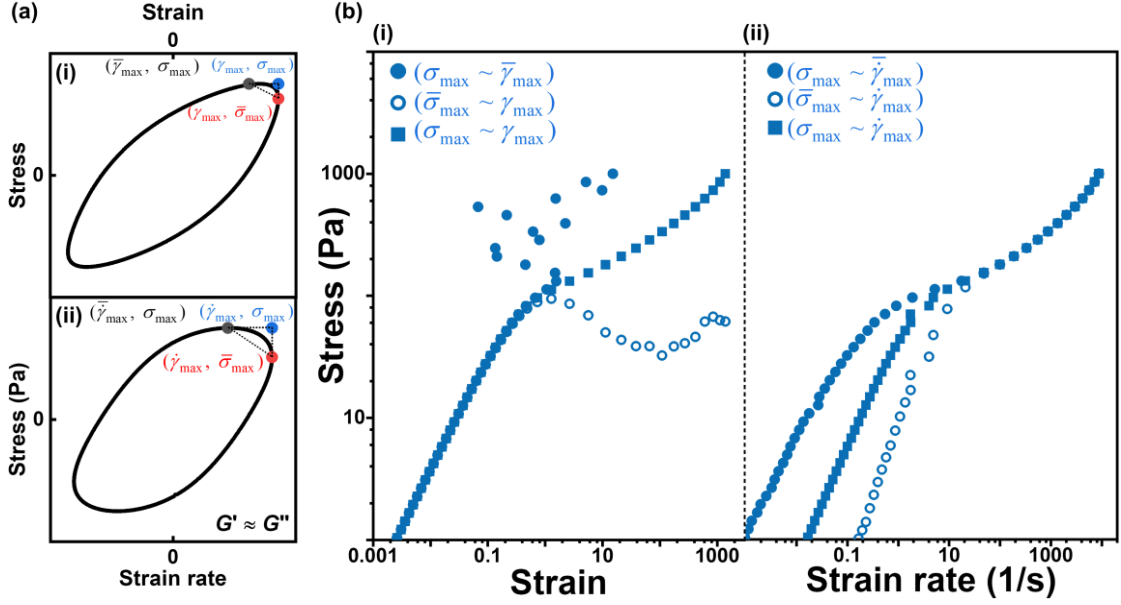

FIG. 11 The principle and representative results of stress bifurcation. (a) The principle of stress bifurcation by using the data of point 46 presenting $G' \approx G''$: (i) the elastic Lissajous curve plotted with the points of $(\bar{\gamma}_{max}, \sigma_{max})$, $(\gamma_{max}, \bar{\sigma}_{max})$, and $(\gamma_{max}, \sigma_{max})$; (ii) the viscous Lissajous curve plotted with the points of $(\bar{\dot{\gamma}}_{max}, \sigma_{max})$, $(\dot{\gamma}_{max}, \bar{\sigma}_{max})$, and $(\dot{\gamma}_{max}, \sigma_{max})$. (b) A series of the above six points from each sweep test point form (i) three $\sigma \sim \gamma$ curves ($\sigma_{max} \sim \bar{\gamma}_{max}$, $\bar{\sigma}_{max} \sim \gamma_{max}$, and $\sigma_{max} \sim \gamma_{max}$) and (ii) three $\sigma \sim \dot{\gamma}$ curves ($\sigma_{max} \sim \bar{\dot{\gamma}}_{max}$, $\bar{\sigma}_{max} \sim \dot{\gamma}_{max}$, and $\sigma_{max} \sim \dot{\gamma}_{max}$).

In detail, the stress bifurcation can be performed according to the following statements and Fig. 11(a). First, the rheological signals of each stress/strain sweep point were transformed into the elastic and viscous Lissajous curves [3, 21, 28]. From the elastic and viscous Lissajous curves, four average parameters were defined. Two parameters ($\bar{\sigma}$) correspond to the decomposed elastic and viscous stresses were used that have been discussed in Sec. II D. The other two parameters refer to the average strain ($\bar{\gamma}$) defined as the arithmetic mean of the total strain at a fixed stress ($\sigma$) and the average strain rate ($\bar{\dot{\gamma}}$) defined in the same way as that of $\bar{\gamma}$. Then, four curves can be obtained: $\bar{\sigma} \sim \gamma$ curve, $\bar{\sigma} \sim \dot{\gamma}$ curve, $\sigma \sim \bar{\gamma}$ curve, and $\sigma \sim \bar{\dot{\gamma}}$ curve. Accordingly, six specific points are found in Fig. 11(a) [28]: the four end points of the $\bar{\sigma} \sim \gamma$, $\bar{\sigma} \sim \dot{\gamma}$, $\sigma \sim \bar{\gamma}$, and $\sigma \sim \bar{\dot{\gamma}}$ curves ($(\gamma_{max}, \bar{\sigma}_{max})$, $(\bar{\gamma}_{max}, \sigma_{max})$, $(\dot{\gamma}_{max}, \bar{\sigma}_{max})$, and $(\bar{\dot{\gamma}}_{max}, \sigma_{max})$) as well as the two points from the stress amplitude $\sigma_{max}$, strain amplitude $\gamma_{max}$, and strain rate amplitude $\dot{\gamma}_{max}$ ($(\gamma_{max}, \sigma_{max})$ and $(\dot{\gamma}_{max}, \sigma_{max})$).

Similarly, a series of the above-mentioned six points can be calculated by investigating all the points of the stress sweep test. As a result, the stress ~ strain curves of $\bar{\sigma}_{max} \sim \gamma_{max}$, $\sigma_{max} \sim \bar{\gamma}_{max}$, and $\sigma_{max} \sim \gamma_{max}$, as well as the stress ~ strain rate curves of $\bar{\sigma}_{max} \sim \dot{\gamma}_{max}$, $\sigma_{max} \sim \bar{\dot{\gamma}}_{max}$, and $\sigma_{max} \sim \dot{\gamma}_{max}$ can be obtained. From the stress sweep test demonstrated in Fig. 1(a), the representative curves of stress bifurcation are plotted as the blue points in Fig. 11(b). In Fig. 11(b.i), the three stress ~ strain curves present a superposition at low stress range and show a deviation at high stress range. However, the three stress ~ strain rate curves behave just the opposite, where the superposition and deviation were observed at high and low stress ranges, respectively. The reason for bifurcation can be interpreted by the following Eqs. 12~14.

The two bifurcations were discussed based on the stress decomposition method of Yu et al. [21]. In the SAOS region, all geometric average curves are straight lines and described as the following equations [21, 28]:

$$\bar{\sigma} = G_{\bar{\sigma}}\gamma \ , \ \sigma = G_{\bar{\gamma}}\bar{\gamma} \ , \ \bar{\sigma} = \eta_{\bar{\sigma}}\dot{\gamma} \ , \ \sigma = \eta_{\bar{\dot{\gamma}}}\bar{\dot{\gamma}} \ , \tag{56}$$

where the slopes of these geometric average lines are represented by introducing the phase angle $\delta$ as follows:

$$G_{\bar{\sigma}} = G' \ , \ \eta_{\bar{\sigma}} = G''/\omega \ , \ G_{\bar{\gamma}} = |G^*|/\cos\delta \ , \ \eta_{\bar{\dot{\gamma}}} = |G^*|/\omega\sin\delta \ . \tag{57}$$

The differences between the slopes of the two stress ~ strain average curves, as well as the slopes of the two stress ~ strain rate average curves, are expressed via Eq. 14:

$$G_{\bar{\gamma}}/G_{\bar{\sigma}} = 1/\cos^2\delta \ , \ \eta_{\bar{\dot{\gamma}}}/\eta_{\bar{\sigma}} = 1/\sin^2\delta \ . \tag{58}$$

For solidlike samples, a good superposition between the $\bar{\sigma}_{max} \sim \gamma_{max}$ and $\sigma_{max} \sim \bar{\gamma}_{max}$ curves appears because $\delta$ is close to zero degree while a significant deviation arises between the $\bar{\sigma}_{max} \sim \dot{\gamma}_{max}$ and $\sigma_{max} \sim \bar{\dot{\gamma}}_{max}$ curves. By contrast, for liquidlike samples, it is clear that the $\bar{\sigma}_{max} \sim \gamma_{max}$ and $\sigma_{max} \sim \bar{\gamma}_{max}$ curves show a big difference, as well as the $\bar{\sigma}_{max} \sim \dot{\gamma}_{max}$ and $\sigma_{max} \sim \bar{\dot{\gamma}}_{max}$ curves present a good overlap, since $\delta$ is close to 90º.

As shown in Fig. 11(b.i), the $\sigma \sim \gamma$ curves superposed at the low stress range, where the strain is linearly dependent on stress, reflecting an ideal linear solidlike behavior. The deviation from linearity happens with the increase in the stress, while the superposition holds, which demonstrates the nonlinear solidlike behavior before the solid-liquid transition. The bifurcation point appears when the stress further increases. Then, $\sigma_{1,s}$ (82.4 Pa) and $\gamma_{1,s}$ (0.485) can be obtained from the first bifurcation point (point 44), which is represented as the critical position to maintaining the solidlike behavior. For the $\sigma \sim \dot{\gamma}$ curves in Fig. 5(a.ii), the change in the curves was opposite of that of the $\sigma \sim \gamma$ curves. The obtained $\sigma_{1,e}$ (154 Pa) and $\dot{\gamma}_{1,e}$ (47.9 (1/s)) from the other bifurcation point (point 48) denote also the critical point above, in which the complete solid-liquid transition happens. The reason for the yield behavior of Carbopol gel is that Carbopol gel contains disorderly arranged microgels to resist external stress at high concentrations [81].

Thus, during the oscillatory sweep test of a YSF, two bifurcations were observed. Accordingly, the bifurcation in Fig. 11(b.i) represents the start yield point and the bifurcation in Fig. 11(b.ii) denotes the end yield point, where a solid-liquid transition region can be further defined as the range between the start yield point and the end yield point. As a result, the start yield stress $\sigma_{1,s}$ and the start yield strain $\gamma_{1,s}$ were obtained from the bifurcation point in Fig. 11(b.i), reflecting the maximum stress and strain to sustain the solidlike behavior. Meanwhile, the bifurcation in Fig.

11(b.ii) provides the end yield stress $\sigma_{1,e}$ and the end yield strain rate $\dot{\gamma}_{1,e}$, indicating that a full solid-liquid transition happens above the minimum stress and strain rate within an oscillatory shear cycle. Thus, stress bifurcation provides an approach to evaluate yielding behavior.

*2. Analytic LAOS approach in stress bifurcation*

The use of the aLAOS approach in stress bifurcation generates two types of methods: algebraic stress bifurcation (Fig. 12) and analytic stress bifurcation (Fig. 13). The two analytic methods aim to promote the efficiency of the previously proposed stress bifurcation. Algebraic stress bifurcation is considered the most efficient way to obtain almost the same start and end yield points as those from stress bifurcation, where only the oscillatory stress/strain sweep data ($G'$, $G''$, and the stress/strain amplitude) are needed (Fig. 12(c)). Therefore, algebraic stress bifurcation will be first and briefly introduced. Then, the analytic stress bifurcation is established to give the reconstructed results with high similarities to the raw results (Fig. 13).

*2.1. Principle of algebraic stress bifurcation*

Algebraic stress bifurcation adapts FT rheology, stress decomposition, and stress bifurcation as the fundaments.

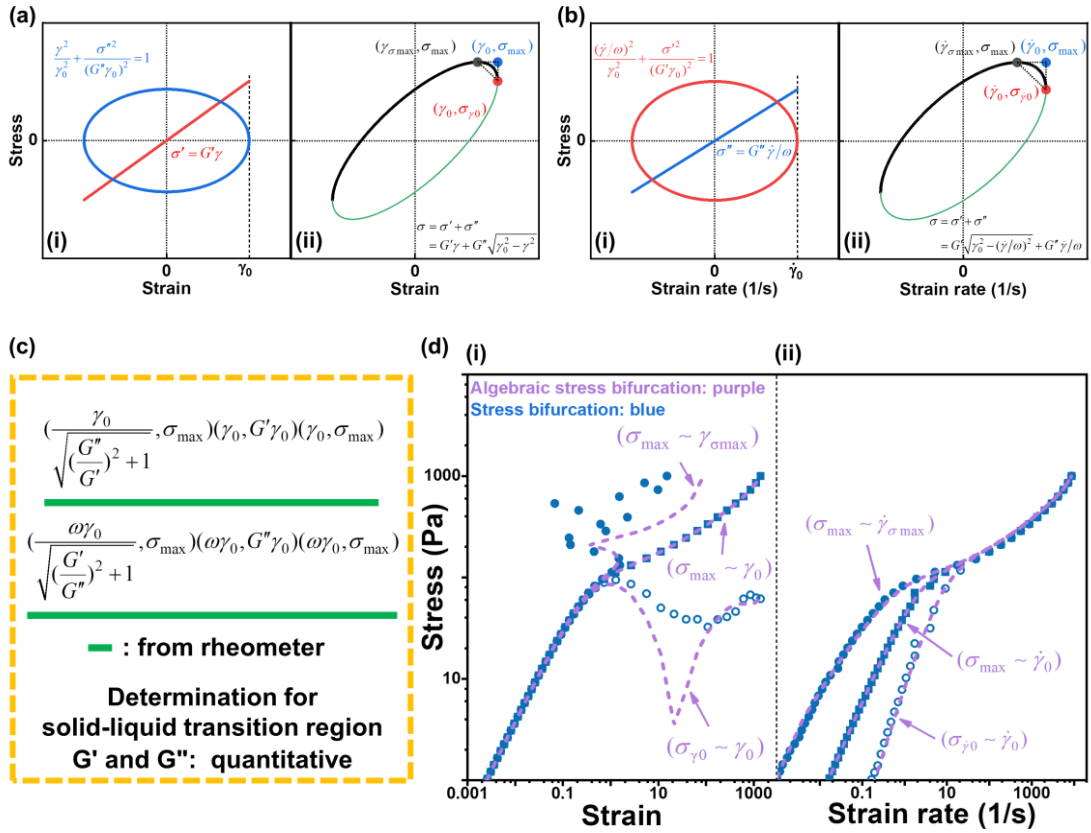

FIG. 12 The principle and representative results of algebraic stress bifurcation. (a) Algebraic stress bifurcation provides the point coordinates of $(\gamma_{\sigma\max}, \sigma_{\max})$, $(\gamma_0, \sigma_{\gamma 0})$, and $(\gamma_0, \sigma_{\max})$ to approximate

the three points of $(\bar{\gamma}_{max}, \sigma_{max})$, $(\gamma_{max}, \bar{\sigma}_{max})$, and $(\gamma_{max}, \sigma_{max})$ from stress bifurcation (the situation of point 46): (i) the algebraic elastic Lissajous curves of a pure elastic sample (red line) and a pure viscous sample (blue ellipse); (ii) the algebraic elastic Lissajous curve constructed by combining the above elastic and viscous responses. (b) The points of $(\dot{\gamma}_{\sigma max}, \sigma_{max})$, $(\dot{\gamma}_0, \sigma_{\dot{\gamma}0})$, and $(\dot{\gamma}_0, \sigma_{max})$ were offered by algebraic stress bifurcation to mimic the three points of $(\bar{\dot{\gamma}}_{max}, \sigma_{max})$, $(\dot{\gamma}_{max}, \bar{\sigma}_{max})$, and $(\dot{\gamma}_{max}, \sigma_{max})$ from stress bifurcation (the situation of point 46): (i) the algebraic viscous Lissajous curves of a pure elastic sample (red line) and a pure viscous sample (blue ellipse); (ii) the algebraic viscous Lissajous curve plotted by involving the two purely elastic and viscous responses. (c) The mathematical framework of algebraic stress bifurcation. (d) A series of the above six points from each sweep test point form (i) three purple $\sigma \sim \gamma$ curves ($\sigma_{max} \sim \gamma_{\sigma max}$, $\sigma_{\gamma 0} \sim \gamma_0$, and $\sigma_{max} \sim \gamma_0$) and (ii) three purple $\sigma \sim \dot{\gamma}$ curves ($\sigma_{max} \sim \dot{\gamma}_{\sigma max}$, $\sigma_{\dot{\gamma}0} \sim \dot{\gamma}_0$, and $\sigma_{max} \sim \dot{\gamma}_0$). Meanwhile, the results from stress bifurcation (blue points) were involved to provide a typical comparison.

In the elastic Lissajous curve, for a pure elastic sample, the strain ($\gamma_{\sigma max}$) of the maximal stress $\sigma_{max}$ is equal to $\gamma_0$. By contrast, for a pure viscous sample, $\sigma_{max}$ is achieved at $\gamma = 0$ [18]. This is demonstrated in Fig. 12(a.i) containing two linear viscoelastic responses, a purely elastic response $\sigma'(\gamma)$ and a pure viscous response $\sigma''(\gamma)$ [6, 18], which is expressed as the following equations:

$$\sigma'(\gamma) = G'\gamma, \quad \frac{\gamma^2}{\gamma_0^2} + \frac{\sigma''(\gamma)^2}{(G''\gamma_0)^2} = 1. \tag{59}$$

After that, from the FT rheology [6] and stress decomposition [20], the algebraic elastic Lissajous curve is constructed in Fig. 12(a.ii) by combining the pure elastic and pure viscous responses in Fig. 12(a.i). Accordingly, three specific points demonstrated in Fig. 12(a.ii) are calculated in the following way, where the point of $(\gamma_0, \sigma_{max})$ can be directly obtained without calculation.

First, the top half of the algebraic elastic Lissajous curve is represented as the following equation:

$$\sigma(\gamma) = \sigma'(\gamma) + \sigma''(\gamma) = G'\gamma + G''\sqrt{\gamma_0^2 - \gamma^2}. \tag{60}$$

Then, the coordinates of the maximum stress $(\gamma_{\sigma max}, \sigma_{max})$ and the maximum strain $(\gamma_0, \sigma_{\gamma 0})$ in Fig. 12(a.ii) in the algebraic elastic Lissajous curve are deduced as follows:

$$\frac{d\sigma(\gamma)}{d\gamma}\bigg|_{\gamma=\gamma_{\sigma max}} = 0, \quad \sigma_{\gamma 0} = \sigma(\gamma_0) = G'\gamma_0. \tag{61}$$

After that, the value of $\gamma_{\sigma max}$ can be obtained from Eq. 61 as the following equation:

$$\gamma_{\sigma max} = \frac{\gamma_0}{\sqrt{(\frac{G''}{G'})^2 + 1}}. \tag{62}$$

Finally, the coordinates of $(\gamma_0, \sigma_{\gamma 0})$ are represented as $(\gamma_0, G'\gamma_0)$ according to Eq. 61. Therefore, the three specific points in Fig. 12(a.ii) were calculated. In the same way, except for the $(\dot{\gamma}_0, \sigma_{max})$ point, the two specific points of $(\dot{\gamma}_{\sigma max}, \sigma_{max})$ and $(\dot{\gamma}_0, \sigma_{\dot{\gamma}0})$ in Fig. 12(b.ii) were obtained from the

following process:

$$\sigma(\dot{\gamma}/\omega) = \sigma'(\dot{\gamma}/\omega) + \sigma''(\dot{\gamma}/\omega) = G'\sqrt{\gamma_0^2 - (\dot{\gamma}/\omega)^2} + G''\dot{\gamma}/\omega, \tag{63}$$

$$\left.\frac{d\sigma(\dot{\gamma}/\omega)}{d(\dot{\gamma}/\omega)}\right|_{\dot{\gamma}=\dot{\gamma}_{\sigma\max}} = 0, \quad \dot{\gamma}_{\sigma\max} = \frac{\omega\gamma_0}{\sqrt{(\frac{G'}{G''})^2 + 1}}, \quad \sigma_{\dot{\gamma}0} = \sigma(\dot{\gamma}_0/\omega) = G''\dot{\gamma}_0/\omega = G''\gamma_0. \tag{64}$$

As a result, the three specific points $(\bar{\gamma}_{\max}, \sigma_{\max})$, $(\gamma_{\max}, \bar{\sigma}_{\max})$, $(\gamma_{\max}, \sigma_{\max})$ in Fig. 11(a.i) and the three specific points $(\bar{\dot{\gamma}}_{\max}, \sigma_{\max})$, $(\dot{\gamma}_{\max}, \bar{\sigma}_{\max})$, $(\dot{\gamma}_{\max}, \sigma_{\max})$ in Fig. 11(a.ii) can be represented as follows:

$$(\gamma_{\sigma\max}(=\frac{\gamma_0}{\sqrt{(\frac{G''}{G'})^2+1}}), \sigma_{\max}), \quad (\gamma_0, \sigma_{\gamma 0}(=G'\gamma_0)), \quad (\gamma_0, \sigma_{\max}), \tag{65}$$

$$(\dot{\gamma}_{\sigma\max}(=\frac{\omega\gamma_0}{\sqrt{(\frac{G'}{G''})^2+1}}), \sigma_{\max}), \quad (\omega\gamma_0, \sigma_{\dot{\gamma}0}(=G''\gamma_0)), \quad (\omega\gamma_0, \sigma_{\max}), \tag{66}$$

In addition, figure 12(a) clearly demonstrates the efficiency of the algebraic stress bifurcation by using only the data of $G'$, $G''$, and the stress/strain amplitude. Furthermore, the six needed points of $(\gamma_{\sigma\max}, \sigma_{\max})$, $(\gamma_0, \sigma_{\gamma 0})$, $(\gamma_0, \sigma_{\max})$, $(\dot{\gamma}_{\sigma\max}, \sigma_{\max})$, $(\dot{\gamma}_0, \sigma_{\dot{\gamma}0})$, and $(\dot{\gamma}_0, \sigma_{\max})$ are visually displayed in Figs. 12(a.ii) and 12(b.ii).

In the same way, a series of the above-mentioned six points can be obtained by studying all the sweep points to form three stress ~ strain curves of $\sigma_{\max} \sim \gamma_{\sigma\max}$, $\sigma_{\gamma 0} \sim \gamma_0$, and $\sigma_{\max} \sim \gamma_0$ (the purple dotted lines in Fig. 12(d.i)), as well as three stress ~ strain rate curves of $\sigma_{\max} \sim \dot{\gamma}_{\sigma\max}$, $\sigma_{\dot{\gamma}0} \sim \dot{\gamma}_0$, and $\sigma_{\max} \sim \dot{\gamma}_0$ (the purple dotted lines in Fig. 12(d.ii)). Then, the first bifurcation point (point 44, Fig. 12(d.i)) can be determined as the start yield point and the other bifurcation point (point 48, Fig. 12(d.ii)) represents the end yield point. The reason for the bifurcation can be interpreted by the following two equations (Eqs. 67 and 68).

For the algebraic stress bifurcation, the physical reasons for the two bifurcation points can be further rationalized. For the upper and lower $\sigma \sim \gamma$ curves in Fig. 12(d.i), the horizontal ($D_h$) and vertical ($D_v$) differences between the points of $(\gamma_{\sigma\max}, \sigma_{\max})$ and $(\gamma_0, \sigma_{\gamma 0})$ can be expressed as the following equation:

$$D_h = (\gamma_0/\sqrt{(G''/G')^2+1})/\gamma_0 = \frac{1}{\sqrt{(\frac{G''}{G'})^2+1}} = G'\gamma_0/\sigma_{\max} = D_v, \tag{67}$$

which indicates that the values of $D_h$ and $D_v$ are close to the value of one for solidlike samples and zero for liquidlike samples, denoting that $(\gamma_{\sigma\max}, \sigma_{\max})$ and $(\gamma_0, \sigma_{\gamma 0})$ points show good overlap and huge differences, respectively.

In the same way, for the upper and lower $\sigma \sim \dot{\gamma}$ curves in Fig. 12(d.ii), the $D_h$ and $D_v$ between the $(\dot{\gamma}_{\sigma\max}, \sigma_{\max})$ and $(\dot{\gamma}_0, \sigma_{\dot{\gamma}0})$ points are calculated as the following equations:

$$D_\mathrm{h} = D_\mathrm{v} = \frac{1}{\sqrt{(\frac{G'}{G''})^2 + 1}} \; . \tag{68}$$

Similarly, the $D_\mathrm{h}$ and $D_\mathrm{v}$ are close to the value of zero for solidlike samples and one for liquidlike samples, reflecting that the $(\dot{\gamma}_{\sigma\max}, \sigma_{\max})$ and $(\gamma_0, \sigma_{\dot{\gamma}0})$ points display significant differences and good overlap, respectively.

Thus, during the solid-liquid transition, two bifurcation points are able to be visually distinguished and determined in Figs. 12(d.i) and 12(d.ii) (purple lines). It is clear that a good agreement was reached between the raw and algebraic results. As can be observed in Fig. 12(d), except for the points at large stress range in the plot of $\sigma \sim \gamma$ curve (Fig. 12(d.i)), a good overlap was shown for both the plots of $\sigma \sim \gamma$ (Fig. 12(d.i)) and $\sigma \sim \dot{\gamma}$ (Fig. 12(d.ii)) curves. The algebraic stress bifurcation evaluates the proximity between the values of $G'$ and $G''$. Then, the start and end yield points can be obtained when $G'$ is close to and away from $G''$ during the stress/strain sweep process, respectively. In other words, the algebraic stress bifurcation gives the start and end points of a solid-liquid transition via obtaining two bifurcation points (thereby solid-liquid transition region is provided). Meanwhile, the start and end points of the solid-liquid transition from the algebraic stress bifurcation are almost the same as those from the stress bifurcation.

The excellent consistency has been discussed in our previously work where the algebraic stress bifurcation has been proposed [82], which can be briefly introduced here by considering the highest harmonic of higher harmonics $I_3$. The $\sigma_{\gamma 0}$ and $\sigma_{\dot{\gamma} 0}$ were obtained from the FT of the distorted stress signal, while $\gamma_{\sigma\max}$ and $\dot{\gamma}_{\sigma\max}$ were calculated from the FT of the distorted strain signal, where the interpolation was applied for the conversion between the distorted stress signal with a perfect sinusoidal strain signal and the distorted strain signal with a perfect sinusoidal stress signal:

$$\sigma(t) = \gamma_0 (G_1' \sin\omega t + G_1'' \cos\omega t + G_3' \sin 3\omega t + G_3'' \cos 3\omega t) \; , \tag{69}$$

$$\sigma_{\gamma 0} = \sigma(t)\Big|_{t=\frac{\pi}{2\omega}} = \gamma_0 (G_1' - G_3') \; , \quad \sigma_{\dot{\gamma} 0} = \sigma(t)\big|_{t=0} = \gamma_0 (G_1'' + G_3'') \; , \tag{70}$$

$$\gamma(t) = \gamma_0 (b_1 \sin\omega t + a_1 \cos\omega t + b_3 \sin 3\omega t + a_3 \cos 3\omega t) \; , \tag{71}$$

$$\gamma_{\sigma\max} = \gamma(t)\Big|_{t=\frac{\pi}{2\omega}} = \gamma_0 (b_1 - b_3) \; , \quad \dot{\gamma}_{\sigma\max} = \frac{d\gamma(t)}{dt}\Big|_{t=\frac{\pi}{2\omega}} = \omega\gamma_0 (-a_1 + 3a_3) \; . \tag{72}$$

For point 44 (the start yield point), considering $I_1$ and $I_3$ ($G_1' = 160$ Pa, $G_1'' = 58.7$ Pa, $G_3' = -2.89$ Pa, $G_3'' = 2.27$ Pa, $b_1 = 0.944$, $a_1 = -0.361$, $b_3 = -0.0106$, $a_3 = 0.0200$, $\gamma_0 = 0.485$), the values of $\sigma_{\gamma 0}$, $\gamma_{\sigma\max}$, $\sigma_{\dot{\gamma} 0}$, and $\dot{\gamma}_{\sigma\max}$ were 79.0 Pa, 0.463, 29.6 Pa, and 0.917. If only $I_1$ was considered, these values were 77.6 Pa, 0.458, 28.5 Pa, and 1.10 with corresponding errors of 1.80%, 1.09%, 3.86%, and 16.6%, respectively. Note that the error in $\dot{\gamma}_{\sigma\max}$ (16.6%) is meaningless for the judgment of the start yield point. When point 48 is concerned (the end yield point, $G_1' = 9.11$ Pa, $G_1'' = 28.1$ Pa, $G_3' = -2.52$ Pa, $G_3'' = -1.52$ Pa, $b_1 = 0.247$, $a_1 = -1.04$, $b_3 = -0.0176$, $a_3 = 0.0931$, $\gamma_0 = 5.67$), the values of $\sigma_{\gamma 0}$, $\gamma_{\sigma\max}$, $\sigma_{\dot{\gamma} 0}$, and $\dot{\gamma}_{\sigma\max}$ were 65.9 Pa, 1.50, 151 Pa, and 47.0. These were 51.7 Pa, 1.40, 160 Pa, and 37.0 by only considering $I_1$, where the corresponding errors were 27.5%, 7.14%, -5.63%, and 27.0%. Note that the errors in $\sigma_{\gamma 0}$ (27.5%) and $\gamma_{\sigma\max}$ (7.14%) are negligible for the determination of the end yield point. However, a big error in $\dot{\gamma}_{\sigma\max}$ (27.0%) arises, which can be interpreted by the $D_h$. $D_h$ is 0.964 by only considering $I_1$, while $D_h$ is equal to 0.963 by introducing $I_1$ and $I_3$ along with an error of 0.104%. Therefore, the error in $\dot{\gamma}_{\sigma\max}$ (27.0%) does not influence the judgment of the end yield point.

In a word, the algebraic stress bifurcation also defines a start yield point and an end yield point along with a determined solid-liquid transition region. The algebraic stress bifurcation allows obtaining the start and the end yield points based only on the data of $G'$, $G''$, and stress/strain amplitude (Eqs. 65 and 66, Fig. 12(c)), which shows a highly promoted efficiency compared with the complex stress bifurcation method. In addition, in practice, the algebraic Lissajous curves based on $G'$ and $G''$ (Figs. 12(a) and 12(b)) are needless since the correlation between each sweep point ($G'$, $G''$) and the corresponding six points of algebraic stress bifurcation has been abstracted (Eqs. 65 and 66). This method allows the application of less specialized software, such as Origin® and Excel, instead of the professional software of MATLAB.

## 2.2. Principle of analytic stress bifurcation

Although the algebraic stress bifurcation based on $G'$ and $G''$ is enough to give quantitative results (the start and end yield points as well as a solid-liquid transition region), the use of the aLAOS approach in stress bifurcation, analytic stress bifurcation, was proposed for providing reconstructed points with more approximate coordinates to address the deviations between the results from the algebraic stress bifurcation and stress bifurcation.

Firstly, the Fourier series of the distorted stress and strain signals were as follows:

$$\sigma(t) = \gamma_0 \sum_{n=1} (G_n'' \cos n\omega t + G_n' \sin n\omega t), \qquad (73)$$

$$\gamma(t) = \gamma_0 \sum_{n=1} (a_n \cos n\omega t + b_n \sin n\omega t). \qquad (74)$$

For giving the results from the analytic stress bifurcation, five parameters are needed including $\sigma_{\gamma 0}$, $\gamma_{\sigma\max}$, $\sigma_{\dot{\gamma} 0}$, $\dot{\gamma}_{\sigma\max}$, and $\dot{\gamma}_0$. Similarly, the $\sigma_{\gamma 0}$ and $\sigma_{\dot{\gamma} 0}$ were calculated via the FT of the distorted stress signal (Eq. 73), as well as $\gamma_{\sigma\max}$ and $\dot{\gamma}_{\sigma\max}$ were provided by the FT of the distorted strain signal (Eq. 74). For the $\dot{\gamma}_0$ value, the analytical solution of $d^2\gamma(t)/dt^2 = 0$ was difficult to be found, where the approximate solution was taken. The principle of the analytic stress bifurcation can be expressed by using the following equations:

$$\sigma_{\gamma 0} = \sigma(t)\Big|_{t=\frac{\pi}{2\omega}} = \gamma_0 \Big( \sum_{n=1,odd} (-1)^{\frac{n-1}{2}} G_n' + \sum_{n=2,even} (-1)^{\frac{n}{2}} G_n'' \Big), \qquad (75)$$

$$\gamma_{\sigma\max} = \gamma(t)\Big|_{t=\frac{\pi}{2\omega}} = \gamma_0 \Big( \sum_{n=1,odd} (-1)^{\frac{n-1}{2}} b_n + \sum_{n=2,even} (-1)^{\frac{n}{2}} a_n \Big), \qquad (76)$$

$$\sigma_{\dot{\gamma} 0} = \sigma(t)\Big|_{t=0} = \gamma_0 \Big( \sum_{n=1,odd} G_n'' + \sum_{n=2,even} G_n'' \Big), \qquad (77)$$

$$\dot{\gamma}_{\sigma\max} = \left.\frac{d\gamma(t)}{dt}\right|_{t=\frac{\pi}{2\omega}} = \omega\,\gamma_0 \left( \sum_{n=1,odd} (-1)^{(\frac{n+1}{2})} n a_n + \sum_{n=2,even} (-1)^{(\frac{n}{2}+1)} n b_n \right), \tag{78}$$

$$\dot{\gamma}_0 = \left.\left|\frac{d\gamma(t)}{dt}\right|\right|_{d^2\gamma(t)/dt^2=0} \approx \dot{\gamma}_{\sigma\max}\Big/\cos(\arcsin(\frac{\gamma_{\sigma\max}}{\gamma_0})). \tag{79}$$

Briefly, in this section, the analytic stress bifurcation was discussed. The equations of $\sigma_{\gamma 0}$, $\gamma_{\sigma\max}$, $\sigma_{\dot\gamma 0}$, $\dot\gamma_{\sigma\max}$, and $\dot\gamma_0$ with simple forms are displayed in Fig. 13(a), which visually displays the simplicity and efficiency of the analytic stress bifurcation, where the values of these parameters are provided based on Fourier coefficients. As a result, the analytic stress bifurcation is a convenient approach since stress bifurcation inherently needs the processing of raw data to calculate a series of the four geometric average curves (the $\bar\sigma\sim\gamma$, $\bar\sigma\sim\dot\gamma$, $\sigma\sim\bar\gamma$, and $\sigma\sim\bar{\dot\gamma}$ curves) by treating the elastic and viscous Lissajous curves of each stress/strain sweep points. The harmonic-based Lissajous curves can be neglected in practice because of the abstracted relationships demonstrated in Fig. 13(a). This method also permits the use of less specialized software instead of MATLAB.

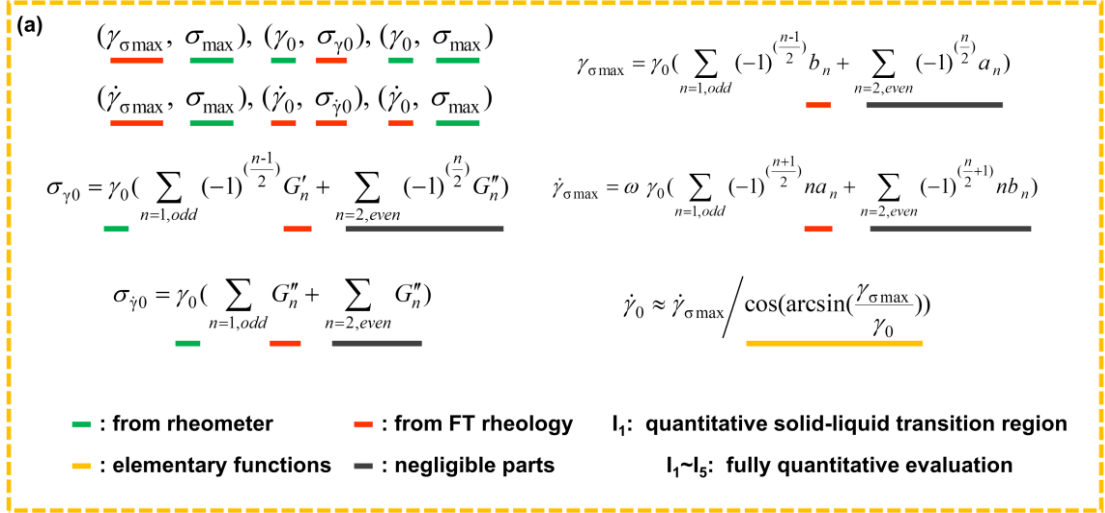

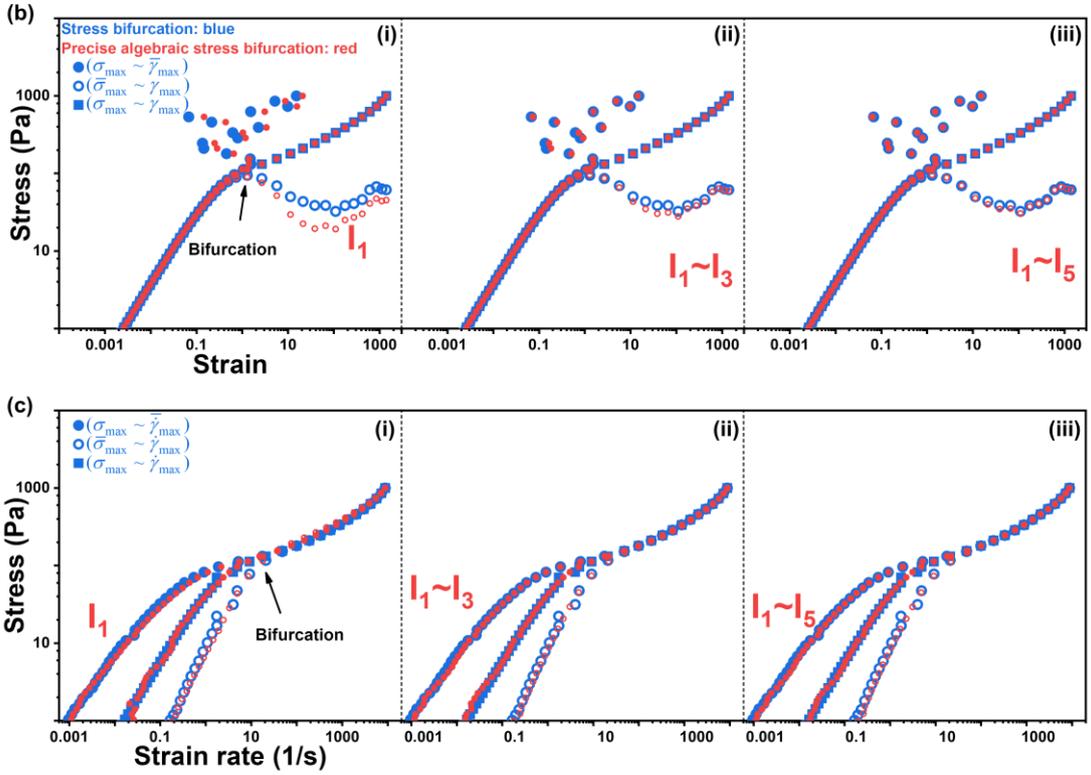

FIG. 13 The principle and representative results of analytic stress bifurcation. (a) The principle of the algebraic stress bifurcation. (b) The plot of raw (three $\sigma \sim \gamma$ curves of $\sigma_{max} \sim \bar{\bar{\gamma}}_{max}$, $\bar{\sigma}_{max} \sim \gamma_{max}$, and $\sigma_{max} \sim \gamma_{max}$) and analytic (three $\sigma \sim \gamma$ curves of $\sigma_{max} \sim \gamma_{\sigma max}$, $\sigma_{\gamma 0} \sim \gamma_0$, and $\sigma_{max} \sim \gamma_0$) stress bifurcation results, where a series of points were obtained by introducing (i) $I_1$, (ii) $I_1 \sim I_3$, and (iii) $I_1 \sim I_5$. (c) The plot of raw (three $\sigma \sim \dot{\gamma}$ curves of $\sigma_{max} \sim \bar{\bar{\dot{\gamma}}}_{max}$, $\bar{\sigma}_{max} \sim \dot{\gamma}_{max}$, and $\sigma_{max} \sim \dot{\gamma}_{max}$) and analytic (three $\sigma \sim \dot{\gamma}$ curves of $\sigma_{max} \sim \dot{\gamma}_{\sigma max}$, $\sigma_{\dot{\gamma} 0} \sim \dot{\gamma}_0$, and $\sigma_{max} \sim \dot{\gamma}_0$) stress bifurcation results, where (i) $I_1$, (ii) $I_1 \sim I_3$, and (iii) $I_1 \sim I_5$. were introduced. Blue lines: raw results. Red lines: reconstructed results.

## 2.3. Whole sweep process

A series of results from the analytic stress bifurcation were generated by using $I_1$, $I_1 \sim I_3$, and $I_1 \sim I_5$ and displayed in Figs. 13(b) and 13(c). More specifically, figures 13(b) and 13(c) show the comparisons between the raw (blue points) and the corresponding reconstructed (red points) results. The $\sigma \sim \gamma$ and $\sigma \sim \dot{\gamma}$ curves are plotted in Figs. 13(b) and 13(c), respectively.

In Figs. 13(b.i) and 13(c.i), by introducing $I_1$, acceptable overlaps appear for both the $\sigma \sim \gamma$ (Fig. 13(b.i)) and $\sigma \sim \dot{\gamma}$ (Fig. 13(c.i)) plots, whereas the points in large stress range in Fig. 13(b.i) just roughly mimic the trend of the raw curves, which shows that $I_1$ can give the quantitative evaluations for the start and end yield points, as well as the solid-liquid transition region (Fig. 13(a)). To handle the arisen deviations in the large stress region, more harmonics were introduced. By introducing $I_2$ and $I_3$, the deviations were greatly eliminated in Fig. 13(b.ii). Meanwhile, the small deviations between the raw and reconstructed results in Fig. 13(c.i) were further corrected in Fig. 13(c.ii). Therefore, the intensities of $I_1 \sim I_3$ were enough for the generation of acceptable results. By further introducing $I_4$ and $I_5$, the reconstructed curves with high similarities to the raw curves were obtained, denoting that 1st~5th harmonics are enough for providing fully quantitative evaluations, regardless of the precise determination of the solid-liquid transition region or all the stress bifurcation points.

In brief, the analytic stress bifurcation also defines a start yield point and an end yield point along with a solid-liquid transition region. The analytic stress bifurcation offers point coordinates with high similarities to the raw results. The results show that $I_1$ can give quantitative evaluations for the start and end yield points. Meanwhile, the intensities of $I_1 \sim I_5$ are enough for generating a fully quantitative evaluation of a solid-liquid transition.

Appendix E provided more information.

## 3. Summary

First, the use of the aLAOS approach in stress bifurcation generates two methods: algebraic stress bifurcation (Sec. II F2.1, Fig. 12) and analytic stress bifurcation (Sec. II F2.2, Fig. 13). Both methods can define the start and end yield points along with a solid-liquid transition region. The two proposed methods promote the efficiency of the stress bifurcation by rejecting the use of geometric average curves obtained from the Lissajous curves. Algebraic stress bifurcation is highly efficient to provide the start and end yield points the same as those from stress bifurcation. Only the $G'$, $G''$, and the stress/strain amplitude are needed (Fig. 12(c)).

Then, analytic stress bifurcation was proposed to handle the arisen deviations between the raw and reconstructed $\sigma \sim \gamma$ curves from algebraic stress bifurcation. Analytic stress bifurcation was proposed based on Fourier coefficients (Fig. 13(a)), where $I_1$ provides the quantitative evaluation for the start and end yield points (Figs. 13(b.i) and 13(c.i)). Analytic stress bifurcation provides $I_1 \sim I_5$-based curves with high similarities to the raw curves (Figs. 13(b.iii) and 13(c.iii)), denoting that the intensities of $I_1 \sim I_5$ are enough for fully quantitative evaluation (Fig. 13(a)).

Furthermore, the two proposed methods allow the application of less specialized software instead of the professional software of MATLAB. Here, since analytic stress bifurcation requires Fourier coefficients, algebraic stress bifurcation is suggested, which is enough for providing the start and end yield points, as well as a solid-liquid transition region. Finally, the aLAOS approach in stress bifurcation is rational and efficient based on Fourier coefficients to offer the start and end yield points instead of calculating geometric average curves from Lissajous curves, which provides another perspective to understand the yielding behavior and solid-liquid transition.

**G. aLAOS in dissipation ratio**

*1. Dissipation ratio in LAOS*

The energy dissipated per unit volume of sample per oscillatory cycle, $E_d = \int_0^{2\pi/\omega} \sigma(t) d\gamma(t)$ [26, 27, 66], can be represented as the loop area in the plot of the elastic Lissajous curve (the blue region in Fig. 14(a)). The elastic Lissajous curve for a perfect plastic response with equivalent stress and strain amplitudes is a rectangular shape (the area within the red square in Fig. 14(a)), where the energy dissipated per cycle is $(E_d)_{pp} = 4\sigma_{max}\gamma_0$. Although the shape of an elastic Lissajous curve is various, the energy dissipated per cycle is only a function of partial $I_1$ (i.e. $G_1''$ and $a_1$). According to previously reported works in the literature [26, 27, 66], $E_d$ is calculated by the equation of $E_d = \pi\gamma_0^2 G_1''$. Therefore, it is clear that $E_d$ is only a function of $G_1''$. However, $G_1''$ correspond to the FT in the strain-controlled situation, where the equation of the stress-controlled condition is not given. Meanwhile, the calculation process from $E_d = \int_0^{2\pi/\omega} \sigma(t) d\gamma(t)$ to $E_d = \pi\gamma_0^2 G_1''$ can be further demonstrated as follows.

For the strain-controlled condition, the situation is:

$$E_d = \int_0^{2\pi/\omega} \gamma_0 \sum_{n=1} (G_n'' \cos n\omega t + G_n' \sin n\omega t) d\gamma_0 \sin \omega t$$
$$= \omega\gamma_0^2 \int_0^{2\pi/\omega} \sum_{n=1} (G_n'' \cos n\omega t \cos \omega t + G_n' \sin n\omega t \cos \omega t) dt \qquad (80)$$

It can be calculated that the integrations $\int_0^{2\pi/\omega} \cos n\omega t \cos \omega t dt$ are all equal to zero value when $n$ is a positive integer and not equal to one value. Meanwhile, the integrations $\int_0^{2\pi/\omega} \sin n\omega t \cos \omega t dt$ are also equal to the zero value when $n$ is a positive integer. Therefore, only the integration $\int_0^{2\pi/\omega} \cos \omega t \cos \omega t dt$ provides a nonzero value. Thus, Eq. 80 can be further calculated:

$$E_d = \gamma_0^2 G_1'' \int_0^{2\pi} \cos^2\theta d\theta = \frac{1}{2}\gamma_0^2 G_1'' \int_0^{2\pi} (1+\cos 2\theta)d\theta = \frac{1}{2}\gamma_0^2 G_1'' \theta \Big|_0^{2\pi} = \pi\gamma_0^2 G_1''. \qquad (81)$$

As a result, at the strain-controlled condition, $E_d$ is only a function of $G_1''$, which essentially originates from the set of orthogonal functions provided by FT. In the same way, the calculation is carried out for the stress-controlled condition as follows:

$$E_d = \int_0^{2\pi/\omega} \sigma_{max} \sin\omega t \, d(\gamma_0 \sum_{n=1}(a_n \cos n\omega t + b_n \sin n\omega t))$$
$$= \sigma_{max}\gamma_0 \int_0^{2\pi} \sum_{n=1}(-na_n \sin n\theta \sin\theta + nb_n \cos n\theta \sin\theta)d\theta \qquad (82)$$
$$= \sigma_{max}\gamma_0 \int_0^{2\pi} -a_1 \sin^2\theta \, d\theta = -\frac{1}{2}a_1\sigma_{max}\gamma_0 \int_0^{2\pi}(1-\cos 2\theta)d\theta = -a_1\pi\sigma_{max}\gamma_0$$

Accordingly, $E_d$ is only a function of $a_1$ at the stress-controlled condition, where the relation between $E_d$ and $a_1$ is $E_d = -a_1\pi\sigma_{max}\gamma_0$.

Therefore, the loop area within an elastic Lissajous curve $E_d$ can be directly obtained by using $G_1''$ (i.e. $E_d = \pi\gamma_0^2 G_1''$) and $a_1$ (i.e. $E_d = -a_1\pi\sigma_{max}\gamma_0$), which can be also identified as the aLAOS approach in dissipation ratio.

After that, the $E_d$ is normalized by $(E_d)_{pp}$, $(E_d)_{pp} = 4\sigma_{max}\gamma_0$, to generate the dissipation ratio $\phi$, $\phi = E_d/(E_d)_{pp}$:

$$\phi = \frac{\pi\gamma_0 G_1''}{4\sigma_{max}} \text{ or } \phi = -\frac{\pi}{4}a_1. \qquad (83)$$

Therefore, it is obvious that the perfect plastic, Newtonian, and pure elastic behaviors correspond to $\phi = 1$, $\pi/4$ ($\approx 0.785$), and 0. The above description is included in Fig. 14(b).

Besides, this method is regarded as "well-behaved" for arbitrary LAOS behavior since the $\sigma_{max}$, $\gamma_0$, and $G_1''$ are always easily determined from the experiment. Thus, the dimensionless dissipation ratio $\phi$ can be applied to all measured LAOS signals.

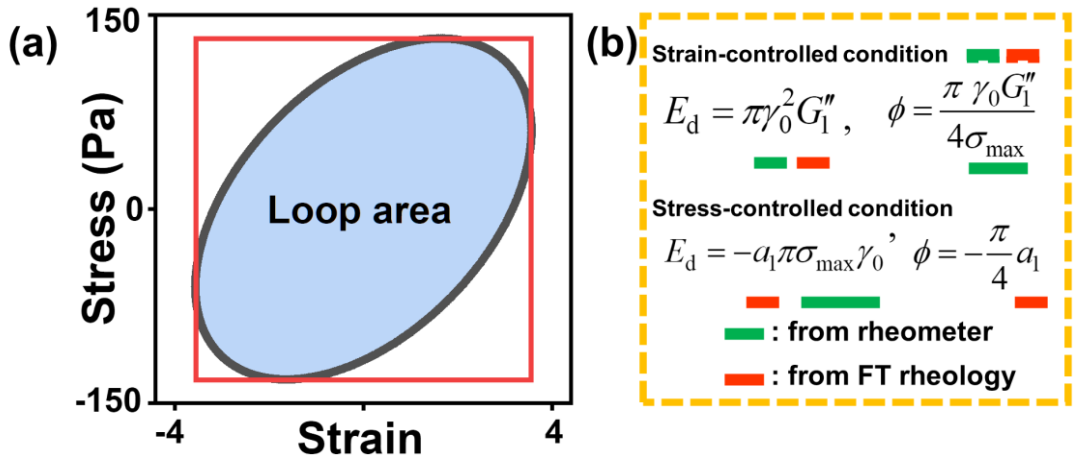

FIG. 14 The principle of the dissipation ratio $\phi$. (a) The energy dissipated per unit volume per cycle $E_d$ represented by the loop area within an elastic Lissajous curve. For a given $\sigma_{max}$ and $\gamma_0$, the maximum dissipated energy is the area of the red rectangle based on the perfect plastic behavior. The elastic Lissajous curve is plotted by using the data from point 47. (b) The dissipation ratio can also be represented as the use of the aLAOS approach in the dissipation ratio.

*2. Whole process*

From the previous section, the relationships among the $E_d$, $\phi$ and Fourier coefficients at both the strain- and stress-controlled conditions are discussed as Eqs. 80~83. Then, the validity of Eq. 83 will be examined by comparing the values from Eq. 83 and the results from the integrations of the elastic Lissajous curves, which is shown in Fig. 15. The blue points refer to the results from the integration, while the red points and purple lines correspond to the results at the strain- and stress-controlled conditions, respectively.

As shown in Fig. 15(a), the $E_d$ value is positively associated with the stress amplitude. Meanwhile, in the region after point 44 (the start yield point determined in Sec. II F), the $E_d$ value increased more significantly with the increase in the stress amplitude than that in the region before point 44. Furthermore, after point 48 (the end yield point determined in Sec. II F), the slope of the curve in Fig. 15(a) obviously decreased and generally remained constant.

As can be seen in Fig. 15(b), the 0.2 wt% Carbopol gel demonstrated an almost purely elastic response in the SAOS region ($\phi \approx 0$) and was close to a perfect Newtonian behavior under extremely large stress amplitudes ($\phi \approx 0.785$). In addition, around the stress amplitude of 200 Pa, the dissipation ratio value exceeded 0.785, indicating the combined plastic and Newtonian behaviors. Furthermore, after point 44, the dissipation ratio $\phi$ sharply increased with the increase in the stress amplitude. After point 48, the $\phi$ value almost remained constant. Therefore, points 44~48 denote the region in which the $\phi$ value was significantly elevated and behaves a "jump". Therefore, from Fig. 15(b), the solid-liquid transition and the inner rheological behavior were visually demonstrated. Furthermore, the blue and red points, as well as the purple lines correspond well with each other, no matter in the SAOS region or in the LAOS region. Therefore, the validity of Eq. 83 was tested.

More information can be found in Appendix F.

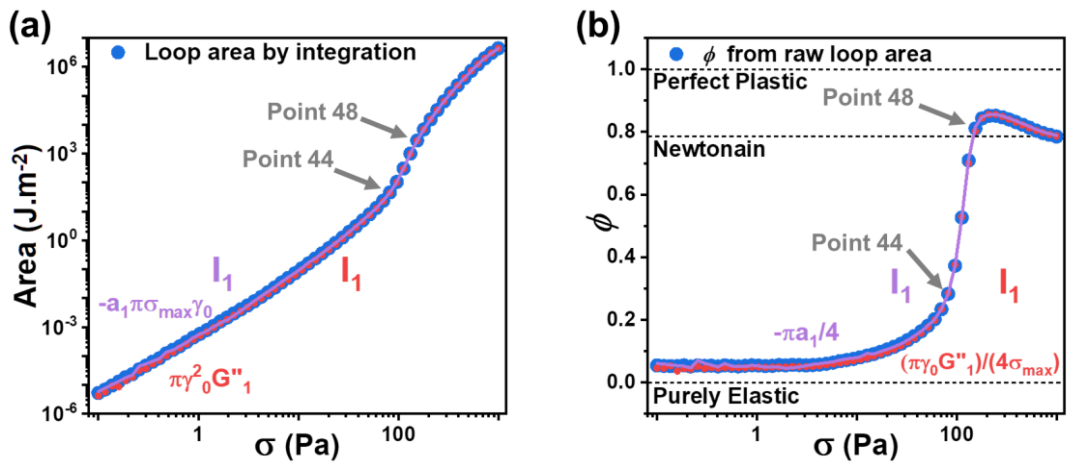

FIG. 15 Comparison between the results from the integrations of the elastic Lissajous curves and the results from Eq. 83. The plots of the (a) $E_d$ and (b) $\phi$ values versus stress amplitude. Blue points: results from integration. Red points: results from Eq. 83 at the strain-controlled condition.

Purple lines: results from Eq. 83 at the stress-controlled condition.

*3. Summary*

In a word, the dissipation ratio and the use of the aLAOS approach in dissipation ratio were discussed. The dissipation ratio was introduced at the beginning of Sec. II G1. Then, the relevant equations were reproved as Eqs. 80~83, where the equations for the stress-controlled condition were derived based on the FT of a distorted strain signal. Meanwhile, the relevant equations were based on Fourier coefficients, showing that the dissipation ratio is directly related to the FT rheology. Therefore, the dissipation ratio method can be considered the use of the aLAOS approach in dissipation ratio.

Then, the validity of Eq. 83 was further proved by the comparison among the blue and red points as well as the purple lines in Fig. 15 to demonstrate the close relationship between the dissipated energy and Fourier coefficients. Meanwhile, both the strain- and stress-controlled conditions were proved to be accessible for the dissipation ratio method. As a result, the use of the aLAOS approach in the dissipation ratio is rational.

**H. aLAOS in transient moduli of SPP**

LAOS (Sec. II A) and FT rheology (Sec. II B) has been introduced. Then, the uses of the aLAOS approach in several classic LAOS methods were proposed and discussed, including the Lissajous curve (Sec. II C), stress decomposition (Sec. II D), $S$ and $T$ ratios (Sec. II E), stress bifurcation (Sec. II F), and dissipation ratio (Sec. II G). Since the Sequence of Physical Processes (SPP) technique was first proposed by Rogers et al. in 2011 [18], SPP has achieved great success and gained wide attention [31, 33-37, 44, 83-87]. Therefore, the following several sections will focus on the SPP method, and this section discusses the use of the aLAOS approach in the transient moduli [34, 37, 85, 86].

*1. Transient moduli in LAOS*

The SPP framework can be typically used to continuously observe elastic and viscous behaviors of complex fluids in LAOS experiments, which provides the transient elastic and viscous moduli, $G_t'(t)$ and $G_t''(t)$ [35, 44, 85, 86]. The transient moduli are regarded as the parameters to conquer the symmetry issues in the FT rheology, Chebyshev coefficients, and stress decomposition [88]. Meanwhile, the transient moduli can be associated with the structural transition in soft glassy samples [36]. The yielding transition can be further distinguished by using the SPP method [34]. Moreover, the transient moduli are well-defined in the complex LAOS region and not strict to the test protocols. These capacities make the SPP approach an absorbed method to investigate the out-of-equilibrium dynamics of complex soft materials [31].

The SPP method considers the rheological responses that occur in a three-dimensional space ($(x, y, z)$) with the axis of the strain ($\gamma$), strain rate ($\dot{\gamma}$), and stress ($\sigma$) [35, 44]. The stress signal is represented as multivariable functions of the strain and strain rate ($\sigma = f(\gamma, \dot{\gamma})$) [88]. Since the rheological response is considered a trajectory in the three-dimensional space, a mathematical tool is thus needed to discuss the inner behaviors. Therefore, the transient moduli are proposed by Rogers et al. [35, 44]. The calculation process is done as follows.

As the rheological responses occurred in the space of $(x, y, z)$, the position of a point ($\vec{P}(t)$, the position vector) is indicated as follows:

$$\vec{P}(t) = (\gamma(t), \dot{\gamma}(t)/\omega, \sigma(t)) = (x, y, z) . \tag{84}$$

After that, three vectors can be obtained including the unit tangent ($\vec{T}$), normal ($\vec{N}$), and binormal ($\vec{B}$) vectors:

$$\vec{T}(t) = \vec{P}'(t)/\|\vec{P}'(t)\|, \ \vec{N}(t) = \vec{T}'(t)/\|\vec{T}'(t)\|, \ \vec{B}(t) = \vec{T}(t) \times \vec{N}(t) . \tag{85}$$

Accordingly, $\vec{T}$ can be viewed as the normalized time derivative of $\vec{P}(t)$ and the direction of velocity. $\vec{N}$ indicates tangent to $\vec{T}$ and the direction of acceleration, and $\vec{B}$ denotes the cross product of $\vec{T}$ and $\vec{N}$. Thus, $\vec{B}(B_\gamma, B_{\dot{\gamma}/\omega}, B_\sigma)$ is provided. Then, the values of $G_t'(t)$ and $G_t''(t)$ are calculated as follows:

$$G_t'(t) = -B_\gamma(t)/B_\sigma(t), \ G_t''(t) = -B_{\dot{\gamma}/\omega}(t)/B_\sigma(t) . \tag{86}$$

Furthermore, $G_t'(t)$ and $G_t''(t)$ can be regarded as the partial derivatives of $\sigma$ versus $\gamma$ and $\dot{\gamma}$, respectively. Considering the physical interpretation of the differential elastic and viscous moduli, $G_t'(t)$ and $G_t''(t)$ demonstrate the transient contribution of both the $\gamma$ and $\dot{\gamma}$ on $\sigma$ separately, whereas other analysis methods lack this ability [35, 44]. Therefore, the concept of the transient moduli is one of the disparities between SPP and other LAOS methods that introduce FT or Lissajous curves in one or two-dimensional space [35, 44]. These LAOS methods are based on secant values ($\sigma/\gamma$ and $\sigma/\dot{\gamma}$) or total derivatives of the stress signal ($d\sigma/d\gamma$ and $d\sigma/d\dot{\gamma}$), where the total derivatives can be expressed as follows:

$$\frac{d\sigma}{d\gamma} = \frac{\partial \sigma}{\partial \gamma} + (\frac{\partial \sigma}{\partial \dot{\gamma}})(\frac{d\dot{\gamma}}{d\gamma}), \ \frac{d\sigma}{d\dot{\gamma}} = \frac{\partial \sigma}{\partial \dot{\gamma}} + (\frac{\partial \sigma}{\partial \gamma})(\frac{d\gamma}{d\dot{\gamma}}) . \tag{87}$$

Therefore, for these LAOS methods, the additional strain rate contribution $(\frac{\partial \sigma}{\partial \dot{\gamma}})(\frac{d\dot{\gamma}}{d\gamma})$ and strain contribution $(\frac{\partial \sigma}{\partial \gamma})(\frac{d\gamma}{d\dot{\gamma}})$ are considered. However, the transient moduli represent the single influence of the strain and strain rate. The SPP method provides a new interpretation of the data from LAOS tests. The SPP analysis can combine the nonlinearity of the samples in LAOS with the linearity in SAOS [88], i.e. the relation between the maximum $G_t'(t)$ near the strain extrema and the $G'$ in the SAOS region. Meanwhile, the SPP method also offers information about the recoverable strain in the material, which indicates the recovered strain when the stress is removed [44, 88].

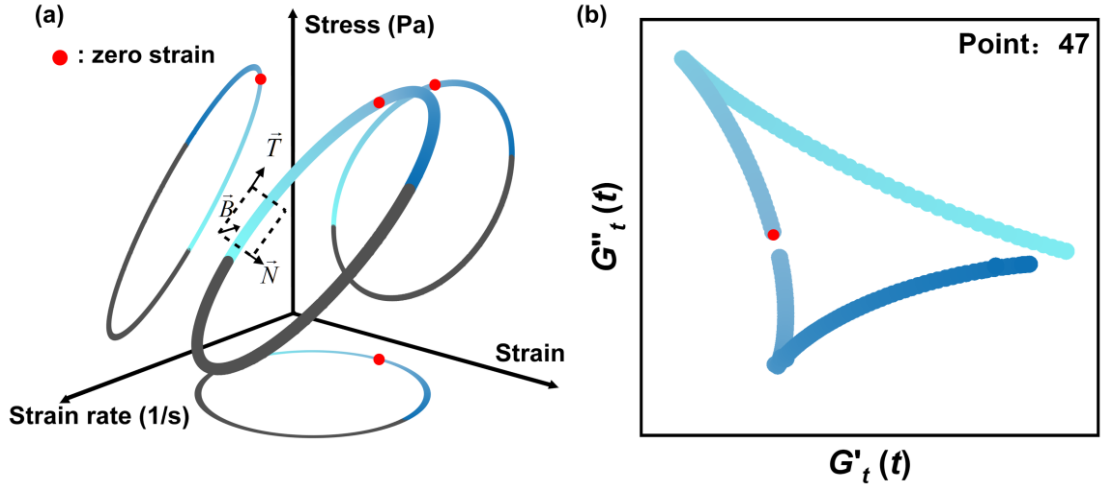

FIG. 16 Schematic illustrations of the transient elastic $G_t'(t)$ and viscous $G_t''(t)$ moduli. (a) The rheological response of point 47 as a trajectory in the three-dimensional space is defined by the strain, strain rate, and stress axis. The projections of this trajectory are also plotted in the stress-strain, stress-strain rate, and strain-strain rate planes. Frenet-Serret Frame (the unit tangent ($\vec{T}$), normal ($\vec{N}$), and binormal ($\vec{B}$) vectors) are denoted with an osculating plane. (b) The Cole–Cole plots of $G_t'(t)$ versus $G_t''(t)$ at point 47. All relevant points are color-mapped.

The rheological behavior demonstrated in Fig. 16(a) is revealed by the movement of $G_t'(t)$ and $G_t''(t)$ in the Cole–Cole plot, as illustrated in Fig. 16(b). The structural transition within the range from the minimum strain to the maximum strain is directly related to the Cole–Cole plot. The increase and decrease in the $G_t'(t)$ value represent the manifestation of stiffening and softening, respectively. Similarly, the behaviors of thickening and thinning can be denoted by the upward and downward movements of $G_t''(t)$ value, respectively. The point at zero strain (the maximum strain rate), regarded as the point presenting the highest viscous contribution without the elastic contribution, is denoted as the red point in Fig. 16. The red point in Fig. 16(b) is within the maximum and minimum $G_t''(t)$ values, reflecting the material is viscoelastic at point 47. During the strain swept from the minimum value to the maximum value, the sample undergoes softening/thickening, stiffening/thinning, and stiffening/thickening sequentially. Thus, continuous observation can be realized by using the transient elastic and viscous moduli, $G_t'(t)$ and $G_t''(t)$.

## *2. Analytic LAOS approach in transient moduli*
## *2.1. Principle*

The use of the aLAOS approach in the transient moduli will be proposed in this section as the following statements.

First, the rheological response refers to Eq. 11: $\sigma(t) = \gamma_0 \sum_{n=1}^{\infty}(G_n'' \cos n\omega t + G_n' \sin n\omega t)$ and $\gamma(t) = \gamma_0 \sin \omega t$. Then, $\vec{P}$, $\vec{T}$, and $\vec{N}$ without normalization are represented as the following equations:

$$\vec{P}(t) = (\gamma(t), \dot{\gamma}(t)/\omega, \sigma(t)) = (\gamma_0 \sin \omega t, \gamma_0 \cos \omega t, \gamma_0 \sum_{n=1}^{\infty}(G_n'' \cos n\omega t + G_n' \sin n\omega t)), \tag{88}$$

$$\vec{T}(t) = \omega \gamma_0 (\cos \omega t, -\sin \omega t, \sum_{n=1}^{\infty}(-n G_n'' \sin n\omega t + n G_n' \cos n\omega t)), \tag{89}$$

$$\vec{N}(t) = \omega^2 \gamma_0 (-\sin \omega t, -\cos \omega t, \sum_{n=1}^{\infty}(-n^2 G_n'' \cos n\omega t - n^2 G_n' \sin n\omega t)). \tag{90}$$

For $\vec{B}(B_\gamma, B_{\dot\gamma/\omega}, B_\sigma) = \vec{T} \times \vec{N}$, $B_\sigma = -\omega^3 \gamma_0^2$ is readily obtained. Then, $G_t'(t)$ and $G_t''(t)$ are expressed as the following equations:

$$G_t'(t) = -\frac{B_\gamma(t)}{B_\sigma(t)} = \frac{B_\gamma(t)}{\omega^3 \gamma_0^2} = G_{t,1}' + G_{t,3}'(t) + G_{t,5}'(t) + \cdots = \sum_{n=1}^{\infty} G_{t,n}'(t), \tag{91}$$

$$G_t''(t) = -\frac{B_{\dot\gamma/\omega}(t)}{B_\sigma(t)} = \frac{B_{\dot\gamma/\omega}(t)}{\omega^3 \gamma_0^2} = G_{t,1}'' + G_{t,3}''(t) + G_{t,5}''(t) + \cdots = \sum_{n=1}^{\infty} G_{t,n}''(t), \tag{92}$$

$$G_{t,n}'(t) = n G_n'(n \sin \omega t \sin n\omega t + \cos \omega t \cos n\omega t) + n G_n''(n \sin \omega t \cos n\omega t - \cos \omega t \sin n\omega t), \tag{93}$$

$$G_{t,n}''(t) = n G_n'(n \cos \omega t \sin n\omega t - \sin \omega t \cos n\omega t) + n G_n''(n \cos \omega t \cos n\omega t + \sin \omega t \sin n\omega t), \tag{94}$$

where the equations of $B_\gamma$ and $B_{\dot\gamma/\omega}$ can be easily obtained from Eqs. 91~94 because $B_\sigma$ is equal to a constant. The validity of Eqs. 91~94 can be easily verified by assuming $n = 1$.

When $n = 1$, $G_t'(t)$ and $G_t''(t)$ can be expressed as:

$$G_t'(t) = G_{t,1}'(t) = G_1'(\sin \omega t \sin \omega t + \cos \omega t \cos \omega t) + G_1''(\sin \omega t \cos \omega t - \cos \omega t \sin \omega t) = G_1', \tag{95}$$

$$G_t''(t) = G_{t,1}''(t) = G_1'(\cos \omega t \sin \omega t - \sin \omega t \cos \omega t) + G_1''(\cos \omega t \cos \omega t + \sin \omega t \sin \omega t) = G_1'', \tag{96}$$

where $G_t'(t) = G_1'$ and $G_t''(t) = G_1''$ are obtained, which accords with the conclusion given by Rogers et al. [33]. Eqs. 91~94 show that the values of $G_t'(t)$ and $G_t''(t)$ are dependent on the intensities of the harmonics and Fourier coefficients. The derived equations present simple forms and directly show the contributions of each harmonic.

## *2.2. Output result from one point*

Figure 17(a) shows the applied elastic and viscous Lissajous curves at point 47. The principle of the aLAOS approach in transient moduli is demonstrated in Fig. 17(b). Figure 17(c) contains the raw and reconstructed transient moduli in the form of the Cole–Cole plot, where different Cole–Cole plots were made by introducing different numbers of higher harmonics.

As shown in Fig. 17(a), the colored parts of the Lissajous curves were applied to generate the raw transient moduli. Figure 17(b) demonstrates the simple and efficient framework of the aLAOS approach in transient moduli. By observing the blue curve shown in Fig. 17(c), the structural transition within the range from the minimum strain to the maximum strain is clearly displayed. Then, the stiffening, softening, thickening, and thinning behaviors thus can be clearly distinguished.

As a result, the continuous observation is reached based on the transient moduli.

In Fig. 17(c), comparisons between the blue and red lines offer a graphical view to show the similarity of the reconstructed results to the raw results. Figure 17(c.i) indicates that by adapting only $I_1$, the reconstructed points all gather in one point that represents the position of $(G'_1, G''_1)$. In other words, the central coordinates of the deltoid can be generally considered as $(G', G'')$, which is consistent with the conclusion given by Rogers et al. [33]. Therefore, the $I_1$-based results cannot express the structural transformation. By introducing $I_2$ and $I_3$, a deltoid emerges in Fig. 17(c.ii), reflecting that the nonlinear behavior is partly shown. Furthermore, as shown in Figs. 17(c.iii) ~ 17(c.vi), when more harmonics are combined, the overlap between the raw and the reconstructed curves is better. As a result, a general criterion is displayed at the bottom of Fig. 17(b). The results in Fig. 17(c) indicate that $I_1$~$I_3$, $I_1$~$I_5$, and $I_1$~$I_7$ are qualified for qualitative, approximate, and quantitative evaluations, respectively.

Briefly, the aLAOS approach in transient moduli is accessible for calculating the reconstructed transient elastic and viscous moduli. The $I_1$~$I_7$-based transient moduli show good overlaps with the raw results. Meanwhile, the validity of Eqs. 91~94 is verified. Figure 17(c) also shows that $I_1$~$I_3$, $I_1$~$I_5$, and $I_1$~$I_7$ are enough for qualitative, approximate, and quantitative evaluations, respectively.

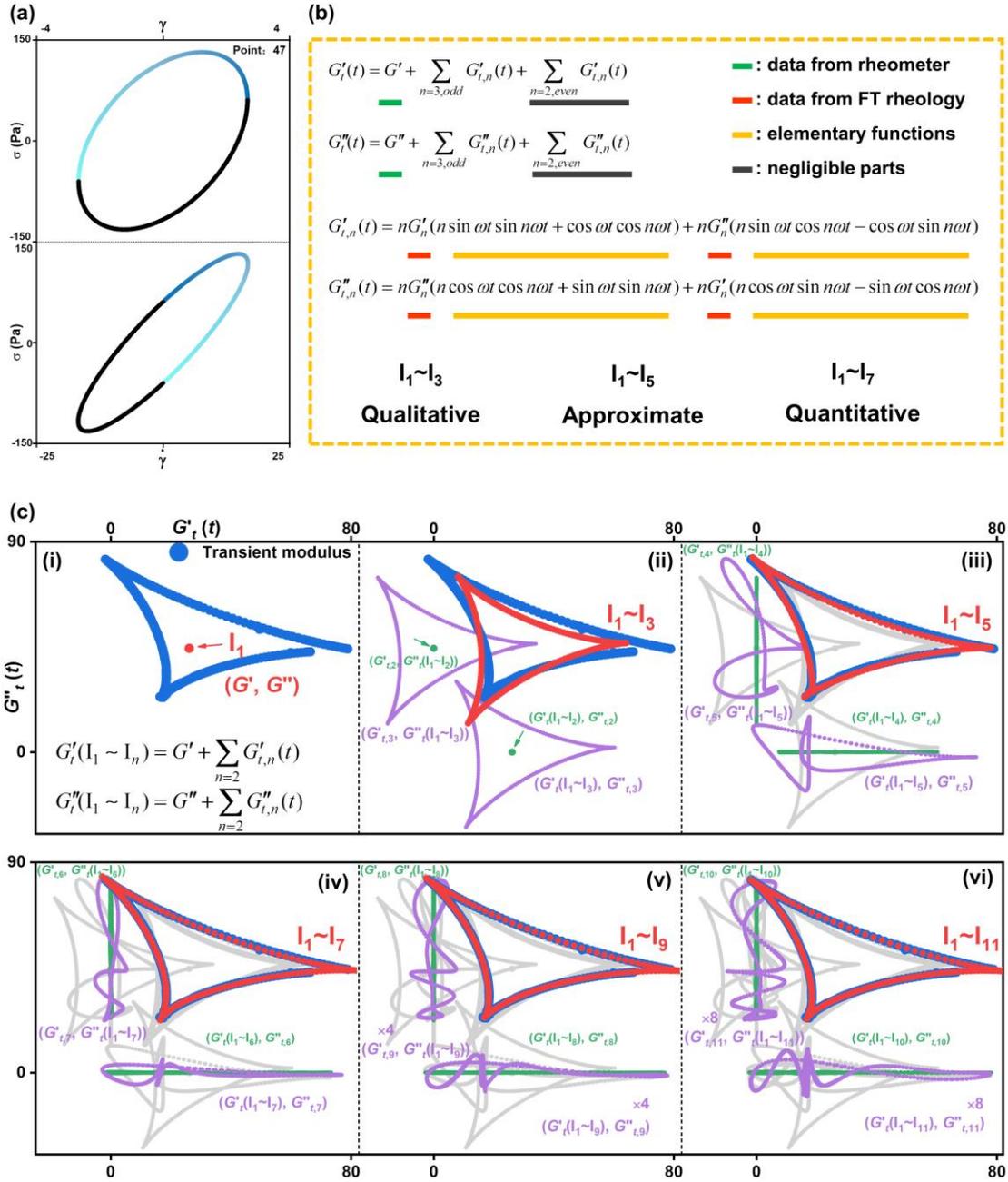

FIG. 17 Comparison between the raw and the reconstructed transient moduli at point 47: (a) the elastic and viscous Lissajous curves with the applied data color-mapped; (b) the principle of the aLAOS approach in transient moduli; and (c) the raw and reconstructed transient moduli in the form of the Cole–Cole plot: (i) $I_1$-based, (ii) $I_1$~$I_3$-based, (iii) $I_1$~$I_5$-based, (iv) $I_1$~$I_7$-based, (v) $I_1$~$I_9$-based, and (vi) $I_1$~$I_{11}$-based Cole–Cole plots. Blue lines: raw results. Red lines: reconstructed results. Purple lines: visualized contributions of odd harmonics. Green lines: visualized contributions of even higher harmonics. Gray lines: the curves that have been shown.

## 2.3. Visualized harmonic contribution

In Fig. 17(c.i), the central red point represents the contribution of $I_1$. As shown in Fig. 17(c.ii), the corresponding information is denoted. For example, the purple expression of $(G'_{t,3}, G''_t(I_1 \sim I_3))$ indicates that $G'_{t,3}$ represents the contribution of $I_3$ and $G''_{t(I_1 \sim I_3)}$ shows the contribution by combining $I_1 \sim I_3$. Meanwhile, the contributions of even harmonics are colored in green. Then, the influence of a single harmonic on the reconstructed curve is successfully demonstrated. Therefore, the contributions of harmonics can be visually displayed in Fig. 17(c) through the purple and green lines.

In Fig. 17(c.ii), the $I_1 \sim I_3$-based purple lines visually show the contributions of $I_3$ to the red reconstructed curve, where $I_1$ dominates the position of the $I_1 \sim I_3$-based deltoid and $I_3$ expands the $I_1$-based red central point in Fig. 17(c.i) to a deltoid in Fig. 17(c.ii). Meanwhile, the contribution of $I_2$ is negligible. The $I_1 \sim I_5$-based red curve in Fig. 17(c.iii) possesses a much closer shape to the blue curve. Therefore, although the amplitude of $I_5$ is lower than $I_3$, the influence of $I_5$ on the reconstructed curve is comparable to $I_3$. This result indicates the crucial contributions of $I_3$ and $I_5$ to the values of $G'_t$ and $G''_t$. In addition, as shown in Fig. 17(c.iii), a high similarity between the raw and $I_1 \sim I_5$-based curves is realized. When $I_6$ and $I_7$ were further introduced, the $I_1 \sim I_7$-based deltoid and the blue deltoid show a good overlap. Meanwhile, the contributions of $I_7$ are still significant (Fig. 17(c.iv)). From the purple curves in Figs. 17(c.v) and 17(c.vi), it is indicated that the contributions of $I_8 \sim I_{11}$ are negligible.

Detailed contributions of higher harmonics to the reconstructed transient moduli can be described. While the contribution of $I_1$ provides the position information and $I_3$ offers the dominant shape information, $I_5$ adjusts the $I_1 \sim I_3$-based curve shape closer to the raw curve (Fig. 17(c.iii)) via the following several effects. The left purple curve in Fig. 17(c.iii) shows that the top corner and the right corner of the $I_1 \sim I_3$-based red deltoid are shifted to the left and right, respectively. The situation of the bottom corner is relatively complex, where the left side is slightly moved to the right, while the right side is significantly moved to the left. Similarly, the bottom purple curve in Fig. 17(c.iii) demonstrates that the top corner and bottom corner of the $I_1 \sim I_3$-based red deltoid are both elevated while the position of the right corner is lowered. Meanwhile, the top side and the bottom side of the deltoid are generally risen and descended, respectively, while the top and the bottom parts of the left side are lifted and reduced, respectively. As a result, a similar shape to the raw curve is obtained by introducing $I_5$.

Furthermore, in Fig. 17(c.iv), $I_7$ decreased the top angle of the deltoid to make it more tapering. On the contrary, $I_7$ increased the bottom angle of the deltoid to let the bottom corner smoother. Meanwhile, the right corner was further shifted to the right. As a result, the $I_1 \sim I_7$-based deltoid can be constructed with high similarities to the raw deltoid.

Briefly, the aLAOS approach in transient moduli can define, calculate, and visually demonstrate the contributions of higher harmonics to the reconstructed transient moduli, which offers another perspective to understand the transient moduli in SPP.

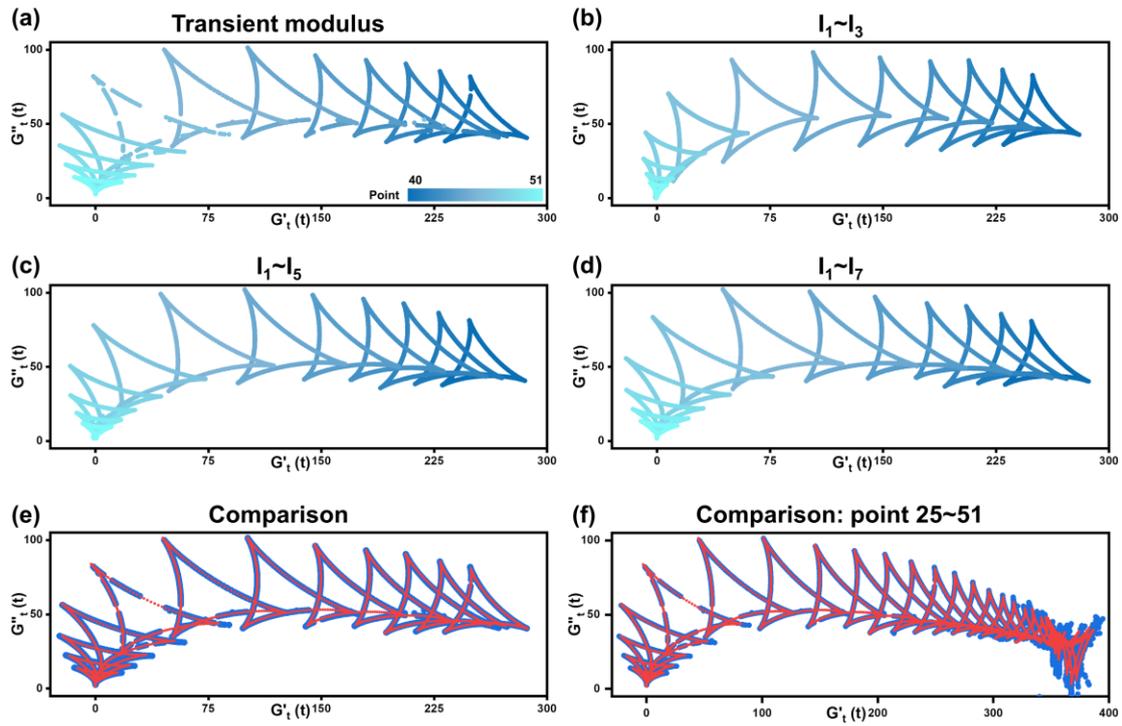

FIG. 18 Comparison between the raw and reconstructed transient moduli. (a) The raw results are in the form of the Cole–Cole plot. (b-d) Reconstructed results by introducing (b) $I_1 \sim I_3$, (c) $I_1 \sim I_5$, and (d) $I_1 \sim I_7$. (e) Comparison between the raw and reconstructed results (points 40~51). (f) Comparison between the raw and reconstructed results (points 25~51). Blue lines: raw results. Red lines: reconstructed results.

## 2.4. Whole process

Figure 18 demonstrates the raw (Figs. 18(a), 18(e), and 18(f)) and reconstructed (Figs. 18(b) ~ 18(f)) values of $G'_t$ and $G''_t$. Different reconstructed results were obtained based on the different numbers of harmonic intensities ($I_1 \sim I_3$, $I_1 \sim I_5$, and $I_1 \sim I_7$). The intracycle structural transitions from point 40 to point 51 are depicted in Fig. 18(a). The Cole–Cole plot shown in Fig. 18(a) presents the conversions in the size and position of the deltoid during the stress sweep, where the central point of each deltoid represents the storage and loss moduli ($G'$, $G''$) of the corresponding stress sweep point [33]. Therefore, the position shift denotes the change in the dynamic moduli. In addition, the size of the deltoid is positively related to the extent of the intra-cycle rheological transition. In the relatively low stress range, the increasing stress promoted the intracycle structural transition. Then, a larger stress value resulted in a shorter time for the reformation of the structure. Therefore, the deltoid size became much smaller. After that, the deltoid of point 51 almost shrank to a single point, which indicated that the rheological behavior of the 0.2 wt% Carbopol gel at point 51 was close to a rheological equilibrium state without intra-cycle rheological transition. Meanwhile, the change in the deltoid size gradually happened within points 40 ~ 51. Moreover, the negative values of $G'_t(t)$ were shown at high stress levels in Fig. 18(a), reflecting that the increase in the strain led to a decrease in the stress within a cycle, which is interpreted by the strain-induced structural breakdown. In addition, $G_t'(t) = 0$ can be attributed to the plastic deformation of the sample [35].

As shown in Fig. 18(b), the $I_1$~$I_3$-based Cole–Cole plot can demonstrate the change in the storage and loss moduli, as well as the structural transformation during the stress sweep. Therefore, the $I_1$~$I_3$-based transient moduli can be applied for the qualitative evaluation. Furthermore, the deviations between the raw and reconstructed results are eliminated by introducing $I_1$~$I_5$ (Fig. 18(c)). Figures 18(a) and 18(c) show no visual difference, indicating that the $I_1$~$I_5$-based transient moduli are enough for the approximate evaluation. However, in Fig. 17(c.iii), the $I_1$~$I_5$-based transient moduli still present deviations from the raw results. Therefore, the $I_1$~$I_7$-based curves are constructed.

$I_1$~$I_7$-based and raw curves were compared with each other in Figs. 18(e) and 18(f). Good superpositions appear between the two kinds of results, no matter for points 40 ~ 51 or points 25 ~ 51. The positions, shapes, and curvatures of the reconstructed results correspond well with the raw results. Therefore, the conclusion in Fig. 17(b) was further verified and proved to be rational by Fig. 18.

To sum up, a series of sequential points (from the SAOS region to the LAOS region) were studied and plotted in Fig. 18. Based on the Fourier coefficients, $I_1$~$I_3$, $I_1$~$I_5$, and $I_1$~$I_7$-based results were constructed, where the $I_1$~$I_7$-based transient moduli showed high similarities to the raw transient moduli in Figs. 18(e) and 18(f). Compared with Fig. 17, figure 18 proves that the aLAOS approach in transient moduli is also applicable on a larger scale.

In Appendix G, more information on analyzing different samples by using the aLAOS approach in transient moduli was provided.

## 3. Summary

In brief, the transient moduli in SPP and the use of the aLAOS approach in transient moduli were introduced and discussed. The principle of the aLAOS approach in transient moduli was proposed in Sec. II H2.1, which was also shown in Fig. 17(b). First, the equations for calculating the transient elastic and viscous moduli (Eqs. 91~94) were given based on the Fourier coefficients, reflecting that the transient moduli in SPP can be closely associated with FT rheology and Fourier coefficients. From Eqs. 91~94, the impact of even harmonics on transient moduli is shown.

Then, figures 17 and 18 indicate that the transient moduli can be precisely determined by combining Fourier coefficients. The aLAOS approach in the transient moduli can provide the reconstructed results with high similarities to the raw results. The general criterion put at the bottom of Fig. 17(b) shows that $I_1$~$I_3$, $I_1$~$I_5$, and $I_1$~$I_7$ are enough for providing qualitative, approximate, and quantitative evaluations, respectively.

Finally, the visualized harmonic contributions were demonstrated as the purple and green curves in Fig. 17(c), which indicated that $I_1$ provided the position information of the deltoids, while no contribution to the structural transformation was observed. Furthermore, $I_3$ and $I_5$ provided the dominant information on the structural transformation. Meanwhile, the visualization reflects how the Fourier coefficients adjust the reconstructed transient moduli curve to simulate the raw transient moduli.

Briefly, the aLAOS approach in transient moduli is rational and shows another perspective to interpret the transient moduli. Combining several Fourier coefficients to provide precise transient moduli instead of treating a group of many points will bring convenience.

**I. aLAOS in the derivative of the transient moduli of SPP**

Based on the transient moduli, the derivatives of the transient moduli were further proposed by Rogers [32, 33, 86]. Therefore, this section focused on the use of the aLAOS approach in the derivatives of transient moduli.

*1. Derivatives of transient moduli in LAOS*

Rogers [32, 33, 86] has indicated that the temporal derivatives of the transient moduli are a useful tool to investigate the continuous rheological behavior change (i.e. stiffening/softening and thickening/thinning). The temporal derivatives of the transient moduli can be expressed as $dG_t'(t)/dt$ and $dG_t''(t)/dt$. The visual shape of the derivatives of a group of transient moduli is a trefoil or a three-petal rose [33]. First, the elastic Lissajous curve of point 47 was demonstrated in Fig. 19(a). Then, the principle of the derivatives of transient moduli is displayed in Fig. 19(b).

After that, the derivatives of transient moduli were plotted ($dG_t''(t)/dt$ versus $dG_t'(t)/dt$) in Fig. 19(c). It is indicated that the phase angle of the third harmonic represents the orientation of the deltoid and trefoil, which shows the quality of the changes [33]. The curve in Fig. 19c denotes in which direction, and how fast, the trajectory in Fig. 19b is traced. The corresponding points were color-mapped. Figure 19c permits visual interpretations of material rheological responses (instantaneously stiffening/softening and thickening/thinning). It also offers the rate value at which the corresponding responses are happening. The nonlinearities can be readily expressed by studying the curve in Fig. 19(c).

$dG_t'(t)/dt > 0$, $dG_t'(t)/dt < 0$, $dG_t''(t)/dt > 0$, and $dG_t''(t)/dt < 0$ represent the stiffening, softening, thickening, and thinning, respectively. From the minimum to maximum strain, the trajectory underwent a sequence of softening/thickening, stiffening/thinning, and stiffening/thickening behaviors. Meanwhile, asymmetric responses were shown with a clear distinction in size between lobes. The softening/thickening portion in Fig. 19c was the largest, indicating the fastest structural transition.

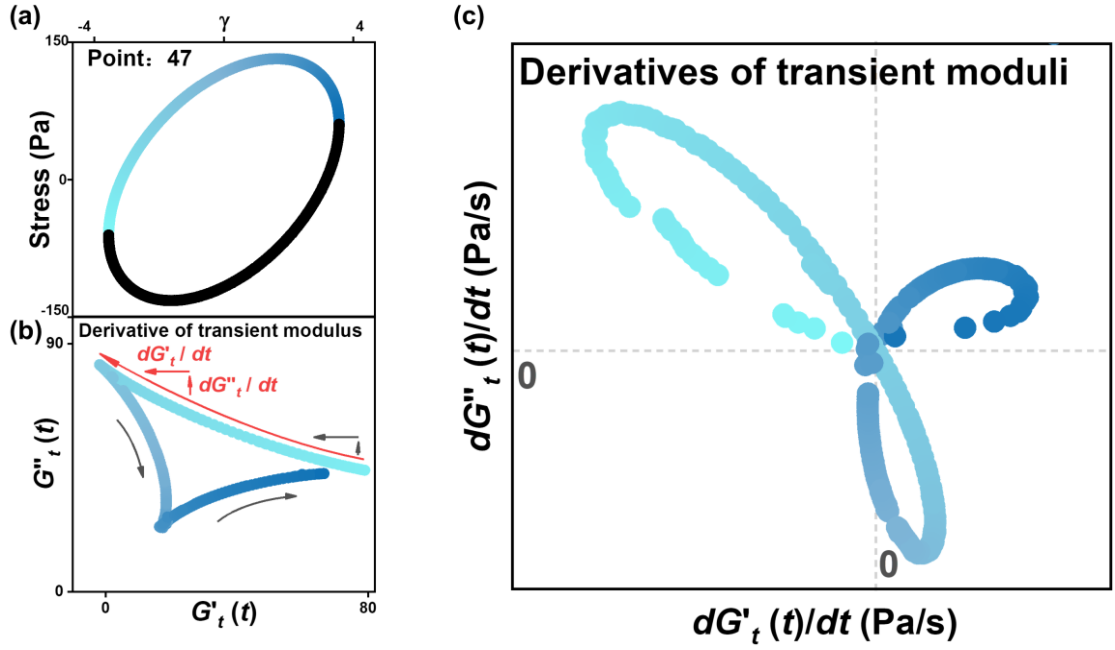

FIG. 19 Schematic illustrations of the derivatives of the transient moduli at point 47: (a) the elastic Lissajous curve with the region of applied points color-mapped; (b) the principle of the derivatives of transient moduli; (c) the parametric plot between time-derivatives of transient elastic and viscous moduli.

## 2. Analytic LAOS method in the derivatives of the transient moduli
### 2.1. Principle

The use of the aLAOS approach in the derivatives of the transient moduli was proposed based on Eqs. 91 ~ 94 as the following statements. The derivatives of $G_t'(t)$ and $G_t''(t)$ ( $dG_t'(t)/dt$ and $dG_t''(t)/dt$ ) can be calculated as follows:

$$\frac{dG_t'(t)}{dt} = \sum_{n=2} \frac{dG_{t,n}'(t)}{dt}, \quad \frac{dG_t''(t)}{dt} = \sum_{n=2} \frac{dG_{t,n}''(t)}{dt}. \qquad (97)$$

Then, the equations $dG_{t,n}'(t)/dt$ and $dG_{t,n}''(t)/dt$ are given as:

$$\frac{dG_{t,n}'(t)}{dt} = \omega n(n^2-1)(G_n'\cos n\omega t - G_n''\sin n\omega t)\sin \omega t, \qquad (98)$$

$$\frac{dG_{t,n}''(t)}{dt} = \omega n(n^2-1)(G_n'\cos n\omega t - G_n''\sin n\omega t)\cos \omega t, \qquad (99)$$

where the two equations can also be expressed as the following equations:

$$\frac{dG_{t,n}'(t)}{dt} = \omega n(n^2-1)\boldsymbol{a}_n \sin \omega t, \quad \frac{dG_{t,n}''(t)}{dt} = \omega n(n^2-1)\boldsymbol{a}_n \cos \omega t, \quad \boldsymbol{a}_n = \begin{vmatrix} G_n' & G_n'' \\ \sin n\omega t & \cos n\omega t \end{vmatrix}. \qquad (100)$$

Therefore, the reconstructed derivatives of transient moduli can be calculated based on Eqs. 97 ~ 100.

To sum up, in this section, the use of the aLAOS approach in the derivatives of transient moduli was proposed, which presents simple forms and shows the contributions of each harmonic. Moreover, the values of $dG_t'(t)/dt$ and $dG_t''(t)/dt$ are dependent on the Fourier coefficients. After that, the reconstructed derivatives of transient moduli can be given by the combination of Fourier coefficients.

*2.2. Output result from one point*

Figure 20 shows the raw and reconstructed derivatives of transient moduli. More specifically, figure 20(a.i) displays the raw derivatives of transient moduli of point 47 that are also shown in Figs. 20(a.ii) ~ 20(a.vi) as the blue lines. Meanwhile, the reconstructed curves were represented as the red curves in Fig. 20(a), where different curves were constructed by introducing different numbers of higher harmonics.

Figure 20(a.i) reflects that the $I_1$-based points all gather in origin (0, 0) because the values of $I_1$-based $G_t'(t)$ and $G_t''(t)$ are constant. By introducing $I_2$ and $I_3$, a trefoil appeared in Fig. 20(a.ii), reflecting that the structural transformation was partly shown. Furthermore, as shown in Figs. 20(c.iii) ~ 20(c.vi), the combination of more harmonics enhanced the superposition between the raw and reconstructed results.

Figure 20(b) visually shows the simple and efficient framework of the aLAOS approach in the derivatives of the transient moduli. Accordingly, the calculation process was greatly simplified by this method. Meanwhile, a general criterion is also displayed at the bottom of Fig. 20(b), which shows that $I_1$~$I_5$ and $I_1$~$I_9$ are enough for qualitative and approximate evaluations, respectively. However, quantitative evaluation cannot be carried out because the raw data has undergone six calculation steps (i.e. Fig. 16(a), Eqs. 85 and 86, Sec. II I1) to obtain the derivatives of transient moduli, where a small error can result in a big difference.

In a word, the aLAOS approach in the derivatives of transient moduli is capable of obtaining the reconstructed derivatives of transient moduli (Fig. 20). The reconstructed curve can show a good overlap with the raw curve. Meanwhile, the validity of Eqs. 97~100 is verified. Furthermore, $I_1$~$I_5$ and $I_1$~$I_9$ can offer qualitative and approximate evaluations, respectively. Therefore, figure 20(b) was proved to be rational.

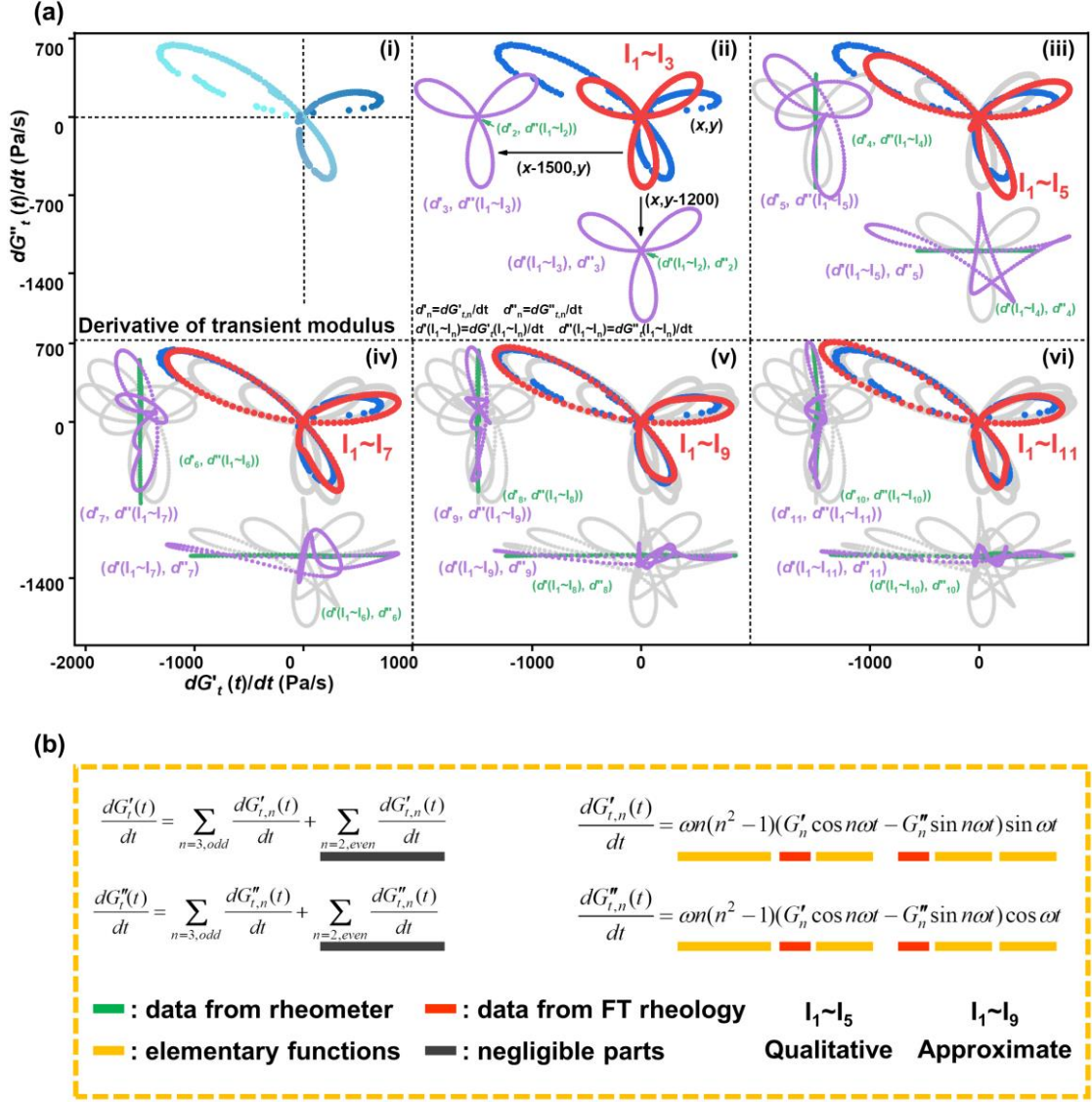

FIG. 20 Comparison between the raw and reconstructed derivatives of the transient moduli at point 47. (a) The raw and reconstructed results in the parametric plots: (i) $I_1$-based, (ii) $I_1\sim I_3$-based, (iii) $I_1\sim I_5$-based, (iv) $I_1\sim I_7$-based, (v) $I_1\sim I_9$-based, and (vi) $I_1\sim I_{11}$-based results. (b) The principle of the aLAOS approach in the derivatives of the transient moduli. Blue lines: raw results. Red lines: reconstructed results. Purple lines: visualized contributions of odd harmonics. Green lines: contributions of even higher harmonics. Gray lines: the curves that have been shown.

## 2.3. Visualized harmonic contribution

As shown in Fig. 20, the contributions of odd and even harmonics were colored purple and green. As can be observed from Fig. 20(a.ii), for example, the expression of ($x$-1500, $y$) indicated that the value of $d'_3$ was decreased by 1500 (translation) to demonstrate the harmonic contributions more clearly while the center of the trefoil was at (0, 0). Then, the impact of a single harmonic on the reconstructed curve can be shown based on Eqs. 97 ~ 100.

Figures 20(a.ii) and 20(a.iii) illustrate that the magnitude of the reconstructed trefoil is dominated

by both $I_3$ and $I_5$. Meanwhile, the contributions of even harmonics are negligible. Although the amplitude of $I_5$ was lower than that of $I_3$, the impact of $I_5$ on the reconstructed result was comparable to $I_3$. Meanwhile, $I_7$ had a significant influence on the shape of the reconstructed trefoil (Fig. 17(c.iv)). Thus, $I_7$ is also important for the adjustment of the curve shape. This result indicates the crucial statuses of $I_5$ and $I_7$ for the structure transformation rate. Furthermore, although the contribution of $I_{11}$ was still significant, a good overlap between the raw and reconstructed curves was reached by introducing $I_1$~$I_9$ as shown in Fig. 17(c.v).

The $I_1$~$I_3$-based trefoil of point 47 was symmetric. However, asymmetric responses appeared in size between lobes. The softening/thickening portion in Fig. 20(a) became larger, while the stiffening/thickening and the stiffening/thinning portions were closer to the thinning and stiffening regions, respectively. Therefore, the $I_1$~$I_3$-based trefoil cannot precisely describe the changing rheological behavior, while the $I_1$~$I_5$-based trefoil is qualified. Furthermore, the $I_1$~$I_9$-based trefoil can show the structure transformation rate.

In short, the aLAOS approach in the derivatives of transient moduli can define, calculate, and visually demonstrate the contribution of each higher harmonic to the reconstructed derivatives of transient moduli, which offers another perspective to understand the derivatives of the transient moduli in SPP.

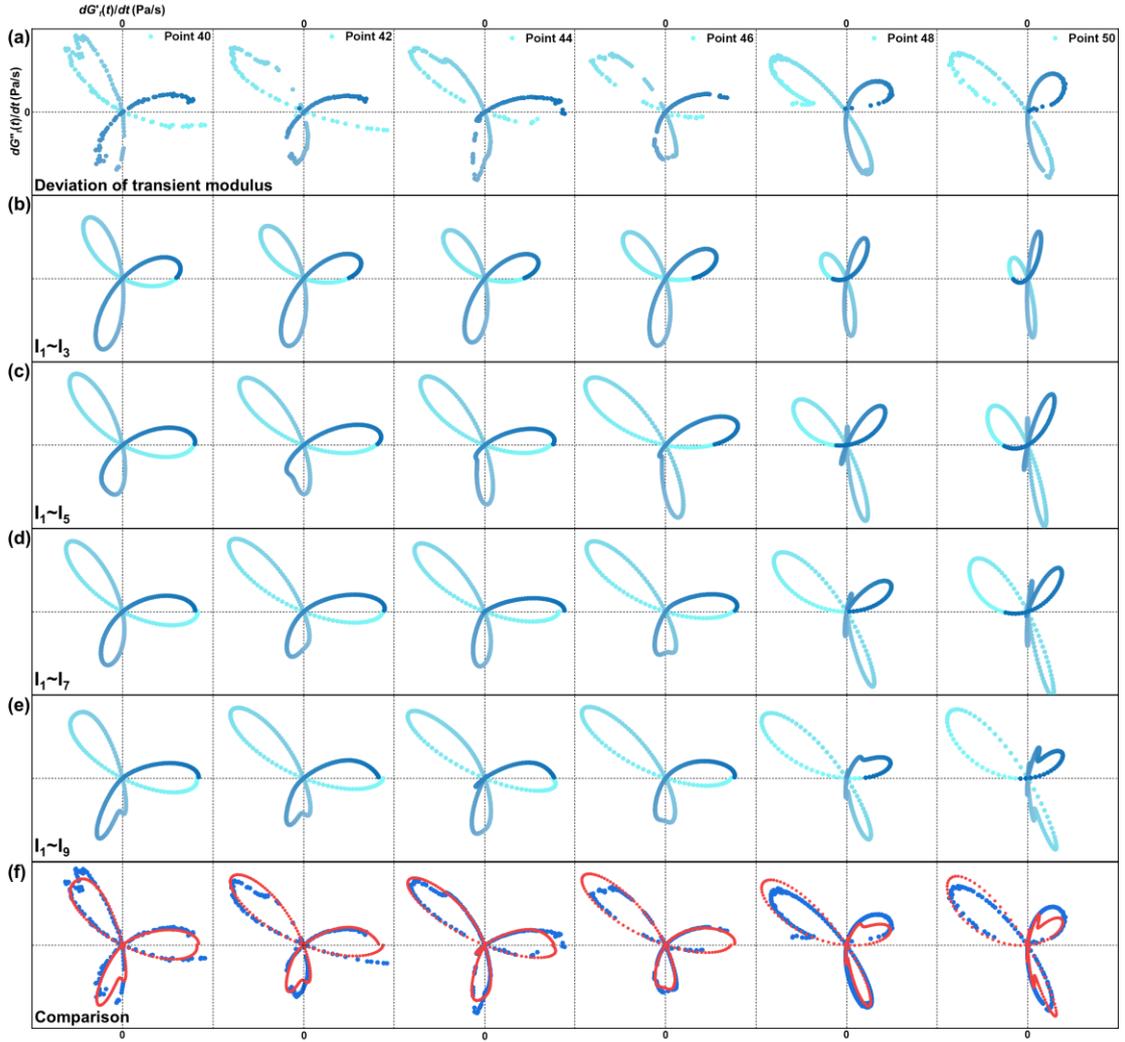

FIG. 21 Comparison between the raw and reconstructed derivatives of the transient moduli by introducing different numbers of higher harmonics. (a) The raw results of points 40, 42, 44, 46, 48, and 50. (b-e) Reconstructed results based on (b) $I_1$~$I_3$, (c) $I_1$~$I_5$, (d) $I_1$~$I_7$, and (e) $I_1$~$I_9$. (f) Comparisons between the raw and $I_1$~$I_9$-based results. Blue lines: raw results. Red lines: reconstructed results.

### 2.4. Whole process

Figure 21 shows the raw and reconstructed curves of $dG'_t(t)/dt$ and $dG''_t(t)/dt$. Different reconstructed results were obtained by using different numbers of harmonic intensities ($I_1$~$I_3$, $I_1$~$I_5$, $I_1$~$I_7$, and $I_1$~$I_9$). In Fig. 21(f), the comparisons between the raw (blue curves) and reconstructed (red curves) results are visually shown.

As shown in Fig. 21(a), the intracycle (from the minimum strain to the maximum strain) and intercycle (points 40, 42, 44, 46, 48, and 50) structural transitions are clearly demonstrated. While point 40 showed the period of stiffening/thinning, softening/thickening, thinning, softening/thinning, and stiffening/thickening from the minimum strain to the maximum strain, points 42, 44, and 46 changed following the sequence of stiffening/thinning, softening/thickening, stiffening/thinning, softening/thinning, and stiffening/thickening. Furthermore, points 48 and 50 presented the consecutive rheological behavior of softening/thickening, stiffening/thinning, and stiffening/thickening. Therefore, as can be ascertained from Fig. 21(a), during the oscillatory test swept from the SAOS region to the LAOS region: (i) more points were in the stiffening region; (ii) the initial stiffening/thinning portion was gradually shrunk and finally disappeared; (iii) the upper left lobe and the upper right lobe became comparatively more significant and smaller, respectively.

After that, the reconstructed values consisting of different numbers of harmonics were demonstrated in Figs. 21(b) ~ 21(e). The $I_1$~$I_5$-based curves of the six points were enough for qualitative evaluation (Fig. 21(c)). Furthermore, the $I_1$~$I_9$-based curves were constructed and show high similarities to the raw results (Fig. 21(e)).

Figure 21(f) compares the red $I_1$~$I_9$-based curves with the blue raw curves. Good superpositions between the raw and reconstructed results were demonstrated. The magnitudes, shapes, and angles (even the start point in cyan and the end point in blue) of the reconstructed results corresponded well with the raw results. Therefore, as can be seen from Fig. 21, the description in Fig. 20(b) was proved to be rational.

To sum up, different sweep points were investigated by the derivatives of transient moduli and the aLAOS approach in the derivatives of transient moduli to generate Fig. 21. The $I_1$~$I_9$-based curves and raw curves showed good overlaps (Fig. 21(f)). The aLAOS approach in the derivatives of transient moduli (Fig. 20(b)) was not only just accessible to a specific point (Fig. 20, point 47) but also applicable on a larger scale.

More information on analyzing different samples by using the aLAOS approach in the derivatives of transient moduli can be found in Appendix H.

### 3. *Summary*

Briefly, the derivatives of the transient moduli in SPP and the use of the aLAOS approach in the derivatives of transient moduli were introduced and discussed. The principle of the aLAOS approach in the derivatives of transient moduli was proposed in Sec. II I2.1, which was also shown in Fig. 20(b). First, specific equations were proposed based on the Fourier coefficients for the derivatives of transient moduli (Eqs. 97~100), indicating that the derivatives of transient moduli are closely related to the FT rheology and Fourier coefficients. Then, figures 20 and 21 indicated that the reconstructed derivatives of the transient moduli can be obtained by using several Fourier coefficients, where the raw and reconstructed results show high similarities. The general criterion at the bottom of Fig. 20(b) indicates that $I_1$~$I_5$ and $I_1$~$I_9$-based curves can be applied for qualitative and approximate evaluations, respectively. Finally, the visualized harmonic contributions reflect that $I_3$ and $I_5$ contain the major information of the structural transformation, while $I_7$ and $I_9$ significantly contribute to the curve shape adjustment.

Briefly, the aLAOS approach in the derivatives of the transient moduli is rational and shows another perspective to interpret the derivatives of transient moduli. Combining several Fourier coefficients to provide reconstructed derivatives of transient moduli instead of treating a group of many points will bring convenience.

**J. aLAOS in other methods of SPP**

In this section, some other methods in SPP [18, 36, 37, 86] will be briefly introduced, which benefits the use of the aLAOS approach in these methods. The order is: cage modulus (before the intracycle yielding [18]), other transient-moduli-based measures (focusing on the intracycle yielding process [36, 37, 86]), and flow curve (after the intracycle yielding [18]).

*1. Apparent cage modulus*

*1.1. Principle of apparent cage modulus*

The apparent cage modulus $G_{\text{cage}}$ can be defined as the following equation [18]:

$$G_{\text{cage}} = d\sigma/d\gamma\big|_{\sigma=0}, \tag{101}$$

which is specific to YSFs. It is apparent that:

$$\lim_{\delta,\gamma_0 \to 0} G_{\text{cage}} = \lim_{\delta,\gamma_0 \to 0} d\sigma/d\gamma\big|_{\sigma=0} = \lim_{\delta,\gamma_0 \to 0} G'. \tag{102}$$

At zero stress value, internal stresses may exist whereas the elastic and viscous contributions are of opposite sign and equal magnitude. Accordingly, $G_{\text{cage}}$ is the transient slope of the elastic Lissajous curve at zero stress value.

*1.2. Analytic LAOS approach in apparent cage modulus*

At the stress-controlled condition ($\sigma(t) = \sigma_{\max} \sin \omega t$ and $\gamma(t) = \gamma_0 (\sum_{n=1} a_n \cos n\omega t + b_n \sin n\omega t)$), $G_{\text{cage}}$ can be calculated as (Fig. 22(b)):

$$G_{\text{cage}} = \frac{d\sigma(t)}{d\gamma(t)}\bigg|_{\sigma=0} = \frac{\sigma_{\max} \cos \omega t}{\sum_{n=1} -na_n \sin n\omega t + nb_n \cos n\omega t}\bigg|_{t=0} = \frac{\sigma_{\max}}{\sum_{n=1} nb_n} \approx \frac{\sigma_{\max}}{\sum_{n=1,odd} nb_n}. \tag{103}$$

After that, the results from Eq. 103 were plotted in Fig. 22(c), which showed small deviations between the raw and reconstructed results. In addition, $I_1 \sim I_3$-based results are accessible for providing the approximate evaluation.

Meanwhile, the other approach was proposed based on the strain-controlled situation ($\sigma(t) = \gamma_0 \sum_{n=1}^{\infty}(G_n'' \cos n\omega t + G_n' \sin n\omega t)$ and $\gamma(t) = \gamma_0 \sin \omega t$) as follows:

$$G_{\text{cage}} = \frac{d\sigma(t)}{d\gamma(t)}\bigg|_{\sigma=0} = \frac{\sum_{n=1} -nG_n'' \sin n\omega t + nG_n' \cos n\omega t}{\cos \omega t}\bigg|_{t=t_0}, \tag{104}$$

where the time $t_0$ of $\sigma(t_0) = 0$ seems difficult to be calculated. Therefore, a situation is considered: the strain is equal to $\sum_{n=1} a_n$ at the stress-controlled condition ($\gamma(t) = \gamma_0(\sum_{n=1} a_n \cos n\omega t + b_n \sin n\omega t)$) when the value of the stress/time is zero. Accordingly, for the strain-controlled condition ($\gamma(t) = \gamma_0 \sin \omega t$), the position of the zero stress with a time of $t_0$ was then determined:

$$\sin \omega t_0 = \sum_{n=1} a_n \text{ or } t_0 = \arcsin(\sum_{n=1} a_n)/\omega, \quad (105)$$

where the values of $t_0$ and $\sin \omega t_0$ were obtained. As a result, based on Eq. 104, $G_{\text{cage}}$ was further obtained. As shown in Fig. 22(d), the raw and reconstructed curves can show good superpositions by introducing $I_1 \sim I_5$. Therefore, $I_1 \sim I_5$-based results are enough for the approximate evaluation. In addition, in the first two panels of Fig. 22(d), the regions, in which the reconstructed values cannot be given, are originated from $|a_1| > 1$ and $|a_1 + a_3| > 1$. Thus, the values of $|a_1|$ and $|a_1 + a_3|$ exceed the region of $y = \sin x$, $y \in [-1, 1]$.

Furthermore, the use of inverse trigonometric functions may bring complexity and the expression of $G_{\text{cage}}$ in the form of fully Fourier coefficients is desired. Thus, based on the conclusion given by Fig. 22(d), equation 104 is further transformed into the following equations:

$$G_{\text{cage}} \approx \left. \frac{-G_1'' \sin \omega t + G_1' \cos \omega t - G_3'' \sin 3\omega t + G_3' \cos 3\omega t - G_5'' \sin 5\omega t + G_5' \cos 5\omega t}{\cos \omega t} \right|_{t=t_0}, \quad (106)$$

$$\sin \omega t_0 \approx \sum_{n=1,odd}^{5} a_n, \quad \cos \omega t = \sqrt{1 - \sin^2 \omega t}, \quad (107)$$

$$\sin 3\omega t = \sin \omega t (3 - 4\sin^2 \omega t), \quad \sin 5\omega t = \sin \omega t (5 - 20\sin^2 \omega t + 16\sin^4 \omega t), \quad (108)$$

$$\cos 3\omega t = \sqrt{1 - \sin^2 \omega t}(1 - 4\sin^2 \omega t), \quad \cos 5\omega t = \sqrt{1 - \sin^2 \omega t}(1 - 12\sin^2 \omega t + 16\sin^4 \omega t). \quad (109)$$

Equation 106 can be further changed as the following equation:

$$G_{\text{cage}} \approx -(G_1'' + 9G_3'' + 25G_5'')\sin \omega t_0 + (G_1' + 3G_3' + 5G_5')\cos \omega t_0 + 4(3G_3'' + 25G_5'')\sin^3 \omega t_0 \\ -12(G_3' + 5G_5')\cos \omega t \sin^2 \omega t_0 + 80(G_5' \cos \omega t - G_5'' \sin \omega t)\sin^4 \omega t_0 \quad (110)$$

Therefore, the aLAOS approach in $G_{\text{cage}}$ can be carried out based in four different ways, i.e. Eq. (103), Eqs. (104) and (105), Eqs. (106) ~ (109), or Eqs. (107) and (110).

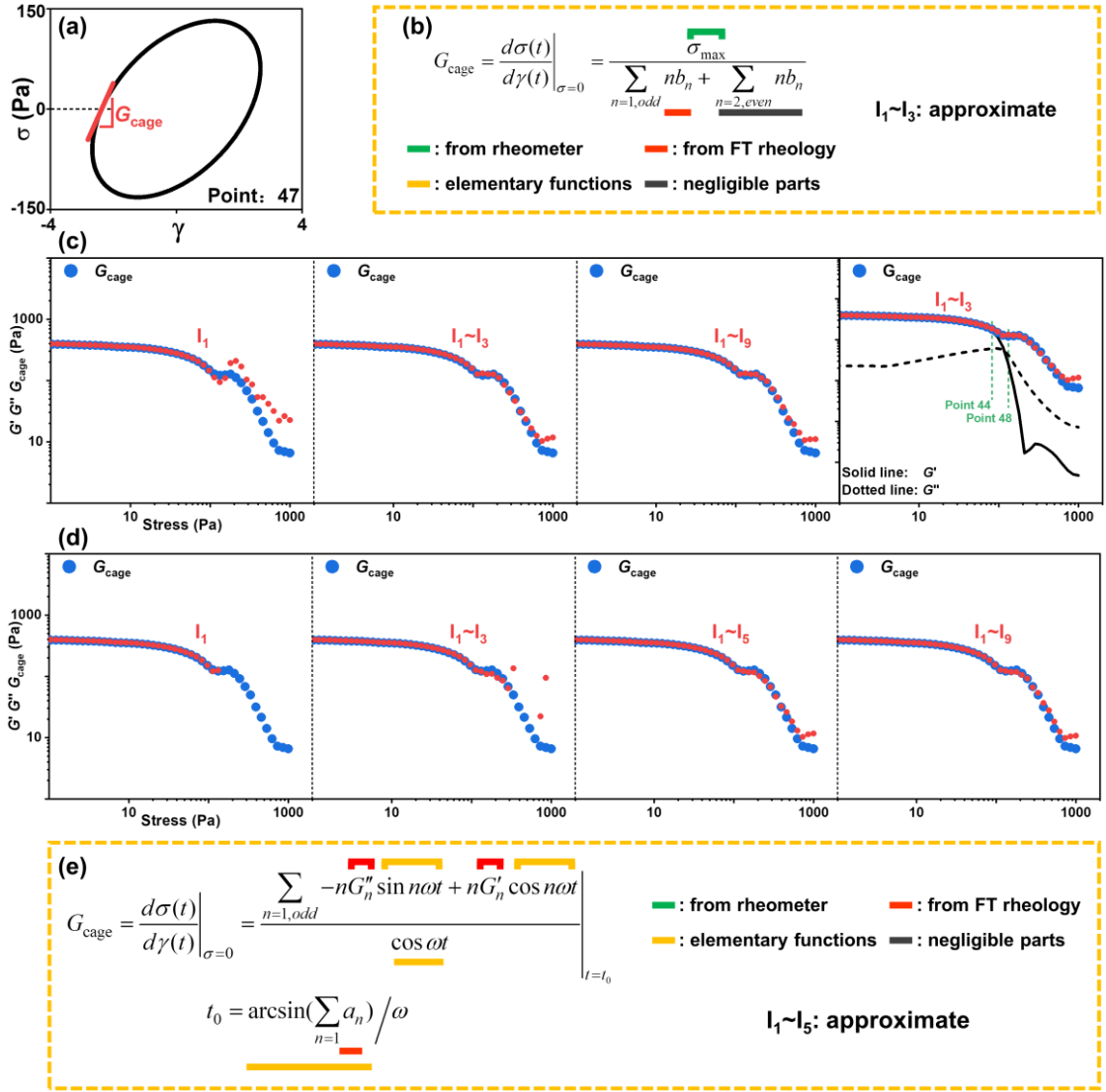

FIG. 22 Comparison between the raw and reconstructed cage modulus $G_{\text{cage}}$. (a) The definition of $G_{\text{cage}}$. (b) The principle and (c) reconstructed results of the $G_{\text{cage}}$ based on the FT of distorted strain signals. (d) The reconstructed results and (e) principle of the $G_{\text{cage}}$ based on the FT of both the distorted strain and distorted stress signals. Blue points: raw results. Red points: reconstructed results.

### 1.3. Whole process

The values of $G_{\text{cage}}$, $G'$, and $G''$ versus stress amplitude were plotted in the last panel of Fig. 22(c). Then, the start (point 44) and end (point 48) yield points from the algebraic stress bifurcation were also denoted. The $G_{\text{cage}}$ and $G'$ showed a good overlap at small stress levels. After point 44, the $G_{\text{cage}}$ revealed the deviation from $G'$. Around point 48, the $G_{\text{cage}}$ value maintained a constant, where big differences were shown between the $G_{\text{cage}}$ and $G'$. The values of $G_{\text{cage}}$ were higher than $G'$ in the LAOS region, reflecting that the intracycle elastic behavior is partly recovered. As a result, the significant elastic behaviors under large stress amplitudes are identified by the $G_{cage}$ measure in SPP, which can not be observed by $G'$ from the rheometric tests. Appendix I provided more information.

## 2. Other transient-moduli-based measures

### 2.1. Phase angle and velocity of phase angle in transient moduli

Based on the time-dependent $G_t'$ and $G_t''$, Rogers et al. proposed the concepts of phase angle $\delta_t = \arctan(G_t''/G_t')$ and the velocity of phase angle $d\delta_t/dt$ [36, 37, 86]. The $\delta_t$ momentarily close to zero represents a nearly perfectly elastic response. The phase angle near $\pi/2$ indicates a predominantly liquidlike response. Meanwhile, the peak of $d\delta_t/dt$ was close to the position at which the stress was changing rapidly with the strain, which was regarded as yielding and the transitions between the elastic and viscous responses by Rogers et al. However, it seems that the peak of $\delta_t$ was close to the yield point from Fig. 7 in their work [36].

### 2.2. Analytic phase angle and velocity of phase angle

Here, the analytic $\delta_t$ value can be easily calculated based on Eqs. 91 ~ 94 by the following equation:

$$\delta_t(t) = \arctan(\frac{G_t''(t)}{G_t'(t)}) = \arctan(\frac{B_{\dot{\gamma}/\omega}(t)}{B_\gamma(t)}) = \arctan(\sum_{n=1}\frac{G_{t,n}''(t)}{G_{t,n}'(t)}) \ , \qquad (111)$$

where $G_{t,n}'(t)$ and $G_{t,n}''(t)$ are expressed by Eqs. 93 and 94. From Eq. 111, the value of $d\delta_t/dt$ can be readily obtained:

$$d\delta_t(t)/dt = d\arctan(\sum_{n=1}\frac{G_{t,n}''(t)}{G_{t,n}'(t)})\Big/dt \ , \qquad (112)$$

where $d\arctan x/dx = 1/(1+x^2)$ can be further introduced. As a result, the uses of the aLAOS approach in $\delta_t$ and $d\delta_t/dt$ values are expressed by Eqs. (111) and (112).

### 2.3. Whole process

Representative results are shown in Fig. 23, where the principles of $\delta_t$ and $d\delta_t/dt$ were illustrated in Fig. 23(a). Then, the $\delta_t$ (Fig. 23(b)) and $d\delta_t/dt$ (Fig. 23(c)) versus strain were plotted by using the data from point 47. In Figs. 23(b) and 23(c), the blue points are the raw results by treating the curve in Fig. 23(a), which shows some noise in $\delta_t$ and a significant fluctuation in $d\delta_t/dt$. It reflects that the raw $\delta_t$ values are hard for quantitative evaluation and the raw $d\delta_t/dt$ values are unable to be analyzed. However, the two analytic curves in Figs. 23(b) and 23(c) with the same color mapping as the curve in Fig. 23(a) are smooth. Meanwhile, the analytic $\delta_t$ values corresponded well with the raw results. Therefore, the aLAOS approach in $\delta_t$ and $d\delta_t/dt$ were verified and proved to be rational.

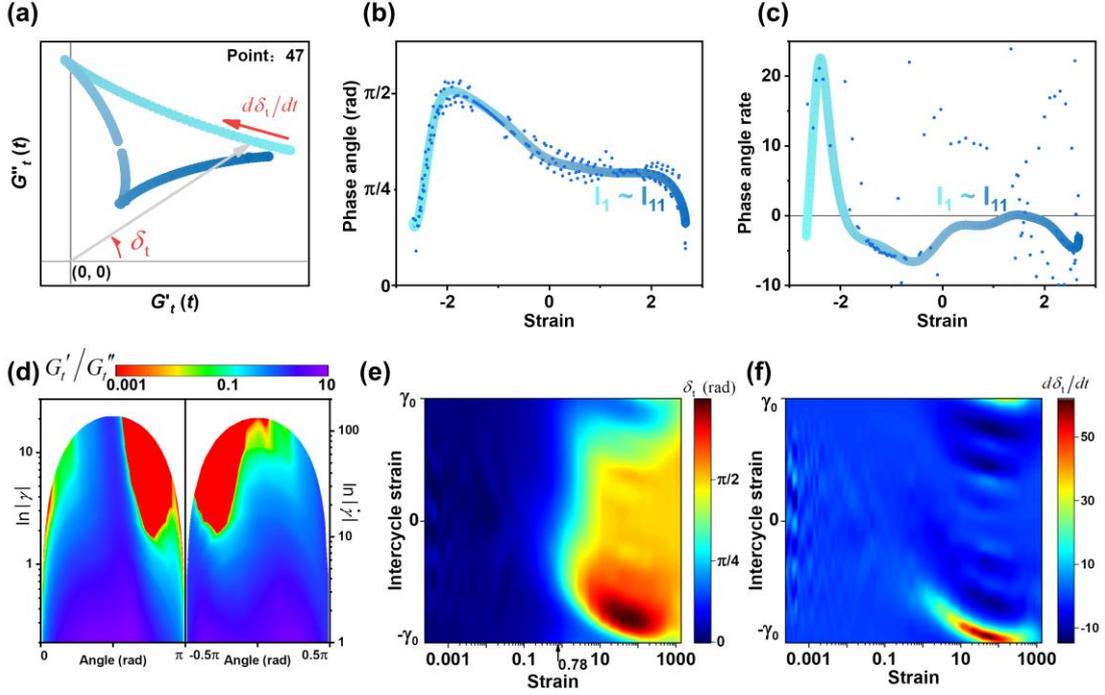

FIG. 23 Transient-moduli-related methods in SPP. (a) Principles of the phase angle $\delta_t = \arctan(G_t''/G_t')$ and the velocity of phase angle $d\delta_t/dt$. (b) The plot of $\delta_t$ versus strain at point 47. (c) The plot of $d\delta_t/dt$ versus strain at point 47. (d) $\ln|\gamma|$ (intracycle strain) and $\ln|\dot{\gamma}|$ (intracycle strain rate) versus angle ($\theta$, $\gamma(t) = \gamma_0 \sin\theta$) to show the $G_t'/G_t''$ ratio with the color mapping (points 41 ~ 50). Contour plots of (e) phase angle $\delta_t$ and (f) phase angle velocity $d\delta_t/dt$. The $\gamma_0 = 0.78$ indicates the first appearance of $\delta_t = \pi/4$ ($G_t' = G_t''$).

Figure 23(d) depicts a series of the intracycle solid-liquid transitions, $G_t'/G_t''$ values at each angle step (equal to the time step). When $G_t'/G_t''$ approached zero, the sample was more liquidlike than solidlike (the red regions). The blue regions represent the values of $G_t'/G_t''$ are much larger than zero, showing the solidlike behavior. Accordingly, the blue and red regions were shrunk and expanded with the increase in the strain amplitude, respectively. This result shows the intracycle and intercycle time-dependent rheological behaviors.

The values of the instantaneous $\delta_t$ and dimensionless $d\delta_t/dt$ are displayed in Figs. 23(e) and 23(f). As can be ascertained from Fig. 23(e), the positions of $\arctan(G_t''/G_t') = 0$ ($G_t' \gg G_t''$), $= \pi/4$ ($G_t' = G_t''$, $\gamma_0 = 0.78$), and $= \pi/2$ ($G_t' = 0$) can be visually observed from the blue, cyan, and orange regions, respectively. Meanwhile, the structure transition velocity was clearly demonstrated in Fig. 23(f). Furthermore, the peaks of $\delta_t$ and $d\delta_t/dt$ were closely related to the intracycle yielding position. Figures 23(e) and 23(f) also visually show that the intracycle yielding position becomes closer to the minimum strain with the increase in strain amplitude.

More information can be found in Appendix J.

To sum up, the two measures of SPP, the phase angle $\delta_t = \arctan(G_t''/G_t')$ and the velocity of phase angle $d\delta_t/dt$, were introduced and discussed in this section. The aLAOS approach in $\delta_t$ and $d\delta_t/dt$ values (Eqs. (111) and (112)) were proved to be rational.

## 3. LAOS-based flow curve
### 3.1. Principle

After the intracycle yielding, the response in the post-yielding region can give the flow curve. which is suggested by Rogers et al. [18]. When the elastic Lissajous curve is without a clear yield point, the curve in the first quadrant is analyzed (Fig. 24(a)). At the strain-controlled condition, the reconstructed flow curve is described as follows:

$$\sigma(t) = \gamma_0 \sum_{n=1} (G_n'' \cos n\omega t + G_n' \sin n\omega t), \quad \dot{\gamma}(t) = \gamma_0 \omega \cos \omega t, \tag{113}$$

which is also demonstrated in Fig. 24(b). Eq. 113 is applied for generating the results in Figs. 24(c) and 24(d). Meanwhile, at the stress-controlled condition, the reconstructed results are given by the following expression:

$$\sigma(t) = \sigma_0 \sin \omega t, \quad \dot{\gamma}(t) = n\omega \gamma_0 \sum_{n=1}^{\infty} (-a_n \sin n\omega t + b_n \cos n\omega t), \tag{114}$$

where the discussion will focus on the commonly used strain-controlled condition. The previously reported works in the literature have shown good agreement between the flow curves from oscillatory tests and steady-shear experiments [18].

### 3.2. Output result from one point

Figure 24(c) demonstrates the raw (blue lines) and reconstructed (red lines) flow curves. From the blue curve, the stress increases when the strain rate was swept from zero to the maximum. By involving the right panel of Fig. 23(d), the transient rheological behavior was changed from more elastic to more viscous from zero to maximum strain rate, which can also be visually observed from the blue curve. By introducing $I_1$, the reconstructed flow curve presents a huge deviation from the raw curve, which is eliminated by further introducing $I_2$ and $I_3$. Furthermore, when more harmonics are added together, the superposition between the raw and reconstructed curves is better (Fig. 24(c)). The $I_1\sim I_5$-based curve can show a good overlap with the raw curve. The results show that $I_1\sim I_3$, $I_1\sim I_5$, and $I_1\sim I_7$-based curves are enough for qualitative, approximate, and quantitative evaluations, respectively.

Meanwhile, the visualization of harmonic contributions was carried out as the purple and green lines in Fig. 24(c), which shows the dominant contribution of $I_1$ to the intensity of the reconstructed curve and the crucial contribution of $I_3$ to the shape adjustment. Compared with the $I_1$-based flow curve, $I_3$ elevates and lowers the low and high strain rate regions, respectively. When $I_5$ was introduced, the flow curve became much closer to the raw curve with high similarity. $I_5$ promoted the point intensities in the low and high strain rate region as well as decreased the point intensities in the middle strain rate region. Therefore, the aLAOS approach can visually demonstrate the harmonic contributions to a flow curve.

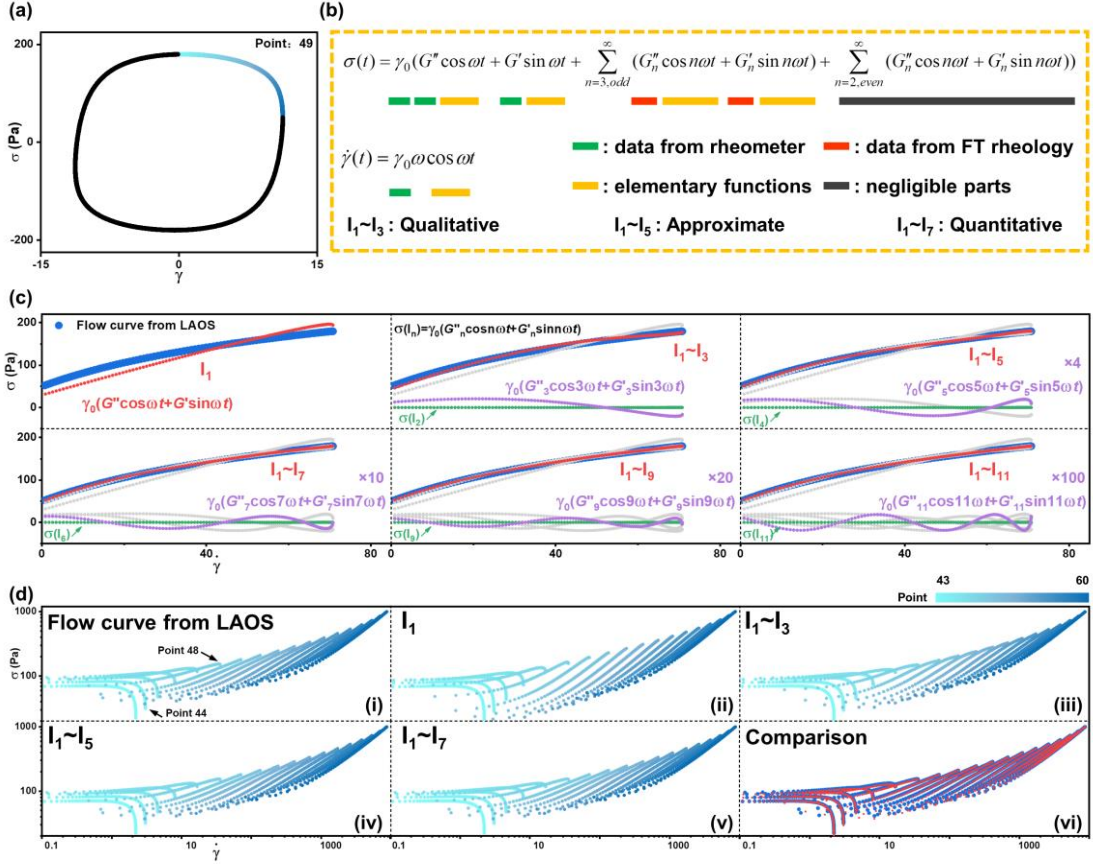

FIG. 24 Flow curve in SPP. (a) The studied region of the elastic Lissajous curve at point 49. (b) Principle. (c) The raw and reconstructed flow curves, including $I_1$, $I_1 \sim I_3$, $I_1 \sim I_5$, $I_1 \sim I_7$, $I_1 \sim I_9$, and $I_1 \sim I_{11}$-based curves. (d) Raw and reconstructed flow curves of points 43 ~ 60: (i) raw curves; (ii~v) reconstructed curves based on different numbers of harmonics; and (vi) the comparison between raw and $I_1 \sim I_7$-based curves, where blue and red lines are raw and reconstructed curves, respectively.

### *3.3. Whole process*

A series of flow curves of points 43 ~ 60 are shown in Fig. 24(d). In Fig. 24(d.i), a higher maximum stress/strain rate will lead to a lower minimum stress, which may be attributed to the significant structural failure under a high stress/strain rate and the short time for the structure recovery. In addition, after the start yield point (point 44), the flow region is significantly expanded. After the end yield point (point 48), the elastic deformation region disappeared. Therefore, the conclusions from the stress bifurcation and flow curve correspond well.

After that, a series of flow curves were constructed based on different numbers of harmonics (Figs. 24(d.ii) ~ 24(d.v)). The differences between the raw and $I_1$-based results are visually shown.

Furthermore, this is eliminated by introducing $I_3$. The $I_1$~$I_5$-based curves show no visual difference from the raw curves and are almost the same as the $I_1$~$I_9$-based curves. The comparison between the raw and $I_1$~$I_5$-based flow curves was made in Fig. 24(d.vi), which indicated the high similarity between the two kinds of results.

More information was provided in Appendix K.

Briefly, the equations for providing reconstructed flow curves (Eqs. 113 and 114) are given based on the Fourier coefficients, showing the close relationship between the LAOS-based flow curve and Fourier coefficients. Then, figure 24 indicates that reconstructed flow curves with high similarities to the raw curves can be obtained by combining Fourier coefficients. Meanwhile, a general criterion is given at the bottom of Fig. 24(b), reflecting that $I_1$~$I_3$, $I_1$~$I_5$, and $I_1$~$I_7$-based flow curves can offer qualitative, approximate, and quantitative evaluations, respectively. In addition, the visualized harmonic contributions are carried out. Therefore, the implementation of the aLAOS approach in the LAOS-based flow curve is rational.

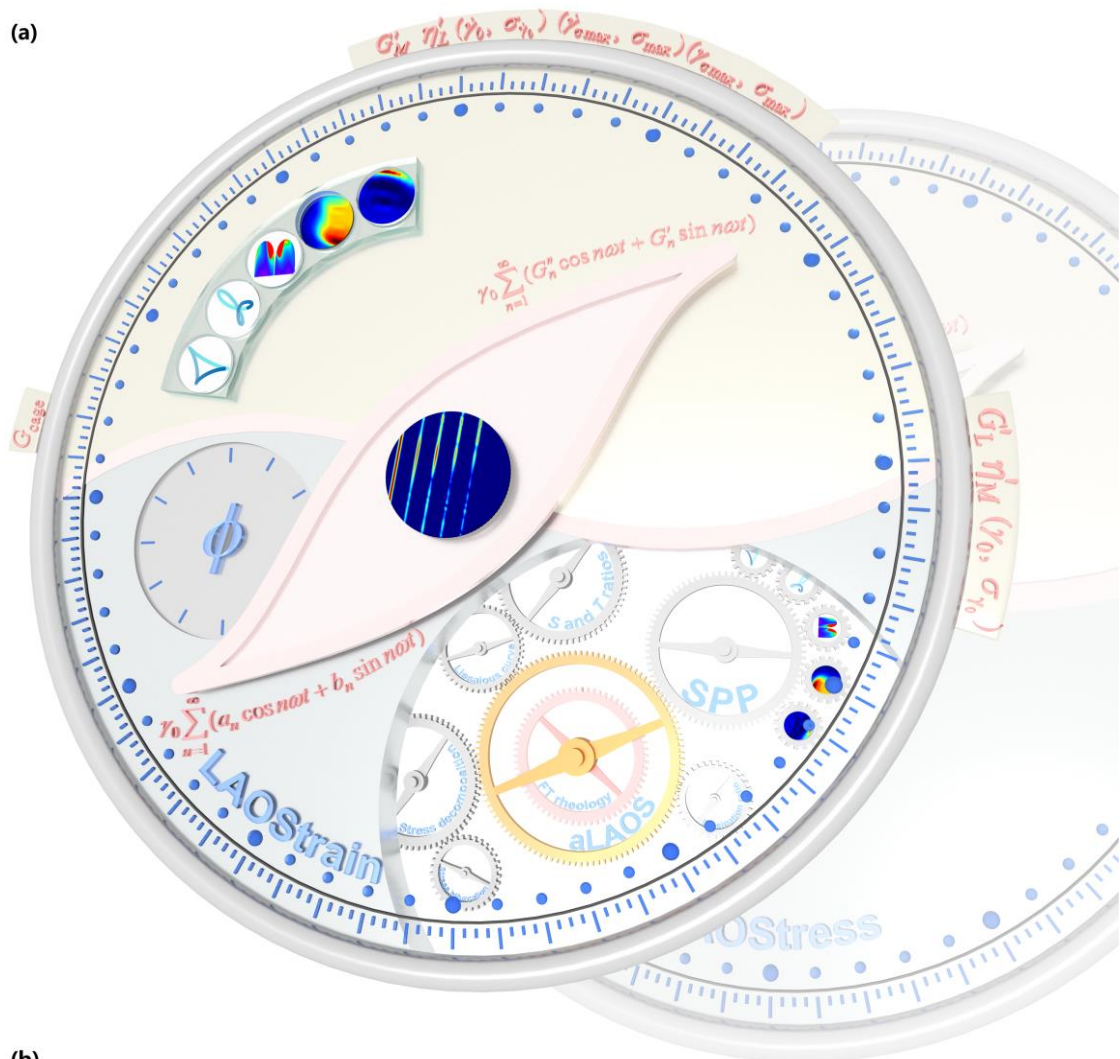

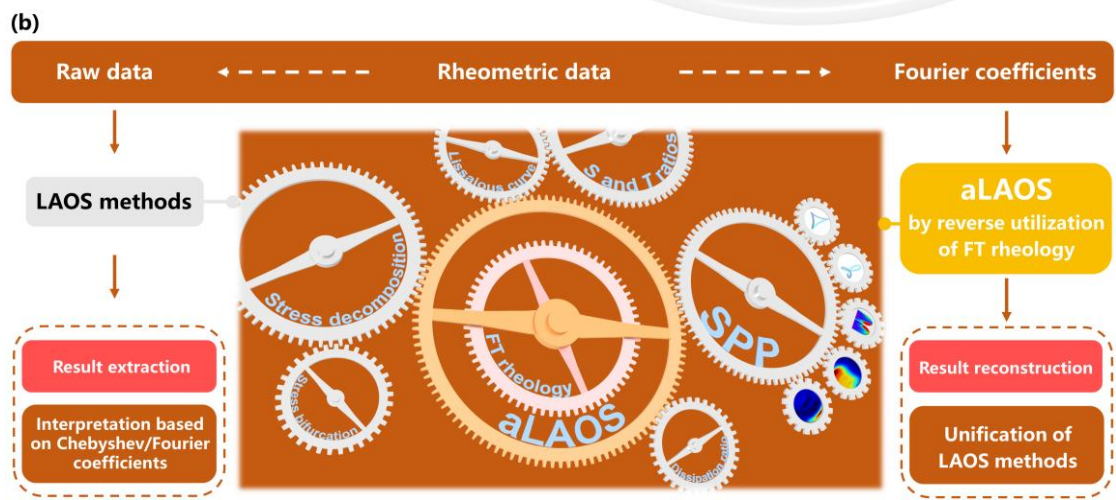

FIG. 25 Schematic illustration of the relationship between the aLAOS approach and LAOS methods. (a) The aLAOS "clock" schematically displays the time-dependent rheological behavior involving classic LAOS methods, including Lissajous curve (see the clock's edge and hand that represent the elastic and viscous Lissajous curves, respectively), FT rheology (see the central shaft of the clock and the two surrounded equations), stress decomposition (see the boundary that divides the dial of the clock into two parts), $S$ and $T$ ratios (see the symbols of $G'_M$ and $\eta'_L$ at 12 o'clock as well as $G'_L$

and $\eta'_M$ at 3 o'clock around the edge), stress bifurcation (see the symbols of $(\dot{\gamma}_0, \sigma_{\dot{\gamma}0})$, $(\dot{\gamma}_{\sigma\max}, \sigma_{\max})$, $(\gamma_{\sigma\max}, \sigma_{\max})$, and $(\gamma_0, \sigma_{\gamma0})$ at the corresponding positions around the edge), dissipation ratio (see the small dial with a central symbol of $\phi$), and SPP (see the arched and raised glass base with five grooves on the dial and the symbol of $G_{\text{cage}}$ around the edge). (b) A further comparison between the aLAOS approach and classic LAOS methods. In the case that the Fourier coefficients are readily generated from the rheometer software, raw data (torque and angle signals) and Fourier coefficients are simultaneously available in tests. Traditionally, the raw data are processed by using different LAOS methods to carry out the result extraction, following the interpretation based on Chebyshev/Fourier coefficients. By contrast, in aLAOS, the obtained Fourier coefficients are regarded as cornerstones to establish the aLAOS by the reverse utilization of FT rheology, thereby deducing the result reconstruction undergoing a simple equation-based calculation process. As a result, the mainstream LAOS methods are unified into one Fourier coefficient-based framework in aLAOS approach.

## III. CONCLUSIONS

In this work, a new mathematical framework called aLAOS approach was proposed to analyze LAOS, thereby generating LAOS data from a different perspective where the FT and FT rheology were reversely utilized to deduce the LAOS results based on trigonometric functions and Fourier coefficients rather than by treating the rheometric raw data of a series of groups containing numerous discrete points (i.e. raw Lissajous curves). The methodology of this proposed approach was first obtaining Fourier coefficients based on raw data by using FT rheology and then generating analytic results, different from the traditionally common consideration of first processing raw data using different LAOS frameworks. Then, raw results were obtained, and finally, the results were interpreted using Chebyshev polynomials/FT rheology.

As shown in Fig. 25(a), a series of classic LAOS methods were used and developed to establish our aLAOS approach, giving the analytical forms of these classic methods by using Fourier coefficients including the Lissajous curve (Sec. II C), stress decomposition (Sec. II D), strain-stiffening $S$ and shear-thickening $T$ ratios (Sec. II E), stress bifurcation (Sec. II F), dissipation ratio (Sec. II G), and SPP (transient moduli (Sec. II H), derivatives of transient moduli (Sec. II I), apparent cage modulus (Sec. II J1), phase angle and velocity of the phase angle of transient moduli (Sec. II J2), and flow curve from LAOS (Sec. II J3)). The FT ensures aLAOS a rigorous mathematical framework since FT is mathematically strict and widely applied in many fields, presenting a set of orthogonal, periodic, derivable, and integrable functions over the infinite interval from $-\infty$ to $+\infty$. The validity of the aLAOS was verified through LAOS tests on several typical YSFs (e.g. Carbopol gel, Laponite suspension, xanthan gum solution, cellulose nanofiber suspension, ketchup, reflective silver paint, and yogurt), showing perfect comparability between the reconstructed results and the raw results obtained using other LAOS methods. In addition, as the distorted signal contains infinite higher harmonics, generally referenced criteria were provided to describe how many harmonics or Fourier coefficients should be introduced to generate qualitative, approximate, and quantitative evaluations.

As shown in Fig. 25(b), in the case that the Fourier coefficients are readily obtained from the existing commercial rheometer software that processes the data from an oscillatory test in real-time via FT, the introduced aLAOS approach can: i) visually show the contribution of arbitrary higher harmonics; ii) unify the major LAOS methods expressed via a single framework based on Fourier coefficients (Fig. 25(a)); and iii) be conveniently applied in various important fields because the given equations are in simple forms and require a few parameters instead of processing many rheometric data points one by one. The aLAOS approach may directly boom further multi-domain applications of LAOS methods.

Some other conclusions can be also stated here:

The introduced classic LAOS methods are shown to be closely related to the Fourier coefficients, where clearly defined equations for these methods are provided. A raw data waveform with many discrete points can be replaced by several continuous, smooth, derivable, and integrable functions. Thus, the aLAOS approach is accessible for many steps of mathematical calculation and allows the probe of rheological responses to arbitrary intermediate timing.

As far as the measurements in LAOS are concerned, more calculation steps are required, and thus higher harmonics need to be introduced. For example, the Lissajous curve method needs only $I_1$ and $I_3$ to give the quantitative evaluation, while the derivatives of the transient moduli can just offer the approximate evaluation by introducing $I_1 \sim I_9$. In addition, two special LAOS methods of stress bifurcation and dissipation ratio require only $I_1$. Furthermore, the influence of even harmonics for these methods is clearly defined by the corresponding equations and proved to be negligible. However, the even harmonic contribution is still important because the raw data from experiments inherently contain even harmonics. Meanwhile, directly ignoring the even harmonics at the beginning of the derivation may not lead to a mathematically rigorous conclusion.

A full expression of the existing stress decomposition methods was given based on Fourier coefficients. The stress decomposition methods are equivalent by eliminating even harmonics. Based on the proposed equations, searching for the partner of every point ($\sigma(\gamma,\dot{\gamma})$) is needless to conquer the issue that the corresponding partner may not always exist. Furthermore, the equations for dissipation ratio, apparent cage modulus, and LAOS-based flow curve methods have been given for both the strain- and stress-controlled conditions. For the transient moduli, $I_3$ and $I_5$ provide the dominant information of structural transformation. Meanwhile, for the derivatives of the transient moduli, the visualized harmonic contributions reflect that $I_3$ and $I_5$ contain the major information of the structural transformation, while $I_7$ and $I_9$ significantly contribute to the curve shape adjustment.

## IV. MATERIALS AND METHODS
### A. Materials

Carbopol 980 powder was bought from Lubrizol/Noveon Consumer Specialties, Shanghai, China. Laponite powder (grade LAPONITE RD) was purchased from BYK Additives & Instruments Co., Ltd, Shanghai, China. Xanthan gum was provided from CP Kelco, Shanghai, China. 2,2,6,6-tetramethylpiperidine-1-oxyl-mediated oxidation cellulose nanofiber (CNF) was purchased from Tianjin Woodelf Biotechnology Co. Ltd., Tianjin, China. Ketchup was purchased from Mccormick (Guangzhou) Food Co., Ltd., Guangzhou, China. Reflective silver paint was provided by BASF (China) Co., Ltd., Shanghai, China. Yogurt (Guanyiru-(daily fresh cheese)) was purchased from Inner Mongolia Mengniu Dairy (Group) Co., Ltd., Inner Mongolia, China. Sodium hydroxide (NaOH) and hydrochloric acid (HCl) were purchased from Titan, Shanghai, China. The water used in all experiments was pretreated with the Milli-Q System (Millipore Corporation, Bedford, MA, USA). All samples and chemicals were used without further purification.

**B. Sample preparation**

All samples were prepared at room temperature. Carbopol gel and Laponite suspension were prepared according to previous literature [28]. 0.2 wt% Carbopol gel was prepared by mixing Carbopol powder and water and stirring (1000 rpm, 15 min). Then, the Carbopol gel was adjusted to pH = 7 by using NaOH. 1.5 wt% Laponite suspension was obtained by dispersing Laponite powder in water (1000 rpm, 10 min). After that, the pH value of the obtained solution was adjusted to 7 by using HCl. The 1.25 wt% xanthan gum solution was prepared according to the procedure of our previous work [3]. Xanthan gum was dissolved in water by violent stirring (12 h). Then, the prepared solution was centrifuged to remove bubbles and obtain a homogeneous solution. 0.65 wt% CNF suspension was obtained by diluting the commercial 1.3 wt% CNF aqueous suspension with water.

**C. Rheological measurement**

All experiments were executed on a stress-controlled rotational rheometer (HaakeMars III, Thermo Fisher, Germany) by using parallel plates (diameter: 60 mm, gap: 0.5 mm). It is noticed that the parallel plate will lead to the shear rate gradient [89] and is applied to study YSFs [36, 90]. Each sample was sealed with low-viscosity silicone oil (10 mPa·s) at 25 °C before the test. Waterproof sandpaper (3 M, particle size: ~20 μm) was applied to mitigate the wall slip effect. The oscillatory stress amplitude sweeps were carried out with different stress ranges at 1 Hz, where the sampling points were 60 for each sample. The minimum and maximum values of the applied stress were regarded as point 1 and point 60. Twenty cycles were set for each sampling point. The raw torque and displacement were recorded as 512 points per cycle. The preshear protocol was: the oscillatory shearing, a strain amplitude of 500%, 300 s. The set rest time was proved to be enough for the equilibrium (the change of $G'$ was within ±5% in 5 h [28]).

**ACKNOWLEDGMENTS**

This work was supported by the National Natural Science Foundation of China (Grant No. 22073062, 21774075).

**CONFLICT OF INTEREST**

The authors declare that they have no competing interests.

**AUTHOR CONTRIBUTIONS**

H.Z. conceived the concept. H.Z. and P.W. designed the processing and mathematical details. H.Z., P.W., J.X. and Z.Z. analyzed the experimental details and co-wrote the manuscript. P.W. carried out most experiments and data analysis. H.Z. supervised the work and revised the manuscript.

**DATA AVAILABILITY**

The data that support the findings of this study are available from the corresponding author upon reasonable request.

**APPENDIX A: FT RHEOLOGY ANALYSIS**

In Sec. II, the principle of the analytic LAOS approach was proposed, verified, and discussed based on the data from the stress-controlled oscillatory shear test of 0.2wt% Carbopol gel at 1 Hz. In this section, this method will be further examined by using six typical samples including 1.5 wt% Laponite suspension, 1.25 wt% xanthan gum solution, 0.65 wt% cellulose nanofiber (CNF) suspension, 100 wt% ketchup, 100 wt% reflective silver paint, and a kind of yogurt (100 wt% Guanyiru-(daily fresh cheese)), at the experiment condition the same as the above test. Then, the sample concentration and test protocol will not be repeatedly denoted in this section.

The FT rheology analyses of the six samples are first carried out, and the representative results are plotted in Fig. 26. Figures 26(a.i), 26(b.i), 26(c.i), 26(d.i), 26(e.i), and 26(f.i) demonstrate the plots of dynamic moduli versus stress (more information please refer to Sec. II B, especially Sec. II B2.2). Then, the nonlinearities of each sample, a series of groups of the obtained Fourier coefficients, were plotted in the x–y projection FT fingerprint in Figs. 26(a.ii), 26(b.ii), 26(c.ii), 26(d.ii), 26(e.ii), and 26(f.ii). All samples show the $I_3/I_1$, $I_5/I_1$, and $I_7/I_1$ values close to zero at low stress amplitudes. After that, the normalized 3rd, 5th, and 7th harmonic intensities ($I_3/I_1$, $I_5/I_1$, and $I_7/I_1$) are depicted in Figs. 26(a.iii), 26(b.iii), 26(c.iii), 26(d.iii), 26(e.iii), and 26(f.iii). These values first increased and then decreased with the increase in the stress amplitude, which denotes the structural transformation in each sample during the sweep process. Finally, the points of $G' \approx G''$ were analyzed in detail, and the results are demonstrated in the remaining panels in Fig. 26. As can be seen, although each point presents $G' \approx G''$, the waveforms are different from each other, resulting in various FT spectra with infinite higher harmonics. Therefore, FT rheology is regarded as a method to distinguish complex soft materials.

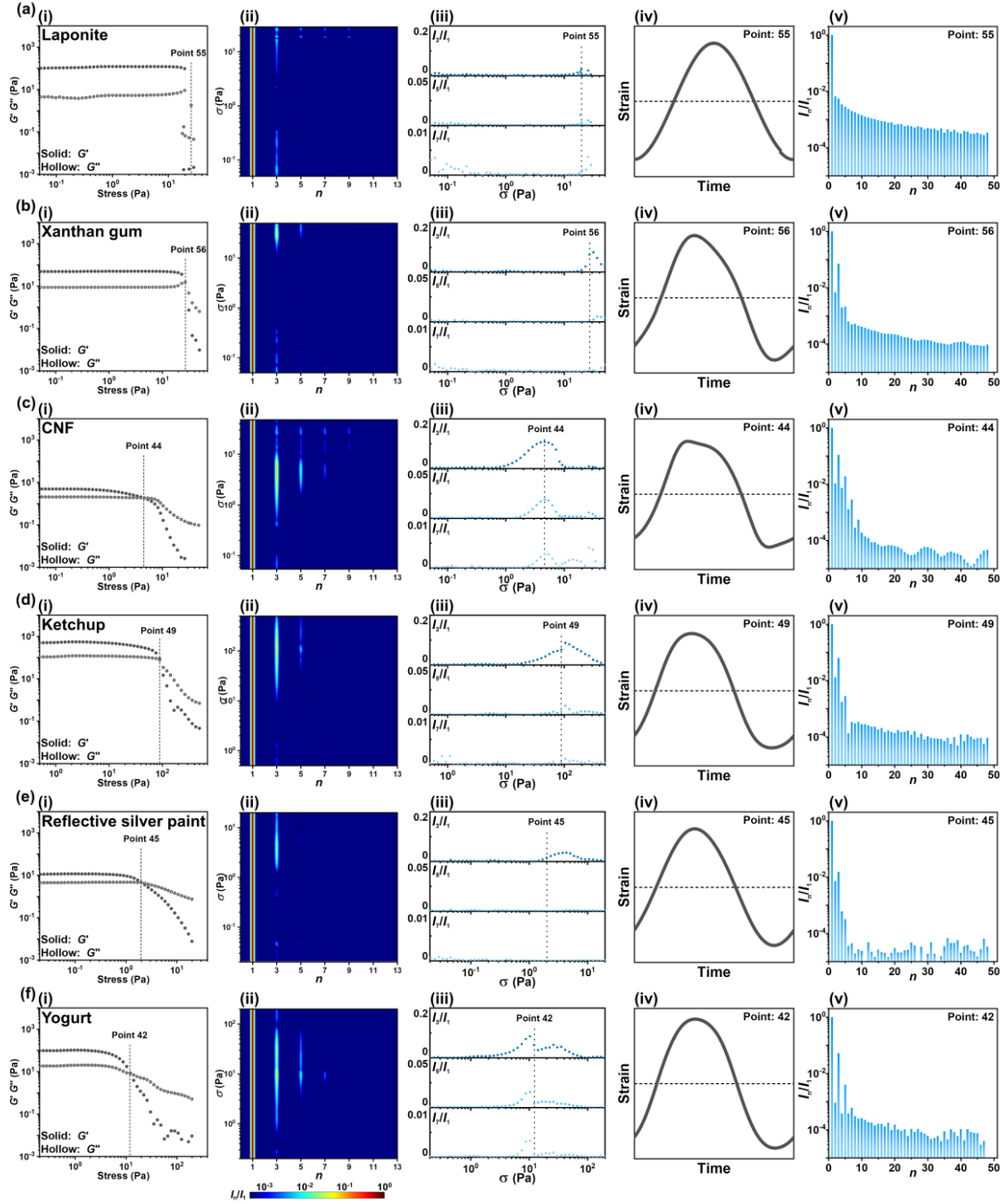

FIG. 26 FT rheology analyses of six typical samples (the first step of the analytic LAOS approach). The results of (a) Laponite suspension, (b) xanthan gum solution, (c) CNF suspension, (d) ketchup, (e) reflective silver paint, and (f) yogurt analyzed by FT rheology: (i) dynamic moduli of the amplitude sweep test; (ii) stress-frequency projection ($\sigma$ versus ($n \cdot \omega$) plot); (iii) the plot of normalized $I_3/I_1$, $I_5/I_1$, and $I_7/I_1$ values versus stress amplitude; (iv) the distorted strain waveform under a sinusoidal stress signal of the corresponding point presenting $G' \approx G''$; and (v) the FT spectrum of the demonstrated distorted strain waveform ($I_n/I_1$ as a function of $n$th harmonic ($I_1 \sim I_{48}$)). The data displayed in this figure will be applied for all afterward discussions in Appendix.

## APPENDIX B: ANALYTIC LISSAJOUS CURVE

After the calculation of many Fourier coefficients (the first step of the analytic LAOS approach),

a series of reconstructed Lissajous curves were thus constructed and compared with the raw Lissajous curves in Figs. 27 and 28 (for detailed information, please refer to Sec. II C). Figure 27 contains the results of the Laponite suspension, xanthan gum solution, and CNF suspension. Figure 28 consists of the results of the ketchup, reflective silver paint, and yogurt. As an example, figure 27(a) has three columns of Lissajous curves, where the first column demonstrates four types of elastic Lissajous curves including the raw results at the strain-controlled condition, as well as the reconstructed results at the stress-controlled condition. The second column corresponds to the raw and reconstructed viscous Lissajous curves at the strain-controlled condition. Similarly, the third column contains the raw and reconstructed viscous Lissajous curves at the stress-controlled condition. More detailed information can be found in Fig. 4.

The rheological behaviors for the six samples are all linear viscoelastic at low stress amplitudes, as demonstrated elliptic shapes of the Lissajous curves. Around the denoted points in the red at the position of $G' \approx G''$, the ellipse opening gradually became larger and smaller for the elastic and viscous Lissajous curves with the increase in the stress amplitude, respectively, along with the increasing nonlinear and pseudoplastic rheological responses. At extremely large stress levels, all six samples present rheological behaviors close to the pure viscous response. Figures 27 and 28 depict that the reconstructed Lissajous curves present high similarities to the raw Lissajous curves throughout the six whole stress sweep ranges, demonstrating the universality of the analytic LAOS approach.

As a series of reconstructed Lissajous curves have shown similarities to the raw Lissajous curves, while the results from classic LAOS methods will be compared with those generated by the analytic LAOS approach.

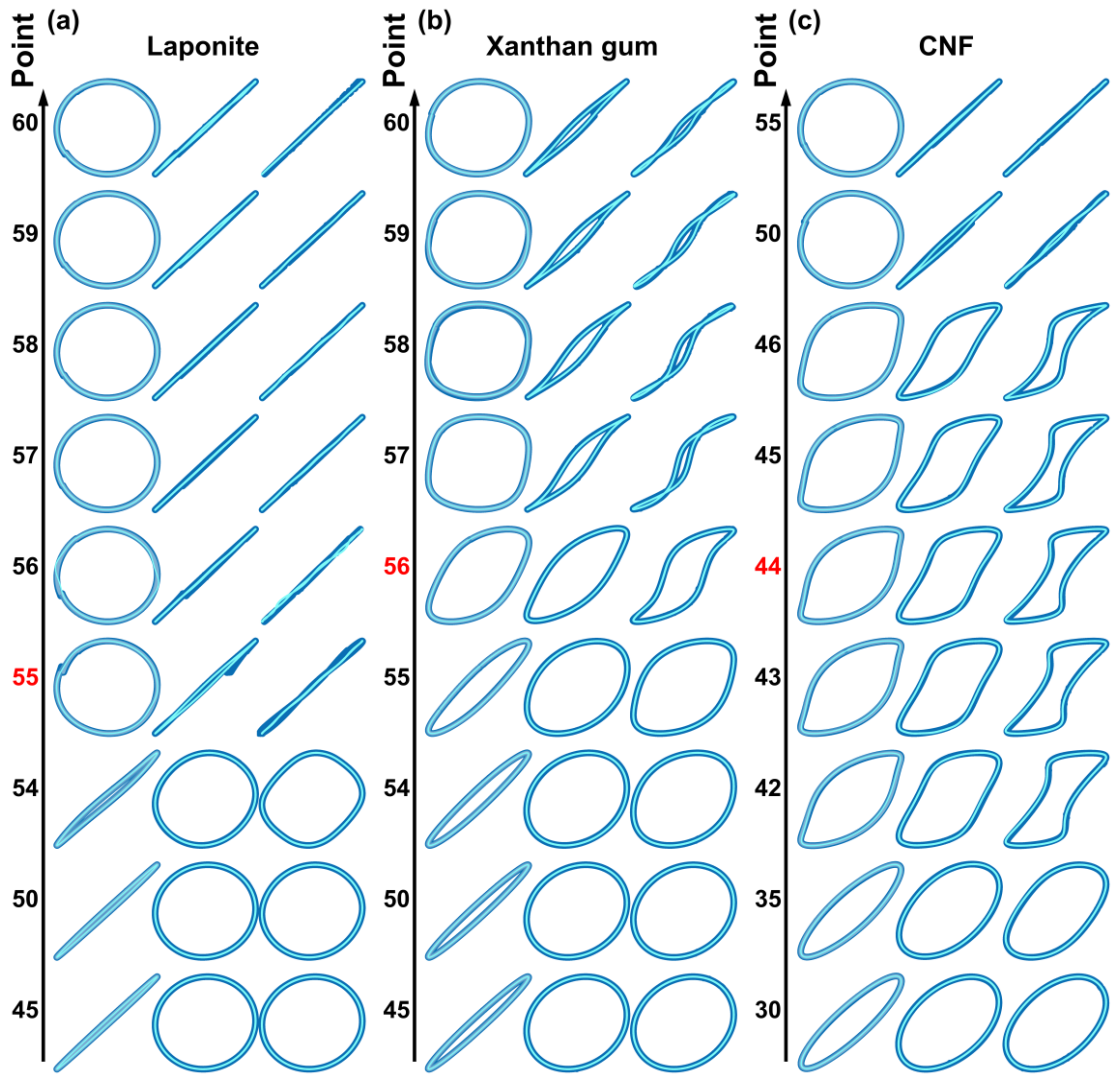

FIG. 27 Comparison between the normalized raw and the reconstructed Lissajous curves at different stress sweep points by introducing the 1st ~ 11th harmonics (red number: the point of $G' \approx G''$): (a) Laponite suspension; (b) xanthan gum solution; and (c) CNF suspension. The corresponding information has been denoted in Fig. 4.

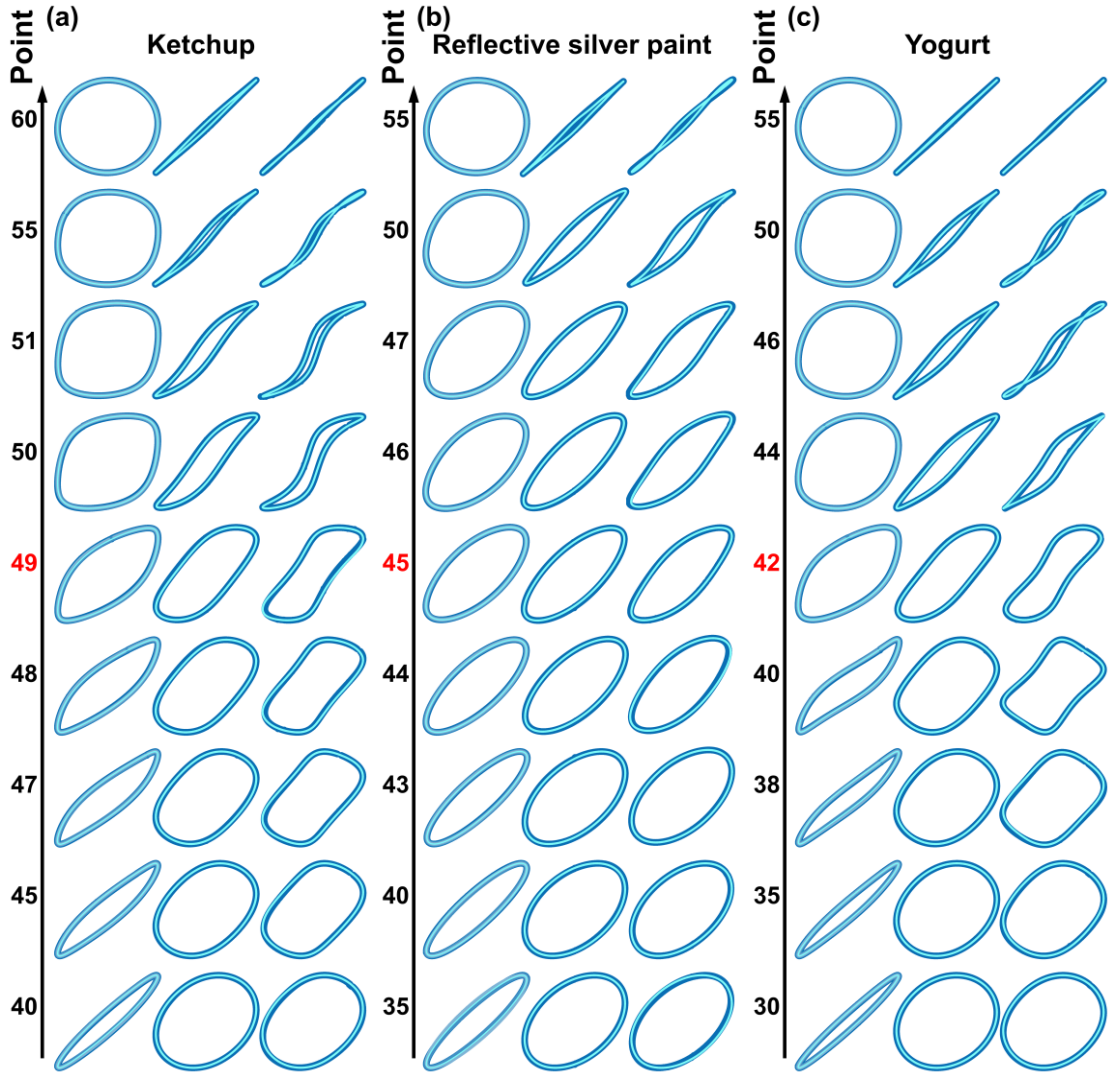

FIG. 28 Comparison between the normalized raw and reconstructed Lissajous curves: (a) ketchup; (b) reflective silver paint; and (c) yogurt. The corresponding information can refer to Fig. 4. Introducing 1st ~ 11th harmonics. Red numbers: points of $G' \approx G''$.

**APPENDIX C: ANALYTIC STRESS DECOMPOSITION**

The raw (blue lines) and reconstructed (red lines) stress decomposition results of the six samples are compared in Fig. 29. It is obvious that good overlaps are shown between the raw and the reconstructed results. Both the intensities and curvatures of these reconstructed results correspond well with the raw elastic and viscous stress curves.

At low stress levels, the elastic and viscous stress curves show linearity. In Fig. 29, the slopes of the two kinds of stress curves decreased with the increase in the stress amplitude, along with the increased intensity of the viscous stress. The manifestation of a solid-liquid transition during the oscillatory stress sweep can be readily distinguished in Fig. 29. Meanwhile, the strain-stiffening ($d^2\sigma'/d^2\gamma > 0$), strain-softening ($d^2\sigma'/d^2\gamma < 0$), shear-thickening ($d^2\sigma''/d^2\gamma > 0$), and shear-thinning ($d^2\sigma''/d^2\gamma < 0$) behaviors can be visually distinguished in Fig. 29. At extremely high stress amplitudes, the linear viscous stress curves indicate the linear viscous behavior of the sample. Meanwhile, the nonlinearities of the elastic stress curves were much larger than the viscous stress curves, indicating that this method is more sensitive to nonlinearity than FT rheology [20]. Except for the visual judgment, the quantitative evaluations of the strain-stiffening, strain-softening, shear-thickening, and shear-thinning behaviors were further discussed in the next section.

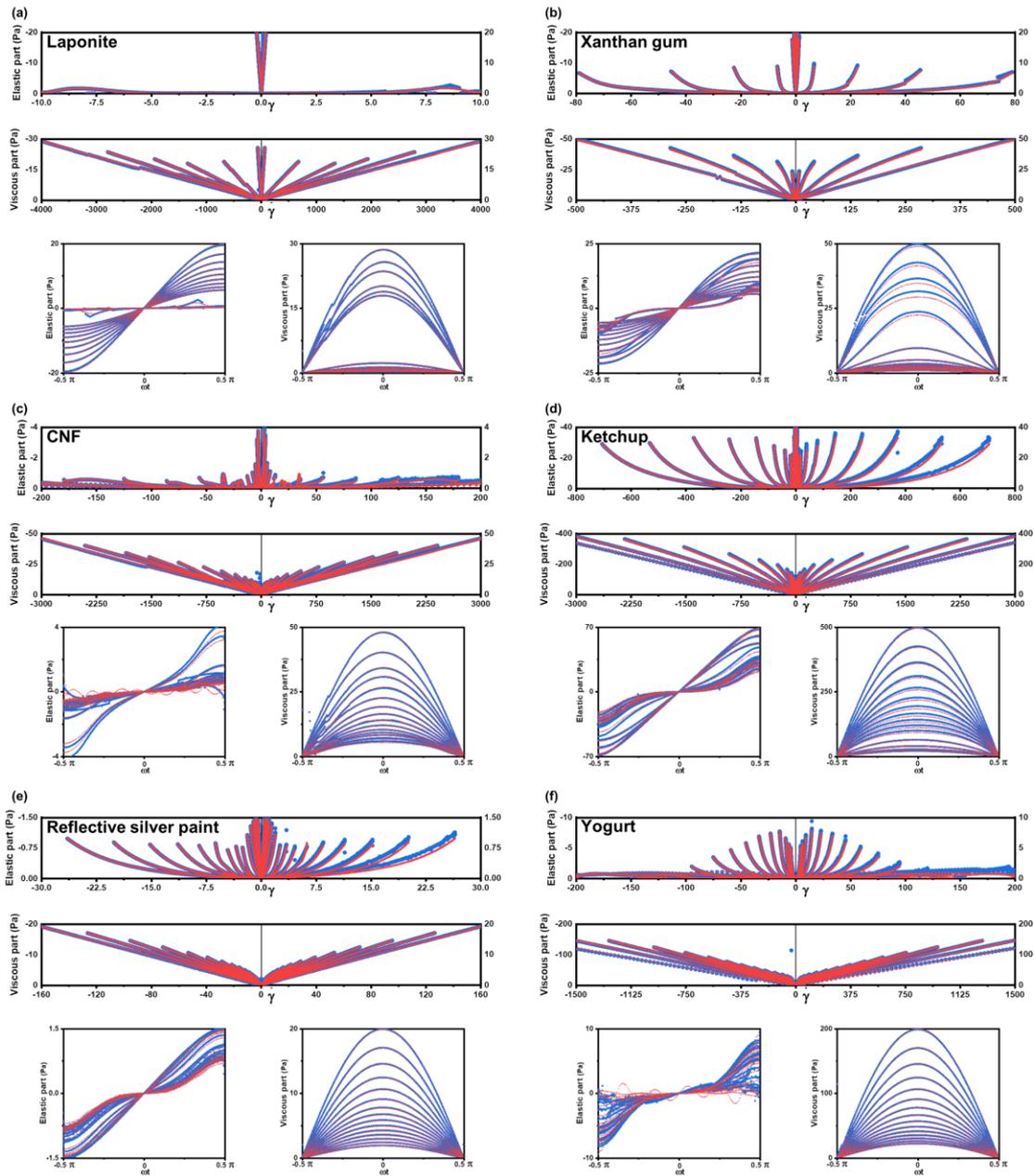

FIG. 29 Comparison between the raw (blue lines) and the reconstructed (red lines) stress decomposition results: (a) Laponite suspension; (b) xanthan gum solution; (c) CNF suspension; (d) ketchup; (e) reflective silver paint; and (f) yogurt. The plots of the elastic stress versus strain, viscous stress versus strain rate, as well as the elastic and viscous stresses versus angle are included. Introducing 1st ~ 9th harmonics.

## APPENDIX D: ANALYTIC *S* AND *T* RATIOS

The raw (blue points) and reconstructed (red points) $G'_M$, $G'_L$, $\eta'_M$, $\eta'_L$, $S$, and $T$ values of the six samples are plotted in Figs. 30 and 31. Good overlaps between the raw and the reconstructed results are shown, no matter for $G'_M$, $G'_L$, $\eta'_M$, and $\eta'_L$ values or $S$ and $T$ values (noting the linear axis for the deviations between the raw and reconstructed values of $\eta'_M$ and $\eta'_L$).

Figure 30 indicates that the six samples show the strain stiffening response. Meanwhile, the Laponite suspension almost shows the linear viscous response (Fig. 30(a)). In addition, the xanthan gum solution (Fig. 30(b)) and reflective silver paint (Fig. 31(b)) demonstrate first the weak shear thickening behavior, then the shear thinning behavior, and finally linear viscous behavior. Furthermore, except for the SAOS region and the region with extremely large stress values, the CNF suspension (Fig. 30(c)) behaves a strong shear thickening response, while the ketchup (Fig. 31(a)) and yogurt (Fig. 31(c)) present the significant shear thinning response.

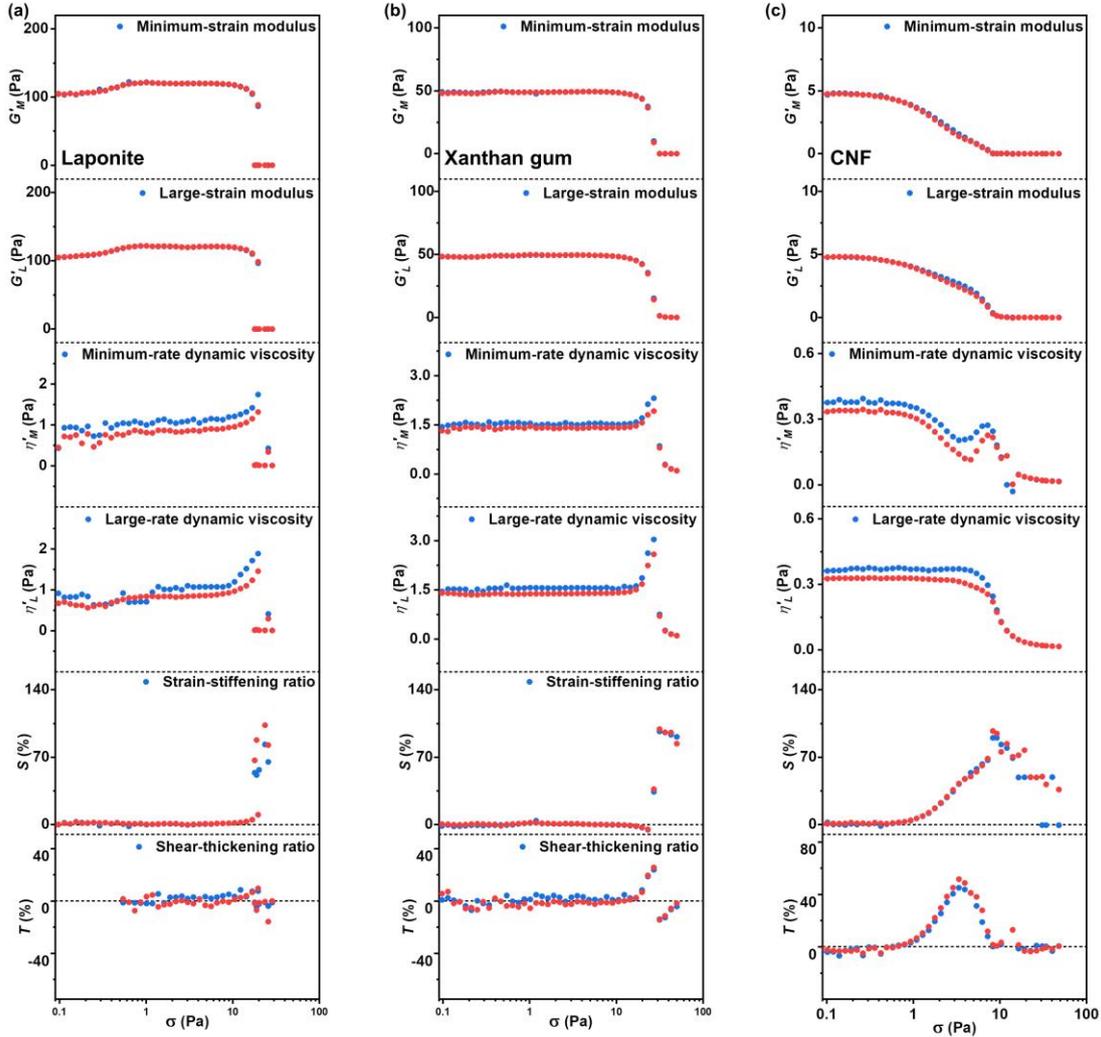

FIG. 30 Comparison between the raw (blue points) and the reconstructed (red points) values of the nonlinear measures, including $G'_M$, $G'_L$, $\eta'_M$, $\eta'_L$, $S$ and $T$ values: (a) Laponite suspension; (b)

xanthan gum solution; and (c) CNF suspension. Introducing 1st ~ 11th harmonics.

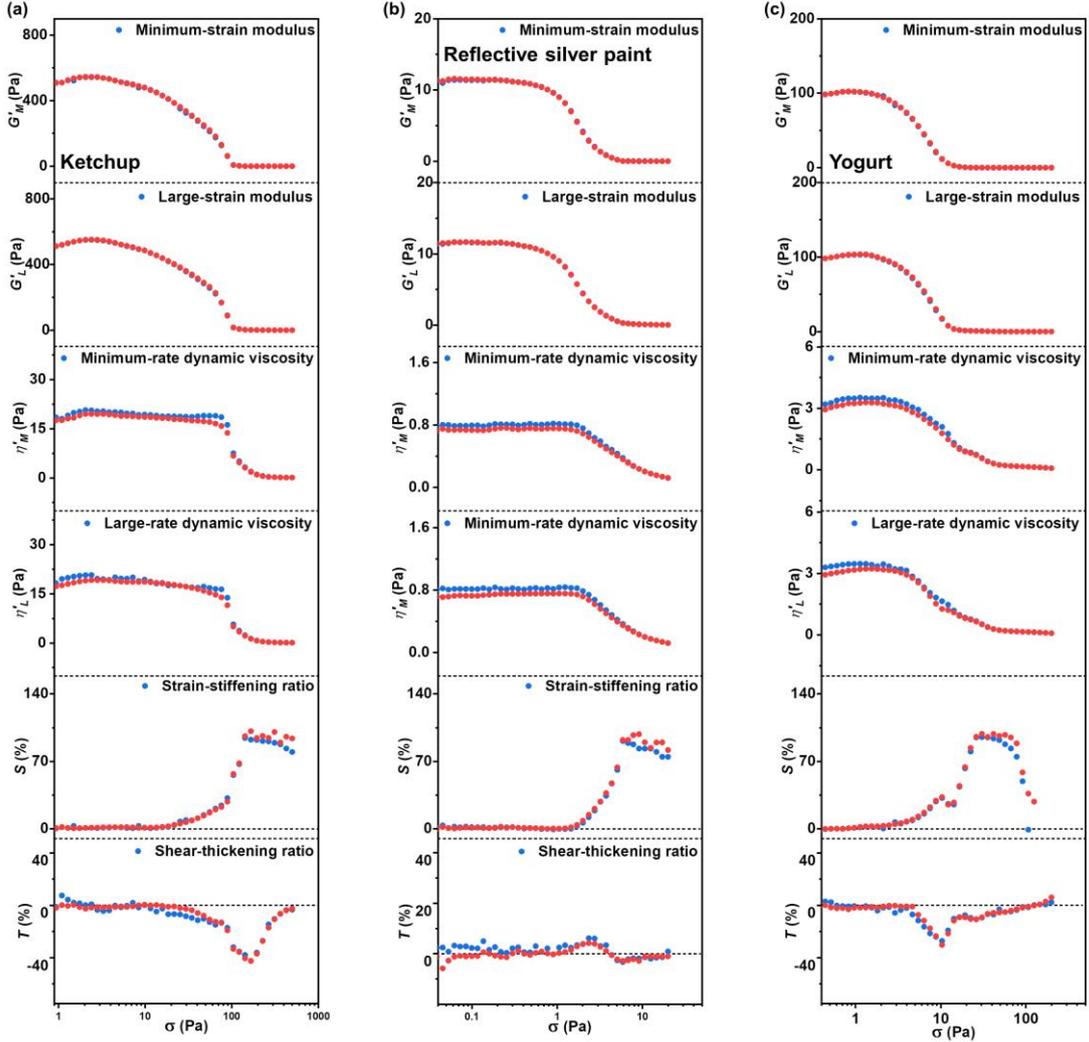

FIG. 31 Comparison between the raw (blue points) and the reconstructed (red points) $G'_M$, $G'_L$, $\eta'_M$, $\eta'_L$, $S$ and $T$ values: (a) ketchup; (b) reflective silver paint; and (c) yogurt. Introducing 1st ~ 11th harmonics.

**APPENDIX E: ALGEBRAIC AND ANALYTIC STRESS BIFURCATIONS**

The raw (blue points) and reconstructed (red points and purple lines) stress bifurcation curves of the six samples are depicted in Fig. 32. The red points and purple lines correspond to the results from the algebraic stress bifurcation and the analytic stress bifurcation methods, respectively. As can be seen, the algebraic stress bifurcation is enough for giving precise bifurcation points (the start and end yield points), while the analytic stress bifurcation is capable of providing curves with high similarities to the raw results.

From Fig. 32, the start and end yield points, as well as the solid-liquid transition regions for the six samples can be determined including points 54 ~ 56 for the Laponite suspension, points 55 ~ 57 for the xanthan gum solution, points 41 ~ 48 for the CNF suspension, points 47 ~ 50 for the ketchup, points 42 ~ 48 for the reflective silver paint, and points 40 ~ 43 for the yogurt, where the corresponding stress values are 20 ~ 19 Pa (the structure has been destroyed), 23 ~ 32 Pa, 2.9 ~ 8.3 Pa, 65 ~ 99 Pa, 1.2 ~ 3.2 Pa, and 8.8 ~ 14 Pa for the six samples in order, respectively.

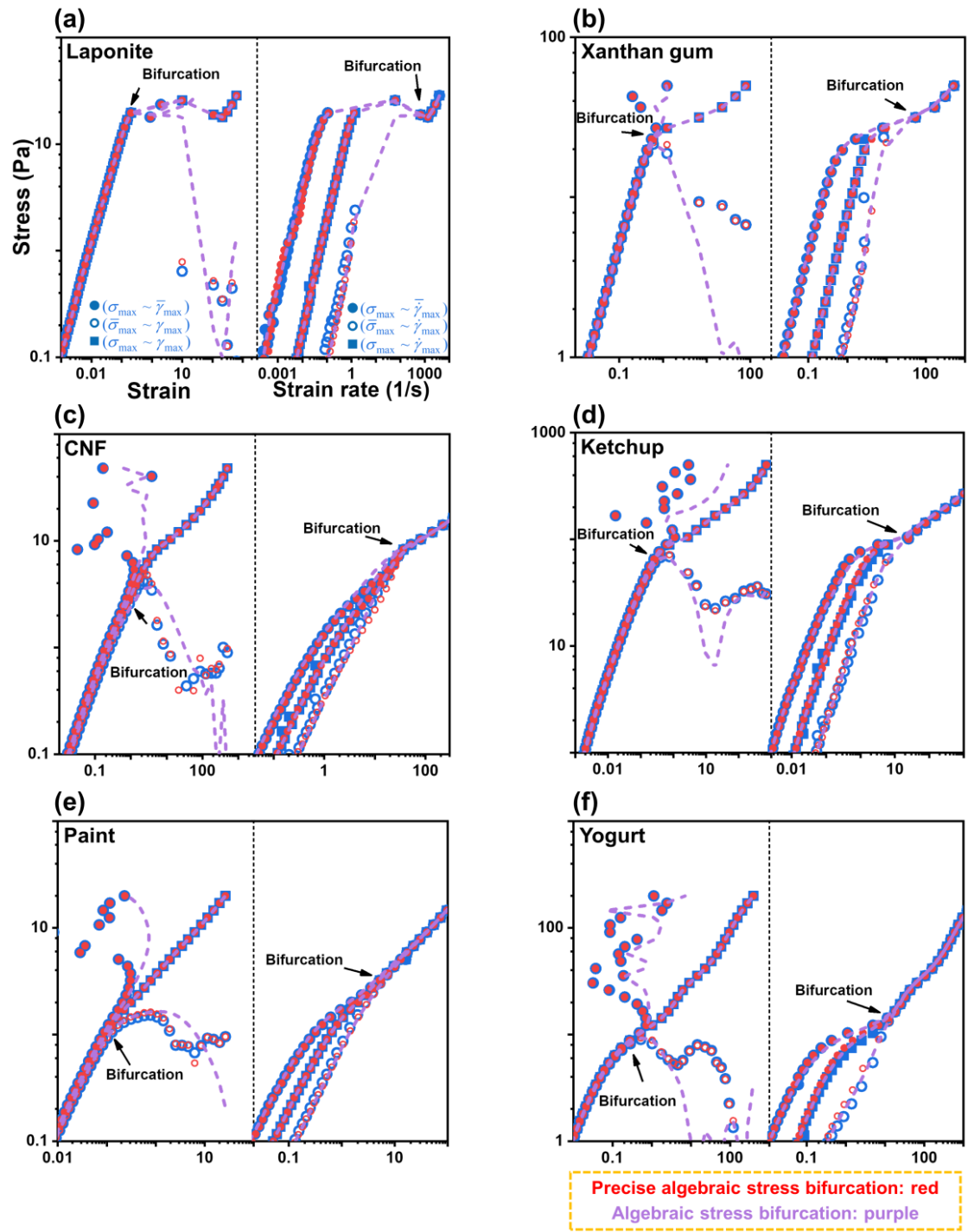

FIG. 32 Comparison between the raw (blue points) and the reconstructed (red points and purple lines) stress bifurcation curves: (a) Laponite suspension; (b) xanthan gum solution; (c) CNF suspension; (d) ketchup; (e) reflective silver paint; and (f) yogurt. Introducing 1st ~ 11th harmonics. Red points: analytic stress bifurcation. Purple lines: algebraic stress bifurcation.

**APPENDIX F: ANALYTIC DISSIPATED ENERGY AND DISSIPATION RATIO**

The raw (blue points) and analytic (red points and purple lines) results of the dissipated energy and dissipation ratio based on the six samples are displayed in Fig. 33. The red points and purple lines refer to the results at the strain- and stress-controlled conditions, respectively. Figure 33 visually shows the high similarities among the three kinds of results including the raw data by treating the elastic Lissajous curve with integration, the analytic data at the strain-controlled condition, and the analytic data at the stress-controlled condition.

It is seen that the dissipation ratios of the six samples at low stress levels follow the order: CNF > reflective silver paint > xanthan gum ≈ ketchup ≈ yogurt > Laponite suspension. In addition, the samples of Laponite suspension, CNF suspension, reflective silver paint, and yogurt show the transitions from viscoelastic behaviors to perfect Newtonian behaviors. However, for xanthan gum solution and ketchup, the transitions follow the order of (viscoelastic behavior)–(combined plastic and Newtonian behavior)–(Newtonian behavior).

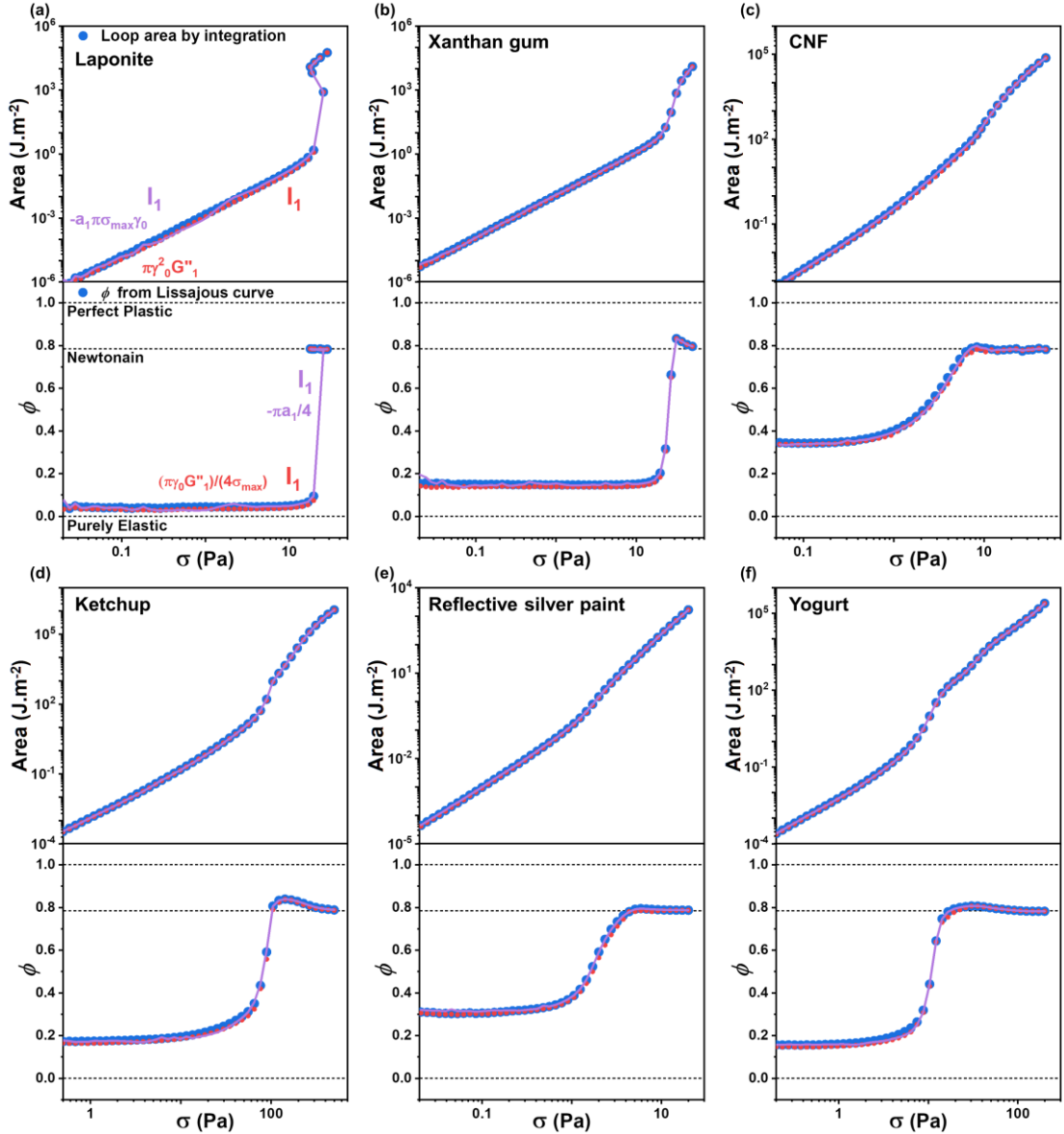

FIG. 33 Comparison between the raw (blue points) and the analytic (red points and purple lines) results of dissipated energy $E_d$ and dissipation ratio $\phi$: (a) Laponite suspension; (b) xanthan gum solution; (c) CNF suspension; (d) ketchup; (e) reflective silver paint; and (f) yogurt. Introducing 1st harmonic. Red points: analytic results at the strain-controlled condition. Purple lines: analytic results at the stress-controlled condition.

**APPENDIX G: ANALYTIC TRANSIENT MODULI**

The raw (blue curves) and the reconstructed (red curves) transient moduli of the six samples are demonstrated in Fig. 34. As an example, figure 34(a.i) shows the structural transformation during the stress swept from the SAOS region to the LAOS region, where figures 34(a.ii) and 34(a.iii) correspond to the LAOS and SAOS regions, respectively. Therefore, the detailed information on the six samples can be sufficiently demonstrated.

The reconstructed results correspond also well with the raw results. However, relatively significant deviations arise between the raw and the reconstructed results of CNF suspension, which may be attributed to the instantaneous rheological behavior originating from the orientation effect of the CNF suspension under deformation. Therefore, this phenomenon implies that FT and FT rheology simulate the dominant information of a waveform, where some potential messages may be lost. However, the analytic LAOS approach is still applicable for treating the CNF suspension sample if the intensity information is insignificant for an investigation.

It can be also observed that the Laponite suspension undergoes the most violent structure transformation and xanthan gum is the second, whereas the reflective silver paint is the mildest. The structures of the ketchup and yogurt sample are changed gradually during the stress sweep. According to the common sense that CNF suspensions possess unique rheological properties, the CNF sample has similar rheological responses within a specific stress range. More information can be found in Sec. II H and the next several sections.

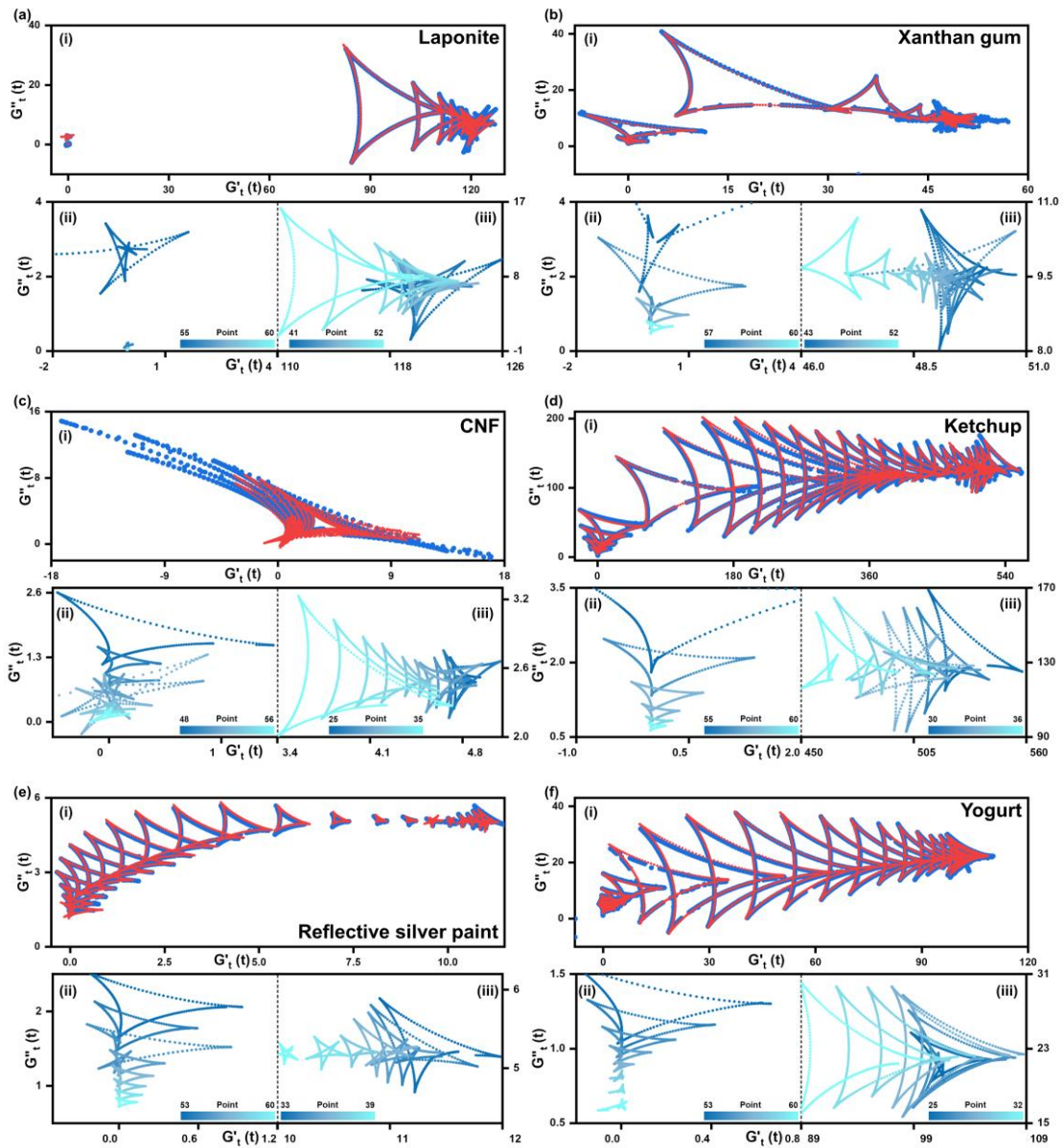

FIG. 34 Raw and reconstructed transient moduli ((a) Laponite suspension; (b) xanthan gum solution; (c) CNF suspension; (d) ketchup; (e) reflective silver paint; and (f) yogurt): (i) comparison; (ii) the reconstructed results in the LAOS region; and (iii) the reconstructed results in the SAOS region. Introducing 1st ~ 7th harmonics. Blue lines: raw results. Red lines: reconstructed results.

**APPENDIX H: ANALYTIC DERIVATIVES OF TRANSIENT MODULI**

The raw (blue points) and the reconstructed (red curves) derivatives of the transient moduli based on the six samples are shown in Fig. 35. Overlaps between the raw and reconstructed results can be realized, reflecting that the analytic LAOS approach can provide the approximate evaluation. However, deviations arise between the raw and reconstructed results. On the one hand, the quantitative evaluation is hard to be offered because the raw data has undergone six calculation steps (i.e. Fig. 16(a), Eqs. 85 and 86, Sec. II I1), leading to sharply amplified errors. On the other hand, the analytic LAOS approach as a precise and quantitative method is also shown and explained. Thus, according to Fig. 35, it can be argued that the raw data provided by our rheometer can only support six steps of data processing. In other words, it is foreseeable that unimaginable deviations will arise between the raw and reconstructed results when the raw data are treated by seven steps of calculation. However, the analytic LAOS approach is accessible for many steps of mathematical calculation.

The instantaneous stiffening ($dG_t'(t)/dt > 0$), softening ($dG_t'(t)/dt < 0$), thickening ($dG_t''(t)/dt > 0$), and thinning ($dG_t''(t)/dt < 0$) behaviors of the six samples will be discussed in our next work because figure 35 lacks color mapping. However, the evolution of the instantaneous rheological behaviors can still be distinguished from Fig. 35. All in all, figure 35 has visually demonstrated the outstanding ability of the analytic LAOS approach.

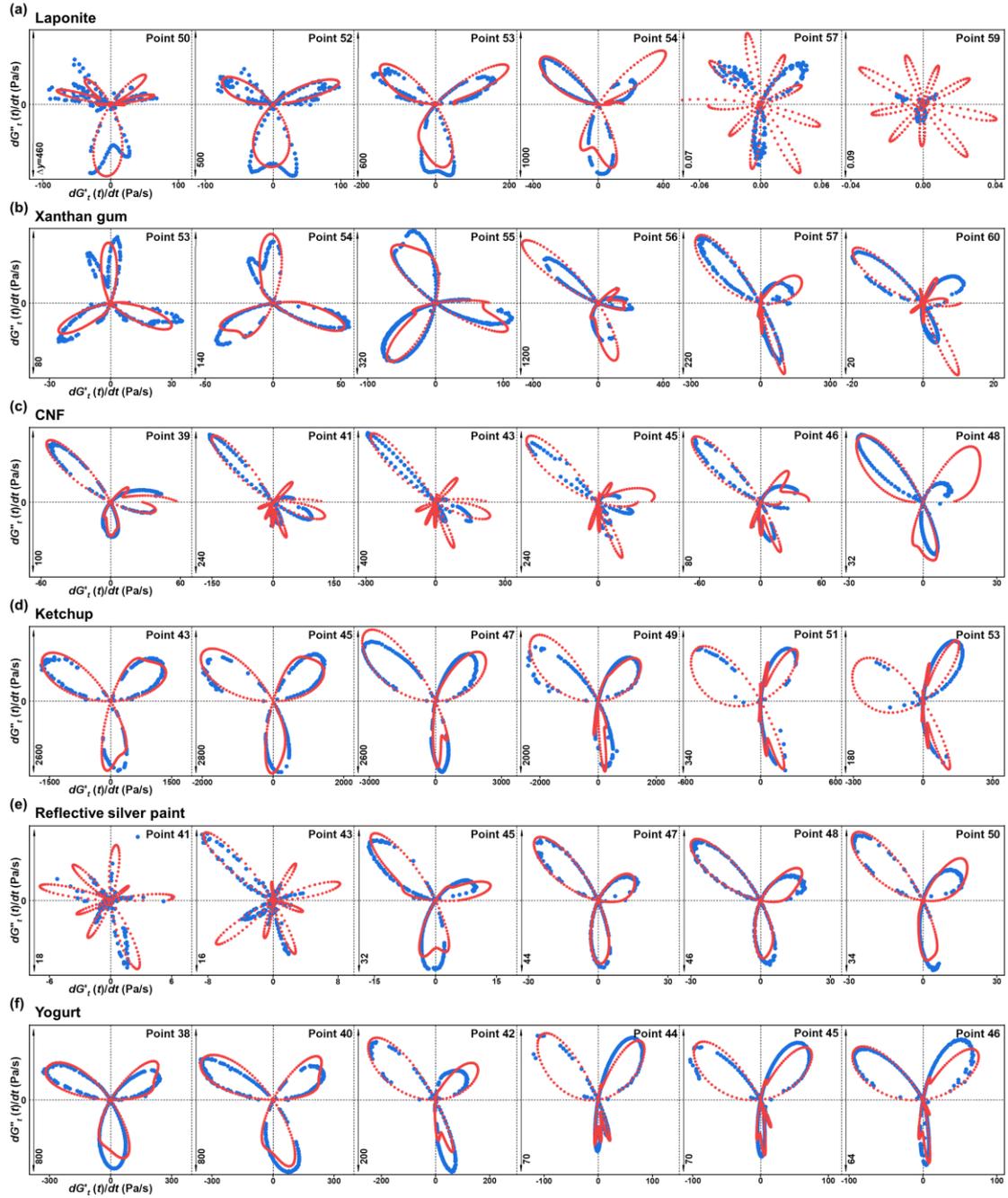

FIG. 35 Comparison between the raw (blue points) and the reconstructed (red points) derivatives of transient moduli: (a) Laponite suspension; (b) xanthan gum solution; (c) CNF suspension; (d) ketchup; (e) reflective silver paint; and (f) yogurt. Introducing 1st ~ 9th harmonics.

**APPENDIX I: ANALYTIC CAGE MODULUS**

The raw (blue points) and the reconstructed (red points and purple lines) results of the cage moduli $G_{\text{cage}}$ based on the six samples are illustrated in Fig. 36. The red points and purple lines denote the results based on the data at the strain- and stress-controlled conditions, respectively. Figure 36 demonstrates the high similarities among the three kinds of results, including the raw $G_{\text{cage}}$ based on the raw elastic Lissajous curve, the reconstructed $G_{\text{cage}}$ values at the strain-controlled condition, and the reconstructed $G_{\text{cage}}$ values at the stress-controlled condition. Meanwhile, the values of $G'$ and $G''$ versus stress are also plotted in Fig. 36.

The $G_{\text{cage}}$ and $G'$ values for the six samples all show good overlaps at low stress level. Then, with the increase in the stress amplitude, the $G_{\text{cage}}$ gradually shows the deviation from $G'$. The values of $G_{\text{cage}}$ were higher than $G'$ in the LAOS region, indicating the intracycle elastic behavior. Thus, the potential elastic behaviors under large stress amplitudes can be identified by the $G_{\text{cage}}$, where $G'$ lacks this ability. It can be noticed that the CNF suspension and reflective silver paint, with the shear thickening behavior determined by Figs. 30 and 31, can present the $G_{\text{cage}}$ values at the LAOS region higher than $G'$ values at the SAOS region. Meanwhile, the reflective silver paint sample presents an impressive structural recovery ability, which may contribute to the application in real practice.

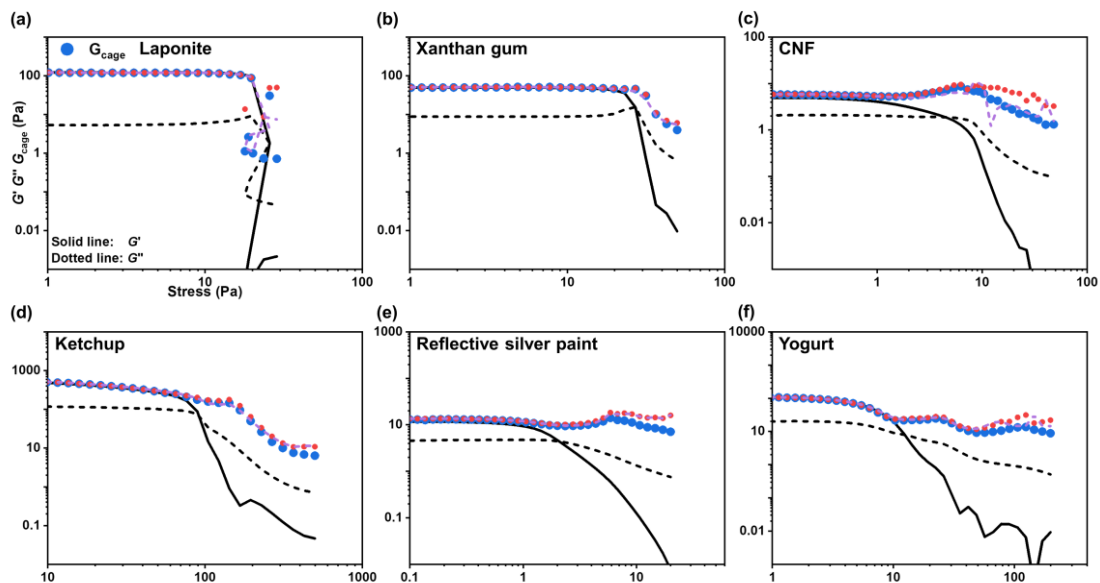

FIG. 36 Comparison between the raw (blue points) and the reconstructed (red points and purple lines) cage moduli $G_{\text{cage}}$: (a) Laponite suspension; (b) xanthan gum solution; (c) CNF suspension; (d) ketchup; (e) reflective silver paint; and (f) yogurt. Introducing 1st ~ 5th harmonics. Red points: reconstructed results based on the FT of the distorted strain signal. Purple lines: reconstructed results based on the FT of both the distorted strain and distorted stress signals.

**APPENDIX J: OTHER TRANSIENT-MODULI-BASED RESULTS**

Figures 37 and 38 show the three transient-moduli-based measures ($G_t'/G_t''$, $\delta_t$, and $d\delta_t/dt$) to expose the inner instantaneous rheological properties of the six samples. As an example, figure 37(a.i) shows a series of the intracycle solid-liquid transitions, $G_t'/G_t''$ values at each angle step (equal to the time step). Meanwhile, the measures of $\delta_t$ and $d\delta_t/dt$ are represented by Figs. 37(a.ii) and 37(a.iii).

When $G_t'/G_t''$ is close to zero, the sample behaves more liquidlike than solidlike (the red regions). It should be noticed that the negative $G_t'/G_t''$ value occurs when $G_t' < 0$ or $G_t'' < 0$, which is also denoted as the red regions in the plots of $\ln|\gamma|$ (intracycle strain) and $\ln|\dot{\gamma}|$ (intracycle strain rate) versus angle. $G_t' < 0$ represents the occurrence of backflow. The blue regions denote that the values of $G_t'/G_t''$ are much larger than zero, showing the solidlike behavior. This plot can show the intracycle and intercycle time-dependent rheological behaviors. In each first panel of Figs. 37 and 38, the blue and red regions are shrunk and expanded with the increase in the strain amplitude, respectively. What's more, the red regions at low strain amplitudes represent $G_t' < 0$ ($G_t'/G_t'' < 0$), which indicates the backflow. It is obvious that the Laponite suspension and CNF suspension present the most violent structural transformation and the most complex rheological behavior. Meanwhile, xanthan gum solution, ketchup, and yogurt show gradual changes in the rheological response.

In each second panel of Figs. 37 and 38, the $\delta_t$ value close to zero indicates a nearly perfectly elastic response. The phase angle around $\pi/2$ represents a predominantly liquidlike response. In addition, the peak of $d\delta_t/dt$ is close to the intracycle yielding behavior. From each second panel of Figs. 37 and 38, the first appearance of $\arctan(G_t''/G_t') = \pi/4$ ($G_t' = G_t''$) for each of the six samples can be determined, which is denoted in the corresponding panel. Meanwhile, the $d\delta_t/dt$ values of the six samples are clearly demonstrated in each third panel of Figs. 37 and 38. Furthermore, the peaks of $\delta_t$ and $d\delta_t/dt$ are closely related to the intracycle yielding position. Figures 37(c.ii) and 37(c.iii) both demonstrate the unique rheological properties of CNF suspension. In addition, figures 38(b.ii) and 38(b.iii) depict the negligible structural transformation of the reflective silver paint during the stress sweep test.

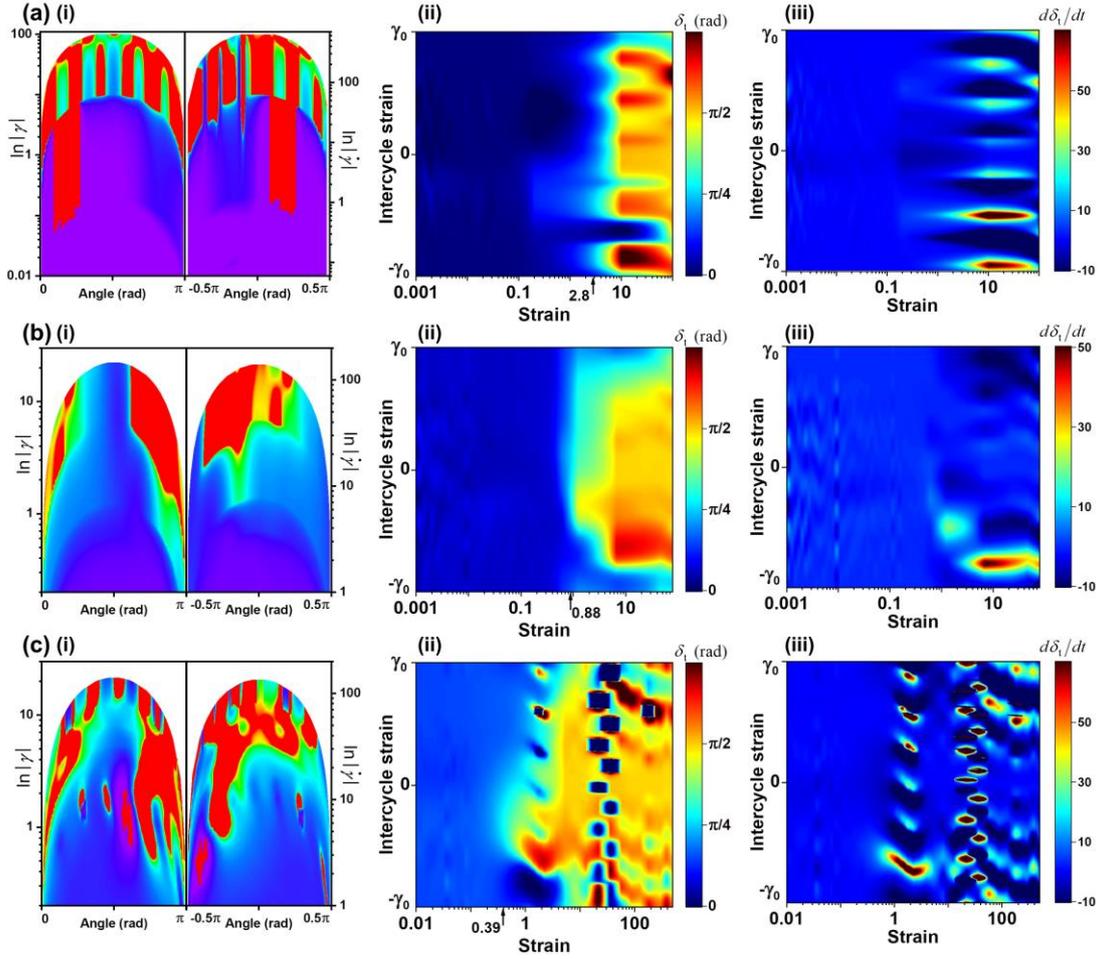

FIG. 37 Transient-moduli-related methods in SPP to treat (a) Laponite suspension, (b) xanthan gum solution, and (c) CNF suspension: (i) $\ln|\gamma|$ (intracycle strain) and $\ln|\dot{\gamma}|$ (intracycle strain rate) versus angle ($\theta$, $\gamma(t) = \gamma_0 \sin\theta$) to show the $G_t'/G_t''$ ratio with the color mapping; (ii) contour plot of phase angle $\delta_t$; (iii) contour plot of the phase angle velocity $d\delta_t/dt$. The denoted strain values represent the first appearance of $\delta_t = \pi/4$ ($G_t' = G_t''$).

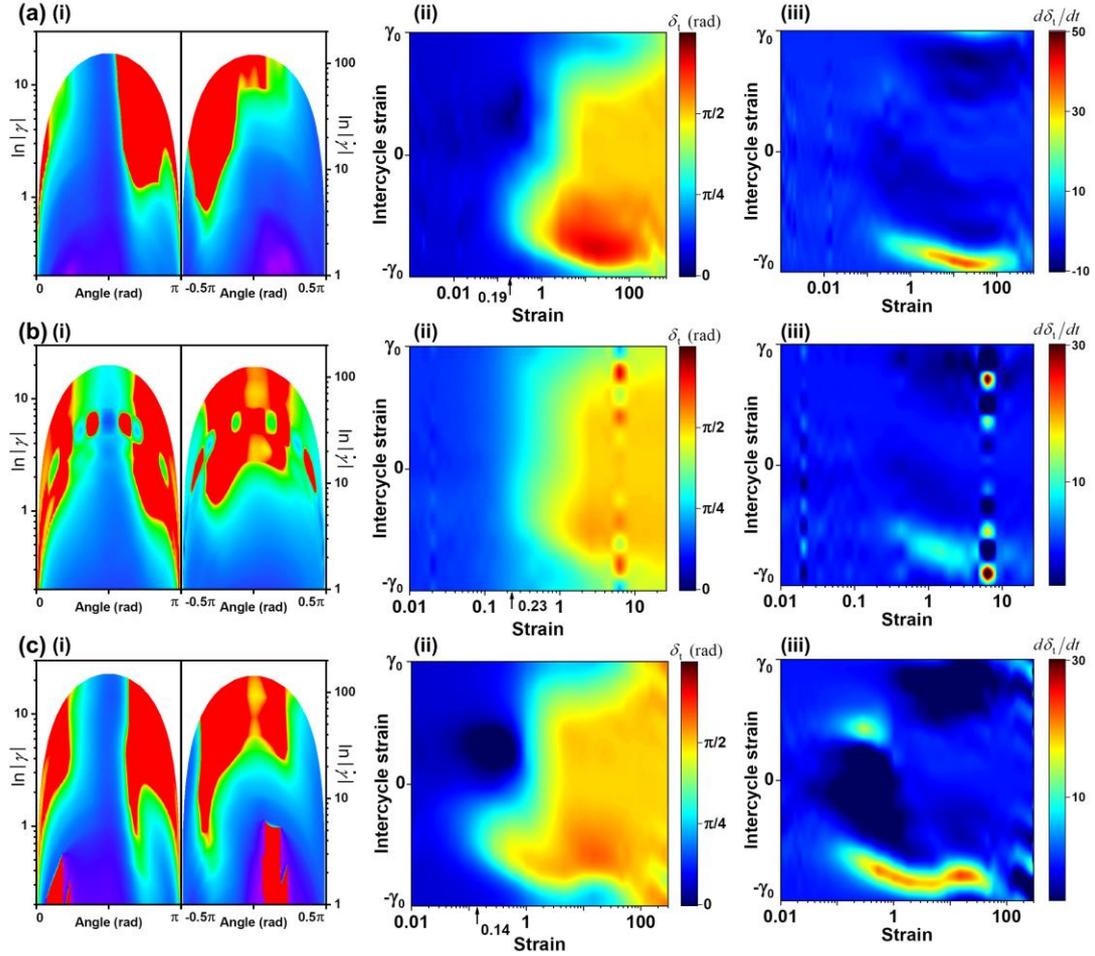

FIG. 38 Transient-moduli-related methods in SPP to treat (a) ketchup; (b) reflective silver paint; and (c) yogurt: (i) $\ln|\gamma|$ and $\ln|\dot{\gamma}|$ versus angle showing the $G_t'/G_t''$ ratio; (ii) contour plot of the phase angle $\delta_t$; (iii) contour plot of phase angle velocity $d\delta_t/dt$.

**APPENDIX K: ANALYTIC FLOW CURVE**

The raw (blue points) and reconstructed (red curves) flow curves from LAOS based on the six samples are illustrated in Fig. 39. Superpositions between the raw and reconstructed results are shown. A higher maximum stress/strain rate results in a lower minimum stress/strain rate, which may be originated from the significant structural failure under a high stress/strain rate along with a short time for the structural recovery.

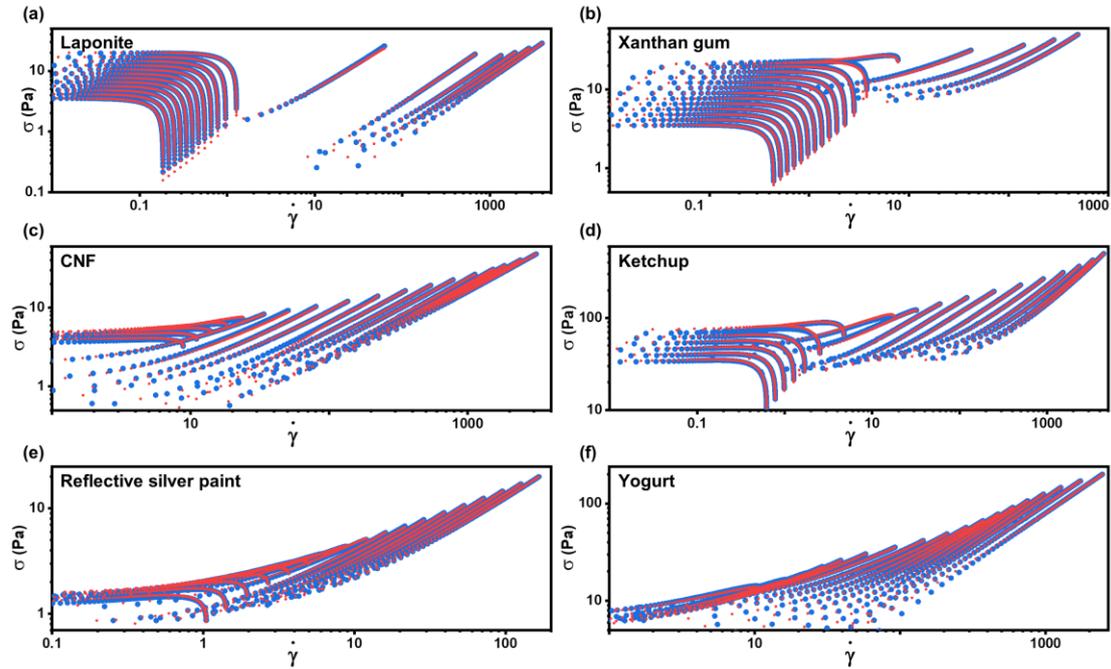

FIG. 39 Comparison between the raw (blue points) and the reconstructed (red points) flow curves: (a) Laponite suspension; (b) xanthan gum solution; (c) CNF suspension; (d) ketchup; (e) reflective silver paint; and (f) yogurt. Introducing 1st ~ 7th harmonics.

**TABLE I.** The deduced equations from the use of aLAOS approach in different LAOS methods and the demonstrated positions of these equations, where the contributions of even harmonics are shown and proved to be negligible.

| aLAOS in different methods | Deduced equation |
| --- | --- |
| Lissajous curve | Eqs. 11 and 12 |
| Stress decomposition | Eqs. 32 and 33 |
| Strain-stiffening and shear-thickening ratios | Eqs. 52 and 53 |
| Stress bifurcation | Eqs. 65, 66, and 75 ~ 79 |
| Dissipation ratio | Eqs. 81, 82, and 83 |
| Transient modulus | Eqs. 91 ~ 94 |
| The derivative of transient modulus | Eqs. 97 ~ 100 |
| Apparent cage modulus | Eqs. 103 ~ 110 |
| Other transient-moduli-based measures | Eqs. 111 and 112 |
| LAOS-based flow curve | Eqs. 113 and 114 |

**NOMENCLATURE**

| Parameter | Interpretation |
|:---:|:---:|
| $a_n$ | The Fourier coefficient of $n$th harmonic |
| $a'_n$ | The Chebyshev coefficient of power series |
| $b_n$ | The Fourier coefficient of $n$th harmonic |
| $b'_n$ | The Chebyshev coefficient of power series |
| $c'_n$ | The Chebyshev coefficient of power series |
| $D_h$ | The horizonal difference between the maximum stress and maximum strain |
| $D_v$ | The vertical difference between the maximum stress and maximum strain |
| $E_d$ | The energy dissipated per unit volume per cycle |
| $e_n$ | Elastic Chebyshev coefficient |
| $G_{cage}$ | The modulus of YSFs at zero stress, Pa |
| $G'$ | Elastic modulus, Pa |
| $G''$ | Loss modulus, Pa |
| $G'_L$ | Large-strain modulus |
| $G'_M$ | Minimum-strain modulus |
| $G'_n$ | $n$th harmonic elastic modulus, Pa |
| $G''_n$ | $n$th harmonic loss modulus, Pa |
| $G'_t(t)$ | Transient elastic modulus, Pa |
| $G''_t(t)$ | Transient viscous modulus, Pa |
| $G_{\bar{\gamma}}$ | The slope of stress-mean strain curve, Pa |
| $G_{\bar{\sigma}}$ | The slope of mean stress-strain curve, Pa |
| $I_n$ | The intensity of $n$th harmonic, Pa |
| $K$ | Consistency coefficient, $Pa \cdot s^n$ |
| $n$ | Non-Newtonian index |
| $S$ | Strain-stiffening ratio |
| $T$ | Shear-thickening ratio |
| $t$ | Time, s |
| $v_n$ | Viscous Chebyshev coefficient |
| $\delta_t$ | The phase angle involving $G'_t(t)$ and $G''_t(t)$ |
| $\gamma$ | Strain |
| $\bar{\gamma}$ | Mean strain |
| $\dot{\gamma}$ | Strain rate, s$^{-1}$ |
| $\bar{\dot{\gamma}}$ | Mean strain rate, s$^{-1}$ |
| $\gamma_0$ | Strain amplitude of ASB |
| $\gamma_{max}$ | Strain amplitude of stress bifurcation |
| $\dot{\gamma}_{max}$ | Strain rate amplitude, s$^{-1}$ |
| $\gamma_{1,s}$ | Start yield strain |
| $\dot{\gamma}_{1,e}$ | End yield strain rate, s$^{-1}$ |
| $(\gamma_0, \sigma_{\gamma 0})$ | The coordinates of the maximum strain in algebraic elastic Lissajous curve |
| $(\gamma_{\sigma max}, \sigma_{max})$ | The coordinates of the maximum stress in algebraic elastic Lissajous curve |
| $(\dot{\gamma}_0, \sigma_{\dot{\gamma} 0})$ | The coordinates of the maximum strain rate in algebraic viscous Lissajous curve |
| $(\dot{\gamma}_{\sigma max}, \sigma_{max})$ | The coordinates of the maximum stress in algebraic viscous Lissajous curve |
| $(\gamma_{max}, \bar{\sigma}_{max})$ | End point of the mean stress–strain curve of the elastic Lissajous plots |
| $(\bar{\gamma}_{max}, \sigma_{max})$ | End point of the stress-mean strain curve of the elastic Lissajous plots |
| $(\bar{\dot{\gamma}}_{max}, \sigma_{max})$ | End point of the stress-mean strain rate curve of the viscous Lissajous plots |
| $(\dot{\gamma}_{max}, \bar{\sigma}_{max})$ | End point of the mean stress-strain rate curve of the viscous Lissajous plots |
| $\eta'_L$ | Large-rate dynamic viscosity |
| $\eta'_M$ | Minimum-rate dynamic viscosity |
| $\eta_{\bar{\gamma}}$ | The slope of stress-mean strain rate curve, Pa s |
| $\eta_{\bar{\sigma}}$ | The slope of mean stress-strain rate curve, Pa s |
| $\sigma$ | Stress, Pa |
| $\sigma_{0,H}$ | Yield stress from the HB model of steady shear, Pa |

| Symbol | Description |
|---|---|
| $\sigma_c$ | Yield stress from creep method, Pa |
| $\sigma_{d,c}$ | Yield stress from the cross over point of $G'$ and $G''$, Pa |
| $\sigma_{d,e}$ | Yield stress from elastic stress method, Pa |
| $\sigma_{d,p}$ | Yield stress from the intersection of two power-law extrapolations at low and high stress ranges in $G'$ vs. stress amplitude, Pa |
| $\sigma_y$ | Yield stress, Pa |
| $\sigma_{max}$ | Stress amplitude, Pa |
| $\sigma_{l,s}$ | Start yield stress, Pa |
| $\sigma_{l,e}$ | End yield stress, Pa |
| $\sigma_{max} \sim \gamma_{\sigma max}$ | Stress amplitude–strain at maximum stress curve of ASB |
| $\sigma_{\gamma 0} \sim \gamma_0$ | Stress at maximum strain–strain amplitude curve of ASB |
| $\sigma_{max} \sim \dot{\gamma}_{\sigma max}$ | Stress amplitude–strain at maximum stress rate curve of ASB |
| $\sigma_{\dot{\gamma} 0} \sim \dot{\gamma}_0$ | Stress at maximum strain rate–strain rate amplitude curve of ASB |
| $\sigma_{max} \sim \bar{\gamma}_{max}$ | Stress amplitude–maximum of mean strain curve of stress bifurcation |
| $\bar{\sigma}_{max} \sim \gamma_{max}$ | Maximum of mean stress–strain amplitude curve of stress bifurcation |
| $\sigma_{max} \sim \bar{\dot{\gamma}}_{max}$ | Stress amplitude–maximum of mean strain rate curve of stress bifurcation |
| $\bar{\sigma}_{max} \sim \dot{\gamma}_{max}$ | Maximum of mean stress–strain rate amplitude curve of stress bifurcation |
| $\sigma'$ | Elastic shear stress, Pa |
| $\sigma''$ | Viscous shear stress, Pa |
| $\phi$ | Dissipation ratio |
| $\phi_n$ | The phase angle of $n$th harmonic, ° |
| $\omega$ | Angular frequency, rad/s |